# GREAT OBSERVATORIES

## the PAST & FUTURE of PANCHROMATIC ASTROPHYSICS

*a report by the*
NASA GREAT OBSERVATORIES SCIENCE ANALYSIS GROUP

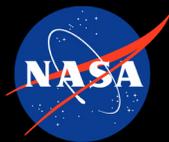

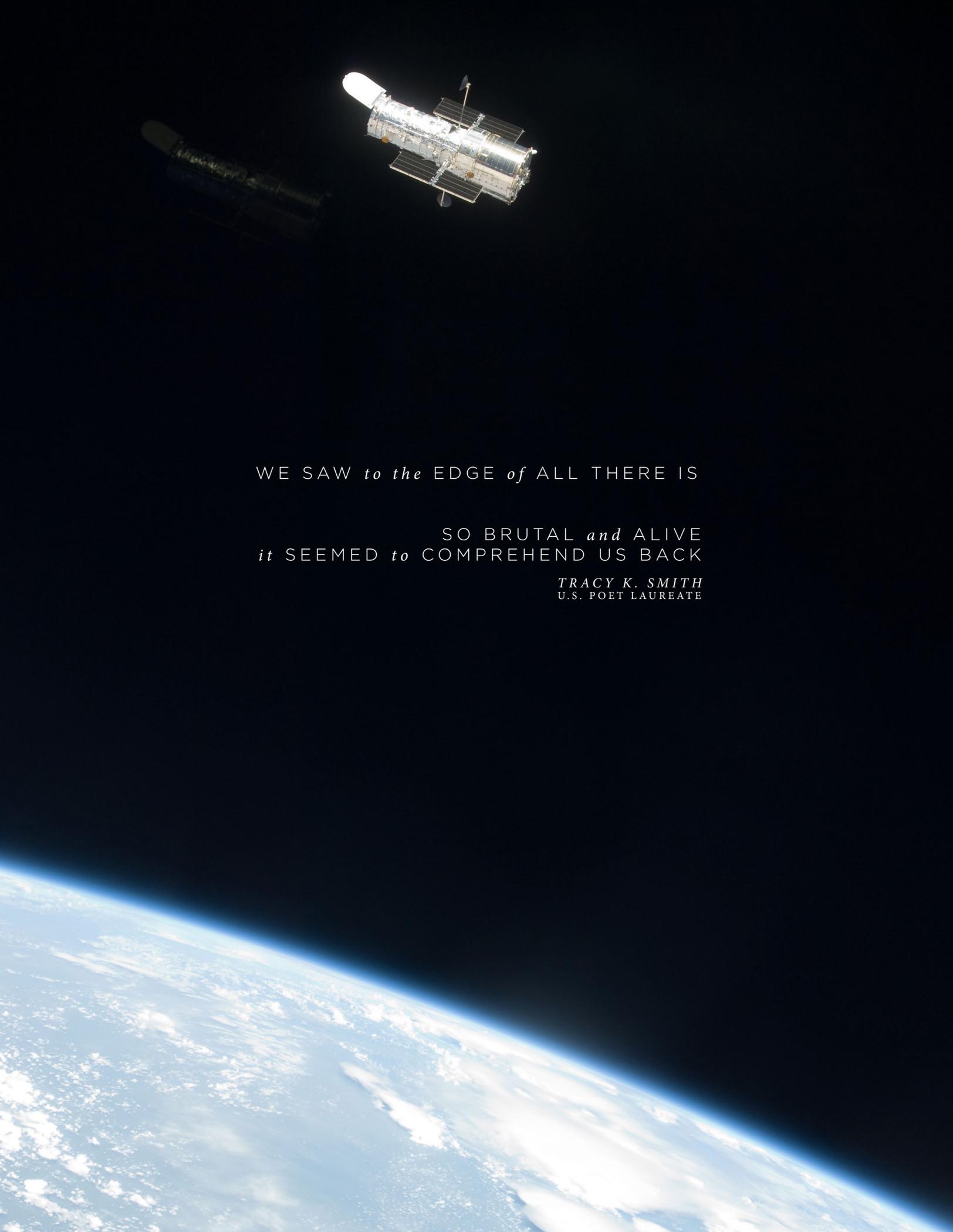

WE SAW *to the* EDGE *of* ALL THERE IS

SO BRUTAL *and* ALIVE
*it* SEEMED *to* COMPREHEND US BACK

*TRACY K. SMITH*
U.S. POET LAUREATE

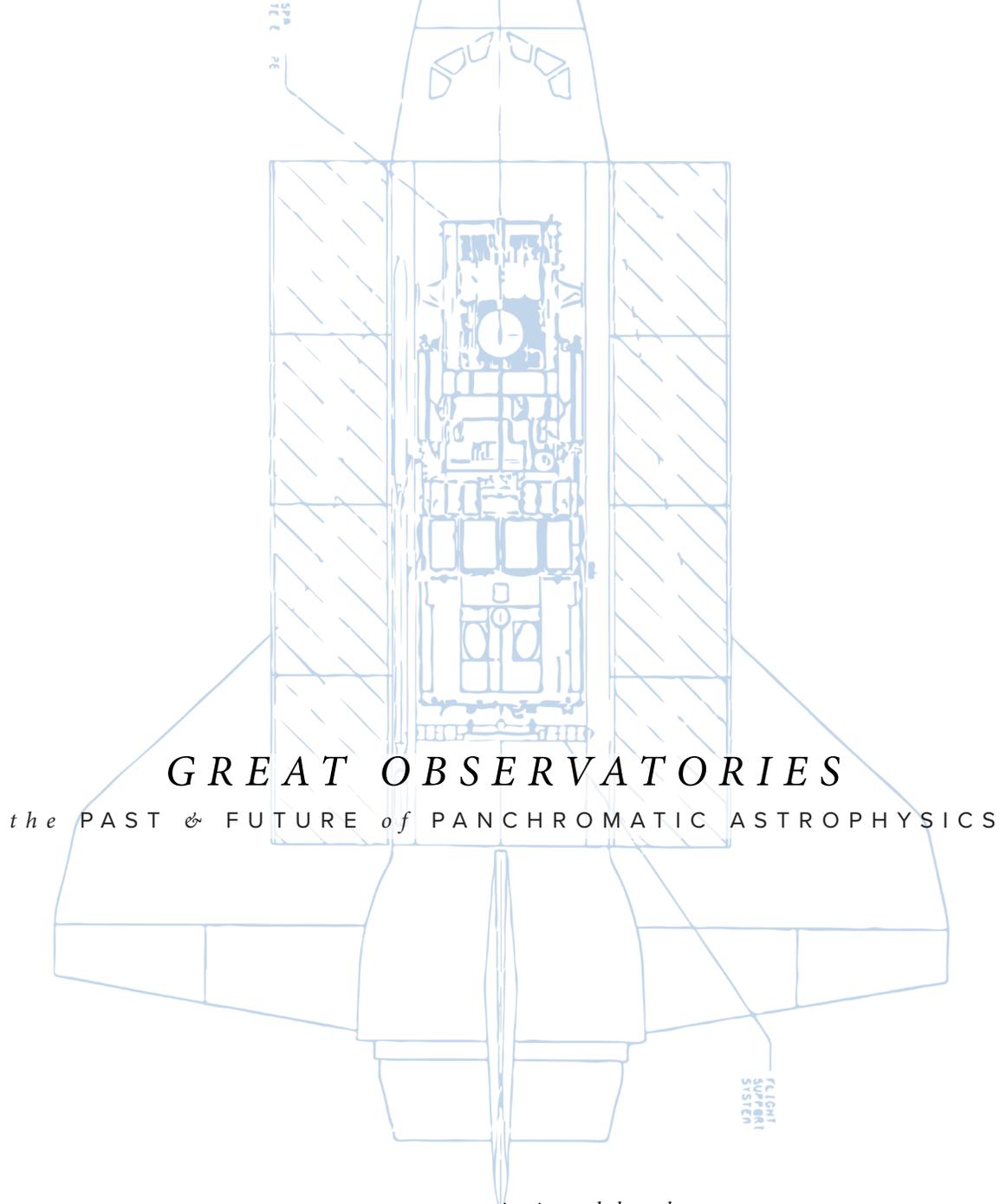

# GREAT OBSERVATORIES
*the* PAST *&* FUTURE *of* PANCHROMATIC ASTROPHYSICS

*a report commissioned by the*

COSMIC ORIGINS, PHYSICS *of the* COSMOS, *and* EXOPLANET EXPLORATION
*PROGRAM ANALYSIS GROUPS*

*for the*
NATIONAL AERONAUTICS AND SPACE ADMINISTRATION

NOVEMBER 2020

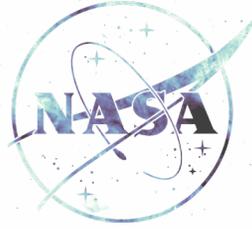

GREAT OBSERVATORIES SCIENCE ANALYSIS GROUP (SAG-10) CO-CHAIRS

| | |
|---|---|
| L. ARMUS | IPAC, California Institute *of* Technology |
| T. MEGEATH | University *of* Toledo |

REPORT CO-AUTHORS *&* SAG-10 MEMBERSHIP

| | |
|---|---|
| L. CORRALES | University *of* Michigan, *Galactic Processes and Stellar Evolution* **co-chair** |
| M. MARENGO | Iowa State University, *Galactic Processes and Stellar Evolution* **co-chair** |
| A. KIRKPATRICK | University of Kansas, *Astrophysics of Galaxy Evolution* **co-chair** |
| J. D. SMITH | University of Toledo, *Astrophysics of Galaxy Evolution* **co-chair** |
| M. MEYER | University *of* Michigan, *Origin of Life & Planets* **co-chair** |
| S. GEZARI | University *of* Maryland, *Fundamental Physics* **co-chair** |
| R. P. KRAFT | Center *for* Astrophysics \| Harvard *&* Smithsonian, *Fundamental Physics* **co-chair** |
| S. McCANDLISS | Johns Hopkins University, *Capabilities, Facilities & Options* **co-chair** |
| S. TUTTLE | University of Washington, *Capabilities, Facilities & Options* **co-chair** |
| M. ELVIS | Center *for* Astrophysics \| Harvard *&* Smithsonian, *Capabilities Facilities & Options* **co-chair** |
| M. BENTZ | Georgia State University |
| B. BINDER | Cal Poly Pomona |
| F. CIVANO | Center *for* Astrophysics \| Harvard *&* Smithsonian |
| D. DRAGOMIR | Massachusetts Institute *of* Technology Kavli Institute, University *of* New Mexico |
| C. ESPAILLAT | Boston University |
| S. FINKELSTEIN | University *of* Texas *at* Austin |
| D. B. FOX | Penn State University |
| M. GREENHOUSE | NASA Goddard Space Flight Center |
| E. HAMDEN | University *of* Arizona |
| J. KAUFFMANN | Haystack Observatory, Massachusetts Institute *of* Technology |
| G. KHULLAR | University *of* Chicago |
| J. LAZIO | Jet Propulsion Laboratory, California Institute *of* Technology |
| J. LEE | IPAC, California Institute *of* Technology |
| C. LILLIE | Lille consulting LLC |
| P. LIGHTSEY | Ball Aerospace |
| R. MUSHOTZKY | University *of* Maryland |
| C. SCARLATA | University *of* Minnesota |
| P. SCOWEN | Arizona State University |
| G. R. TREMBLAY | Center *for* Astrophysics \| Harvard *&* Smithsonian |
| Q. D. WANG | University *of* Massachusetts |
| S. WOLK | Center *for* Astrophysics \| Harvard *&* Smithsonian |

# TABLE of CONTENTS

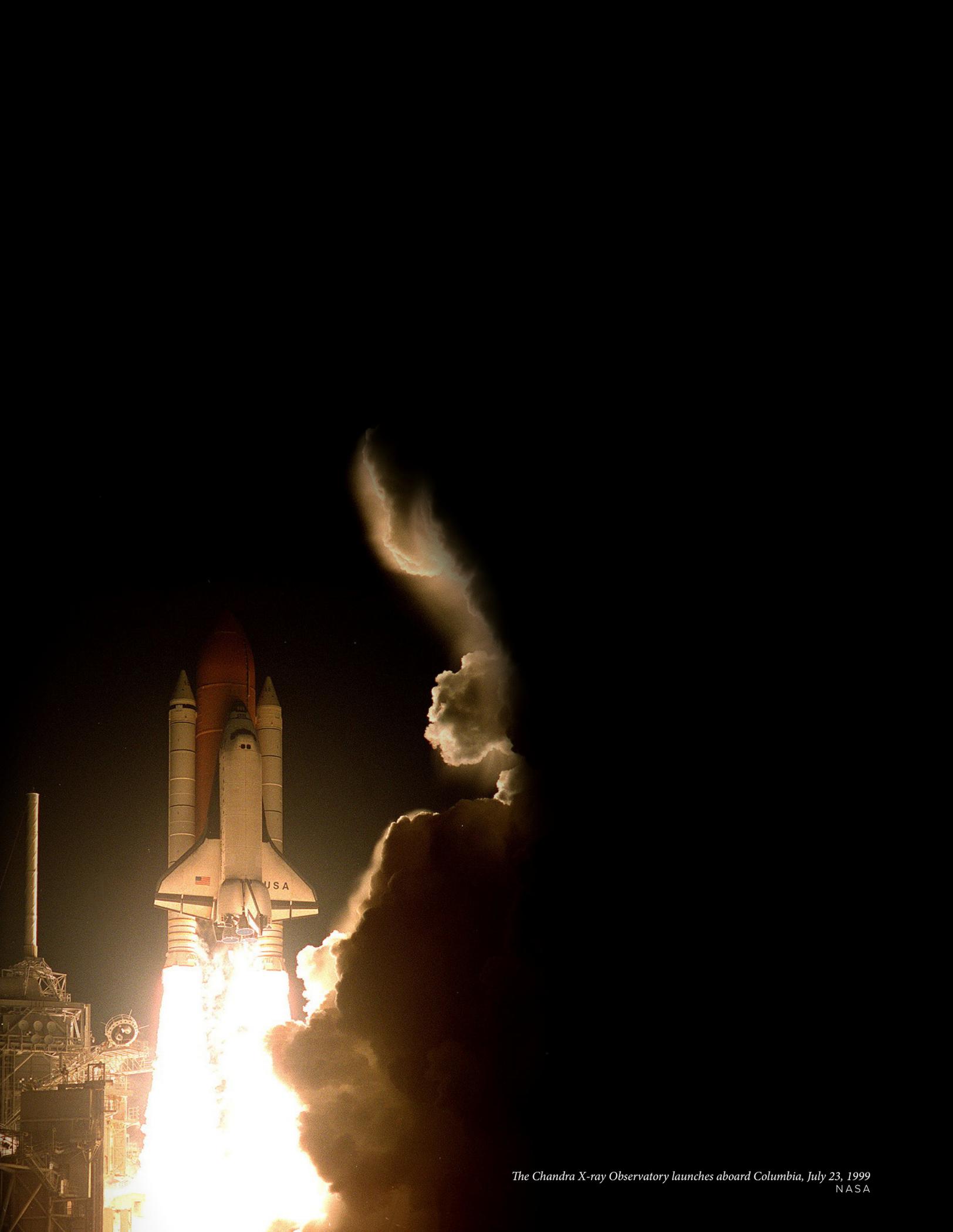

*The Chandra X-ray Observatory launches aboard Columbia, July 23, 1999*
NASA

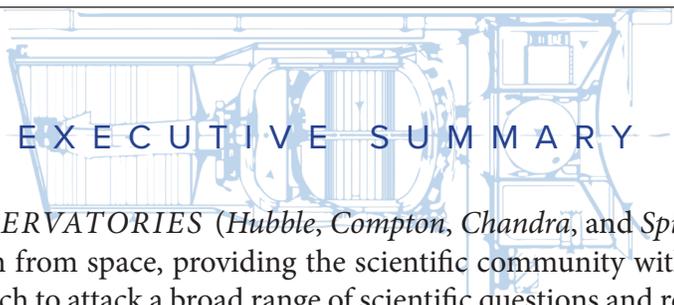

# EXECUTIVE SUMMARY

NASA's *GREAT OBSERVATORIES* (*Hubble, Compton, Chandra,* and *Spitzer*) opened up the electromagnetic spectrum from space, providing the scientific community with a flexible and powerful suite of telescopes with which to attack a broad range of scientific questions and react to a rapidly changing scientific landscape. The Great Observatories, and the missions that followed, established community access to wavelength bands that are either inaccessible or highly compromised from the ground. The achievement of a panchromatic view of the sky led to the current Golden Age of astronomy, in which individual observatories are utilized as a part of a system providing essential access to the Sub-mm, IR, Visual, UV, X-ray and Gamma-ray wavelength regimes.

This report analyzes the importance of multi-wavelength observations from space during the epoch of the Great Observatories, providing examples that span a broad range of astrophysical investigations organized into four areas: Galactic Processes and Stellar Evolution, Astrophysics of Galaxy Evolution, Origin of Life and Planets, and Fundamental Physics. In each area, this report also discusses key questions for the next two decades that demand multi-wavelength measurements from space, providing a summary of the capabilities required in each area. Examples of the panchromatic science enabled by the Great Observatories, and the key future questions that require similar capabilities, are listed here.

### EXAMPLES *of* PANCHROMATIC SYNERGY *enabled by the* GREAT OBSERVATORIES

- First detection and characterization of exoplanet atmospheres
- Detection of multiple planetesimal belts around nearby stars
- Characterization of the composition of primordial, planet forming disks
- Detection of dust formation in SN 1987a
- Star Formation laws in molecular clouds.
- Characterization of young clusters and sites of massive star formation in our Galaxy and the LMC
- Discovery of suppressed star formation in the Milky Way's central molecular zone and detection of activity around Sgr A*
- Discovery of a main sequence for star forming galaxies across cosmic time
- Discovery of the co-evolution of black holes and stellar bulges in galaxies
- The characterization of AGN feedback in massive galaxies and clusters
- The first detection of z~10 galaxies
- Discovery and characterization of galaxies in the epoch of re-ionization
- Placement of strong constraints on the properties of dark matter and the dark energy equation of state
- Characterization of neutron star mergers and confirmation of kilonovae.
- Establishment of the current tension between SN1a and CMB derived Hubble Constant
- Determination of the nature of high energy transients

Working together, the Great Observatories enabled unique science and fueled a rapid pace of discovery and understanding by establishing commensurate and concurrent capabilities across the electromagnetic spectrum. Through combinations of sensitivity, angular resolution, mapping speed and spectral resolution, the Great Observatories collectively studied an exceptionally broad range of phenomena, much of which was not even envisioned at the time of their launch. The ability to observe phenomena at multiple wavelengths concurrently and sample different temperature regimes led to the rapid development and validation of as-trophysical models. Time varying phenomena such as supernovae, young star outbursts, gamma ray bursts, and the first signals from a gravitational wave event, were studied across the electromagnetic spectrum. The pace and adaptability of the science done with the Great Observatories was further enabled by well funded General Observer programs that ensured that the community could respond quickly to new discoveries and pursue fresh areas of investigation.

> **FUTURE QUESTIONS *that* REQUIRE PANCHROMATIC CAPABILITIES**
>
> - Can we find evidence for organics and biosignatures in the atmospheres of Earth-like exoplanets?
> - How do planetary systems form from protoplanetary disks and create habitable worlds?
> - How does star formation and the initial mass function depend on environment?
> - How do massive binaries evolve, drive stellar evolution and interact with their environments?
> - How much of a star's mass is accreted during episodic outbursts?
> - What accelerates cosmic rays?
> - What drives turbulence in the interstellar medium?
> - Where does most of the interstellar dust form and how do its properties vary with environment?
> - How do stars form in the early universe?
> - How was the universe re-ionized?
> - How did the first black holes form and how do they co-evolve with galaxies?
> - What are the electromagnetic counterparts to gravitational wave sources?
> - What new insights will be revealed about the physics of compact objects and the structure of the Universe from transient observations?
> - What is the nature of dark matter?
> - How has dark energy evolved over cosmic time?

As the existing Great Observatories age, or are decommissioned, access to the electromagnetic spectrum from space is diminishing, with an accompanying loss of scientific capability, and the potential to significantly impede progress in astrophysics. *Spitzer* will be decommissioned in January 2020, with the *James Webb Space Telescope* (*JWST*) bringing new capabilities to the near and mid-IR. *Compton* was decommissioned in 2000 and was partially replaced by *Fermi* in 2008, which itself is past its designed lifetime. *Chandra* and *Hubble* are 20 and 29 years old, respectively, with *Hubble* having been serviced five times, the last in March 2009. The future performance and lifespan of these observatories is unclear. Upcoming or approved space-based facilities will only partially fill the impending wavelength gaps, leaving in place a significant loss of scientific capability, slowing our ability to further develop astrophysical models, and eroding the expertise needed to develop technologies for future missions. Newly discovered phenomena may have to wait decades for observations in critical wavelength regimes. Time variable astronomical events could lack coverage in crucial spectral regimes.

However, there is an opportunity to learn from the success of the Great Observatories, and use emerging technologies to expand access to the electromagnetic spectrum from space to tackle some of the most pressing astrophysical questions of the next decade. The Great Observatories spanned nearly an order of magnitude in cost, yet they functioned together to redefine astrophysics, primarily because they provided the astronomical community with a concurrent set of powerful space telescopes with highly commensurate capabilities that spanned rich regions of the electromagnetic spectrum. The growing archives from the Great Observatories also facilitated novel, panchromatic science, providing direct and contextual data across the electromagnetic spectrum for newly discovered objects and phenomena and setting the baseline for time variable domain astronomy. Similarly, smaller missions successfully supported studies with the Great Observatories, producing science gains through targeted observations in key wavelength bands or through large area surveys of the sky, using newly developed technologies and observing techniques.

Within the current budget envelope, innovative technologies and strategic mixes of flagship and Probe-scale missions, with robust general observer programs, can continue to be used to effectively maintain a similar level of panchromatic coverage. Maintaining concurrent panchromatic capabilities across multiple flagship observatories, however, requires mission longevity. The Great Observatories demonstrated that missions could be operated effectively over multi-decadal timespans. In the case of *Hubble*, they also showed that servicing could be used to maintain, and upgrade, capabilities. Servicing or in-orbit construction may be viable routes for establishing long-term, panchromatic coverage. Finally, it is clear that strong international partnerships will continue to play a vital role in the design, operations and ultimate scientific success of future space observatories, providing key contributions that enhance capabilities and enable access for the US and international astronomical community.

The scientific legacy of the Great Observatories has demonstrated the importance of sensitive, panchromatic observations for progress in astrophysics, as well as the ability of NASA and its partners to provide concurrent and sustained access to a large part of the electromagnetic spectrum from space. The Great Observatories became a deliberate NASA agency program that transcended individual missions and wavelength regimes. This legacy points the way to a future where panchromatic capabilities are not just maintained but enhanced, and the remarkable growth in our understanding of the Universe continues through the development of the next generation of space observatories that will inspire further giant leaps in astrophysics in the coming decades.

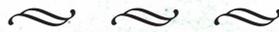

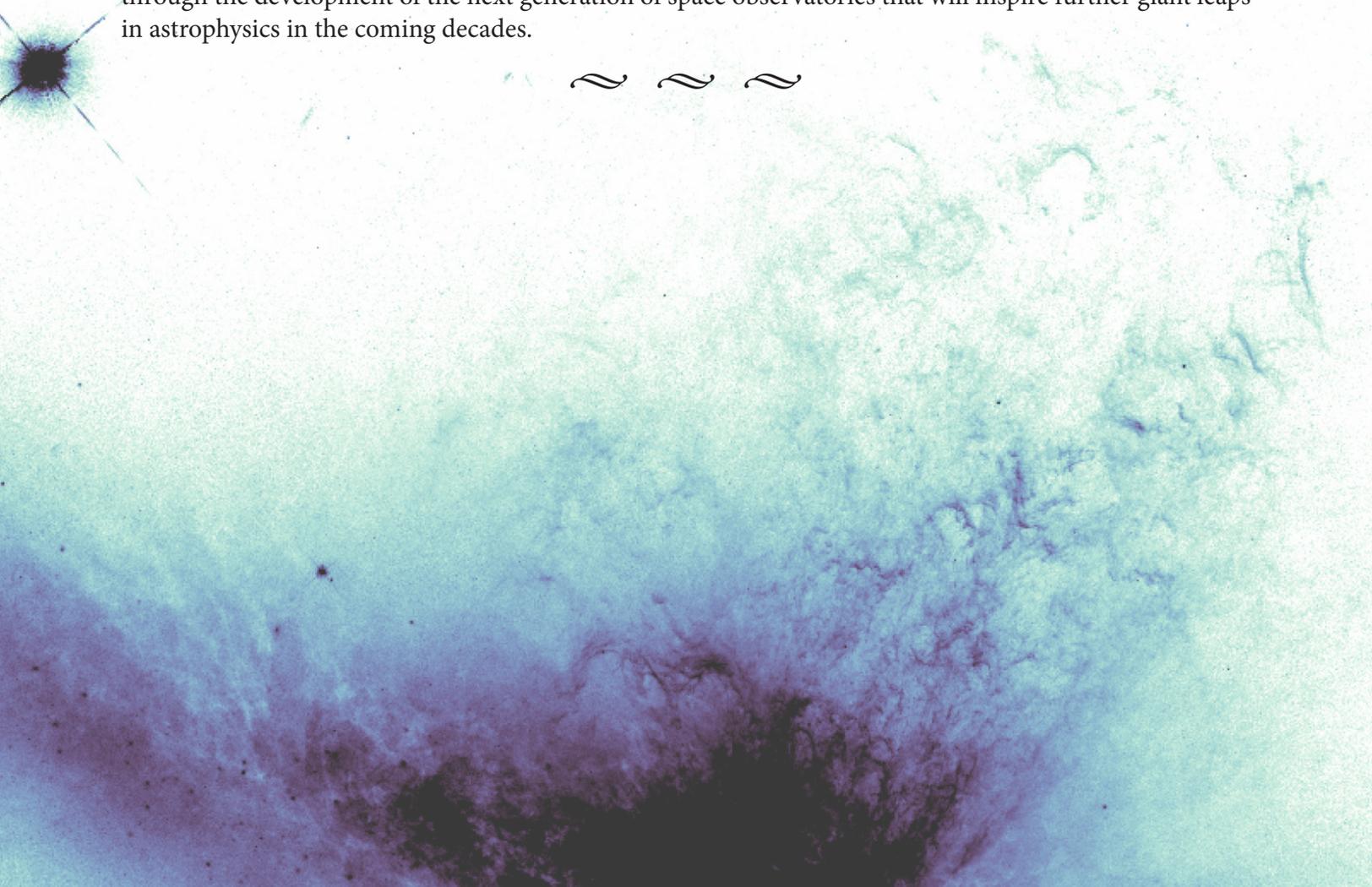

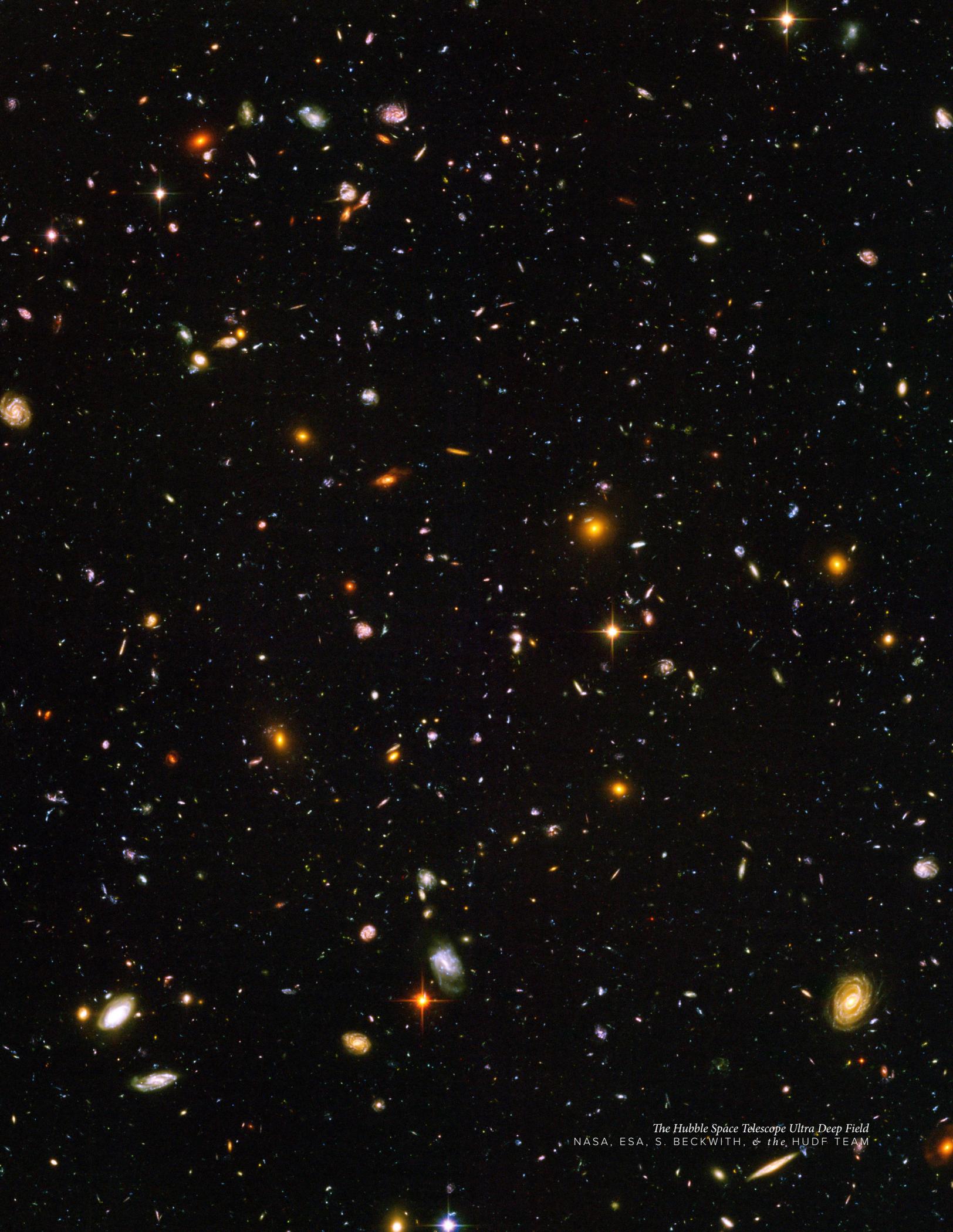

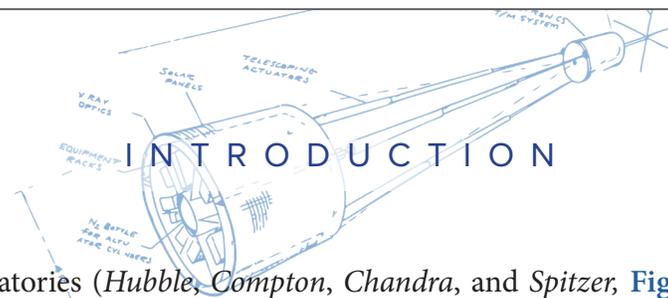

# INTRODUCTION

NASA's Great Observatories (*Hubble, Compton, Chandra,* and *Spitzer,* Fig. 1-1) have opened up the electromagnetic spectrum from space, providing sustained access to wavelengths not visible, or greatly compromised, from the ground due to Earth's atmosphere. The first, *Hubble*, was launched in 1990, and two of the four (*Hubble* and *Chandra*) are still operating today. Each of these observatories delivered large gains in sensitivity, angular resolution, mapping speed and/or spectral coverage. Together, they have provided the scientific community with a flexible and powerful suite of telescopes capable of addressing broad scientific questions, and reacting to a rapidly changing scientific landscape. Through regular peer-reviewed proposal calls open to the community, this has become a central feature of modern astrophysics, where objects are now routinely observed across the electromagnetic spectrum from the ground and space. It has also become the basis upon which multiple generations of students and post-doctoral scholars have built their careers. However, the concept of the Great Observatories was not an inevitable outcome of a system where communities vied and competed for a share of the limited resources available for new missions.

## *the* GREAT OBSERVATORIES

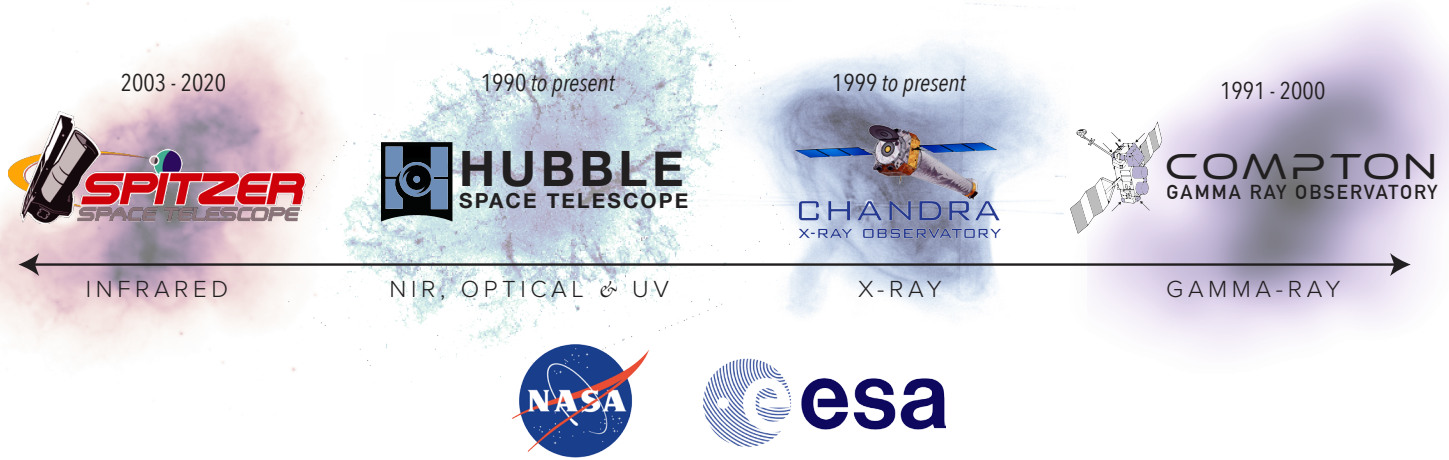

**Fig. 1—1. The Great Observatories.** *Spitzer, Hubble, Chandra,* and *Compton*, arranged according to the part of the electromagnetic spectrum they observe.

The concept of the Great Observatories took shape in the late 1970s as scientists and NASA administrators recognized that fundamental strides in astrophysics required access to the entire electromagnetic spectrum, well beyond what could be accessed from the ground, and any single space observatory could deliver. The article "*The Number of Class A Phenomena Characterizing the Universe*" (Harwit, 1975) served as inspiration first for Frank Martin and later Charlie Pellerin, who succeeded Martin as Astrophysics Division director in 1983 and initiated the study of the Great Observatory concept. By that time, *Hubble* and *Compton* were already approved, and the key issue was how to get support and funding for *AXAF* and *SIRTF* (later *Chandra* and *Spitzer*; both highly ranked by the 1980 Decadal review), which would open up the X-ray and Infrared windows, respectively, so that they could be launched and be operational well before the *HST* and *CGRO* missions were over. The Astrophysics Council, formulated by Pellerin in 1985 and chaired by Harwit, was charged with sketching out a total astrophysics program that would require all four observatories.

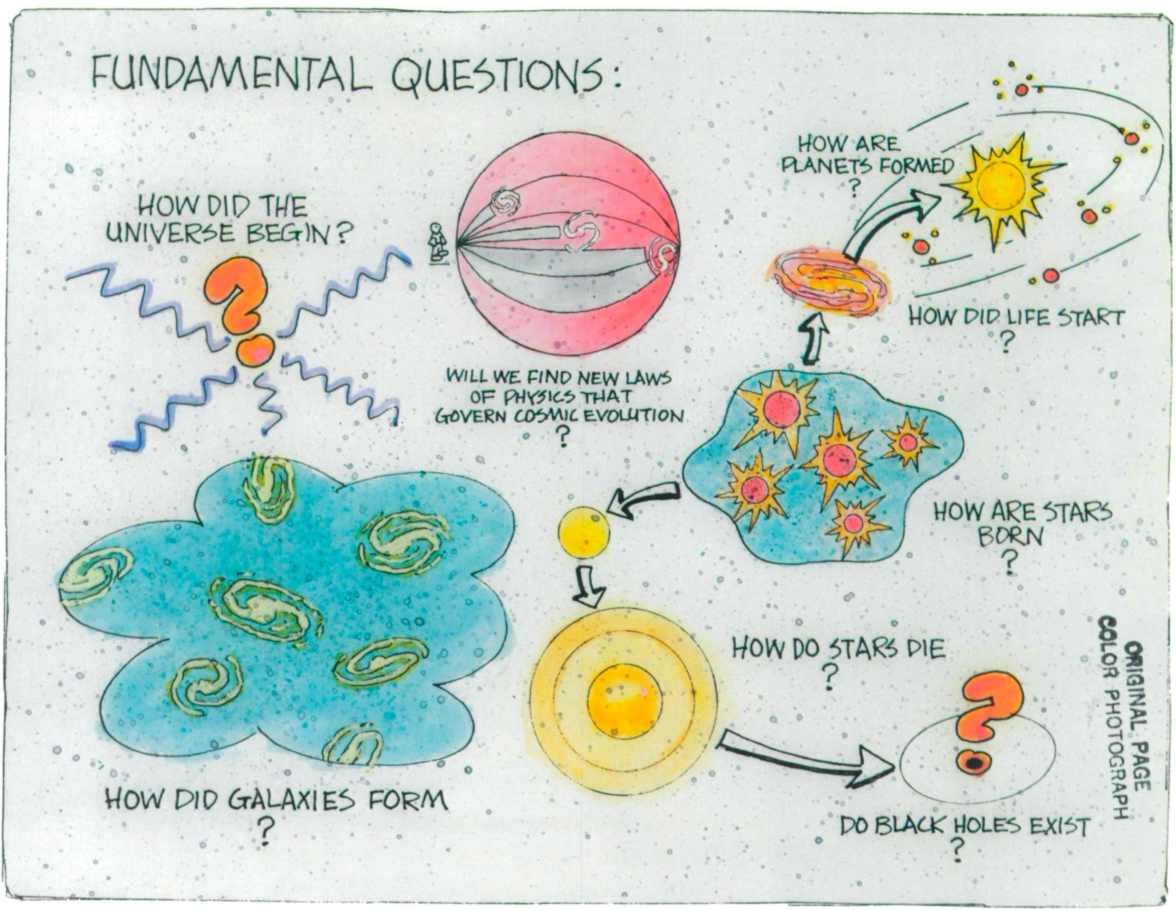

**Fig. 1—2. Fundamental Questions for the Great Observatories.** *Sketch from the brochure "Great Observatories for Space Astrophysics" 1985, prepared under the auspices of the NASA Astrophysics division, Charles J. Pellerin, Jr. Director, by Martin Harwit and Valerie Neal.*

The council produced and released a brochure entitled "*Great Observatories for Space Astrophysics*" that captured the basic concepts and key questions in a highly readable form. In 1986, the Great Observatories Planning Group, led by Harvey Tananbaum, produced a set of slides and a booklet entitled "*New Windows on the Universe: The NASA Great Observatories*" that was used to promote the concept, and gain public and congressional support (Fig. 1-2).

Now, as the existing Great Observatories age or are decommissioned, and the community's access to these wavelengths is diminishing, it is time to consider the lessons of the Great Observatories and how access to the electromagnetic spectrum from space can be continued into the future. This report analyzes the importance of multi-wavelength observations from space, and examines the options available for maintaining panchromatic capabilities in the coming decades. This report is divided into two main sections. The first, highlights examples of the impactful science achieved with the Great Observatories, noting where this has been enhanced through observations with other space and ground-based observatories. We then consider the scientific landscape of the next decade, and discuss areas where panchromatic coverage achieved through space-based observatories is necessary to address key astrophysical problems. This section is divided into four broad categories representing the focus of each of the four SAG-10 science working groups: Galactic Processes and Stellar Evolution, Astrophysics of Galaxy Evolution, Origin of Life and Planets, and Fundamental Physics.

Throughout, we discuss two main paths by which *panchromatic observations* impact astrophysics: by allowing for *concurrent* studies of phenomena in multiple wavelength regimes, and by providing *commensurate* capabilities across the electromagnetic spectrum. The second section, derived from the fifth SAG-10 working group, Capabilities and Facilities, outlines the space landscape for the coming decades as it currently exists, and identifies the gaps in wavelength coverage that are anticipated over the next 10-20 years as current spaced-based observatories age or are decommissioned. We then identify the likely scientific impacts in terms of loss of discovery space, the ability of the community to address key questions, and the flexibility for the community to react to a rapidly evolving scientific landscape. Finally, we examine some options for filling these gaps and achieving pan-chromatic, concurrent coverage of the electromagnetic spectrum from space in the next two decades. These options are not meant to be comprehensive, and a full analysis of their viability and applicability in the next decade extends well beyond the charter for this SAG. However, they are briefly described here to serve as a guideline or menu for the types of investigations that could lead to achieving the kinds of breakthrough science delivered by the Great Observatories. We also note that the analysis here focuses on the capability to cover the electromagnetic spectrum by future observatories. Opportunities for advances in multi-messenger astrophysics are being analyzed in detail by a separate SAG. Finally, this SAG-10 report complements the Astrophysics Roadmap, "*Enduring Quests, Daring Visions*", in that it specifically concentrates on the importance of multi-wavelength observations and panchromatic capabilities for progress in astrophysics.

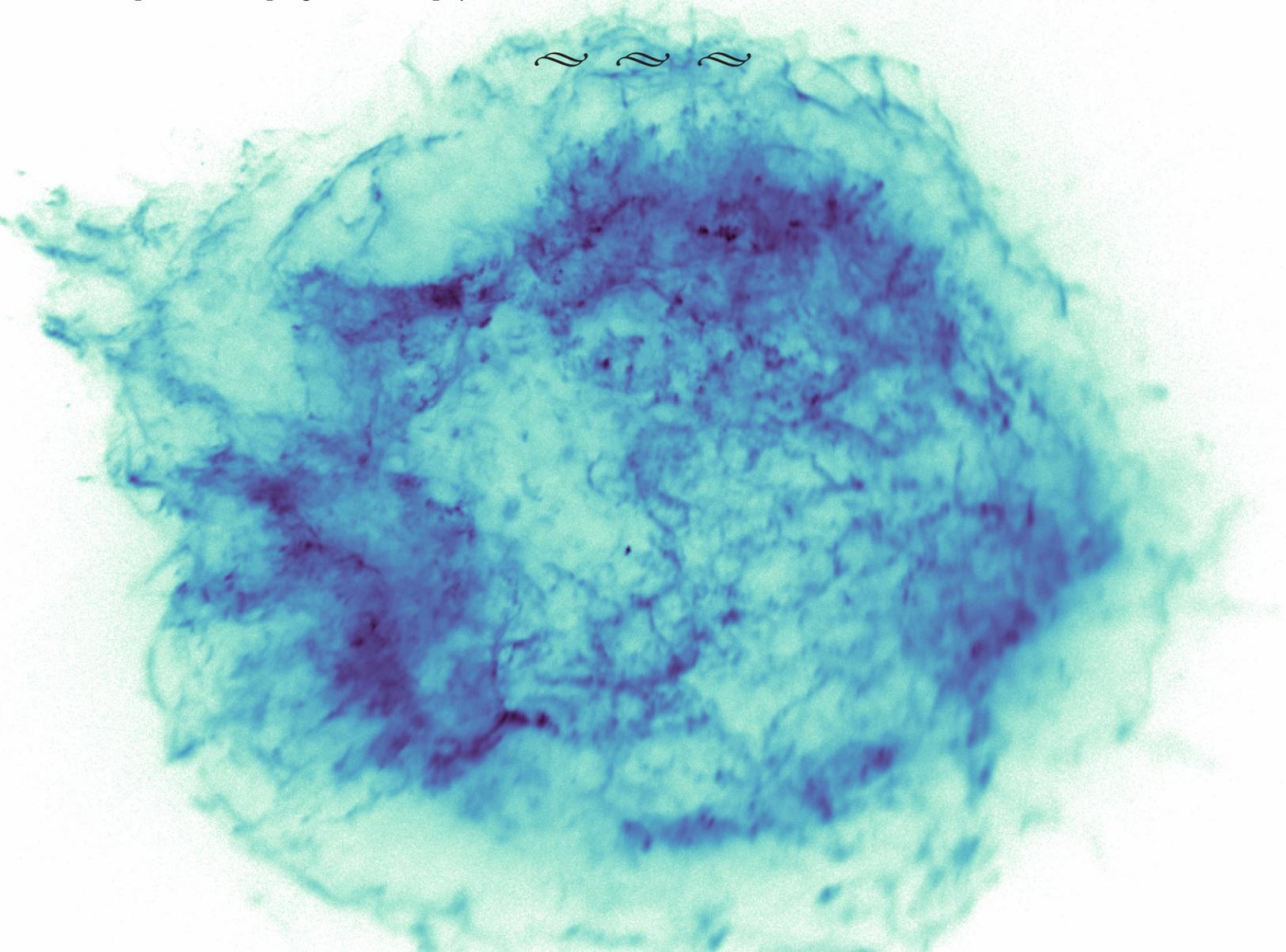

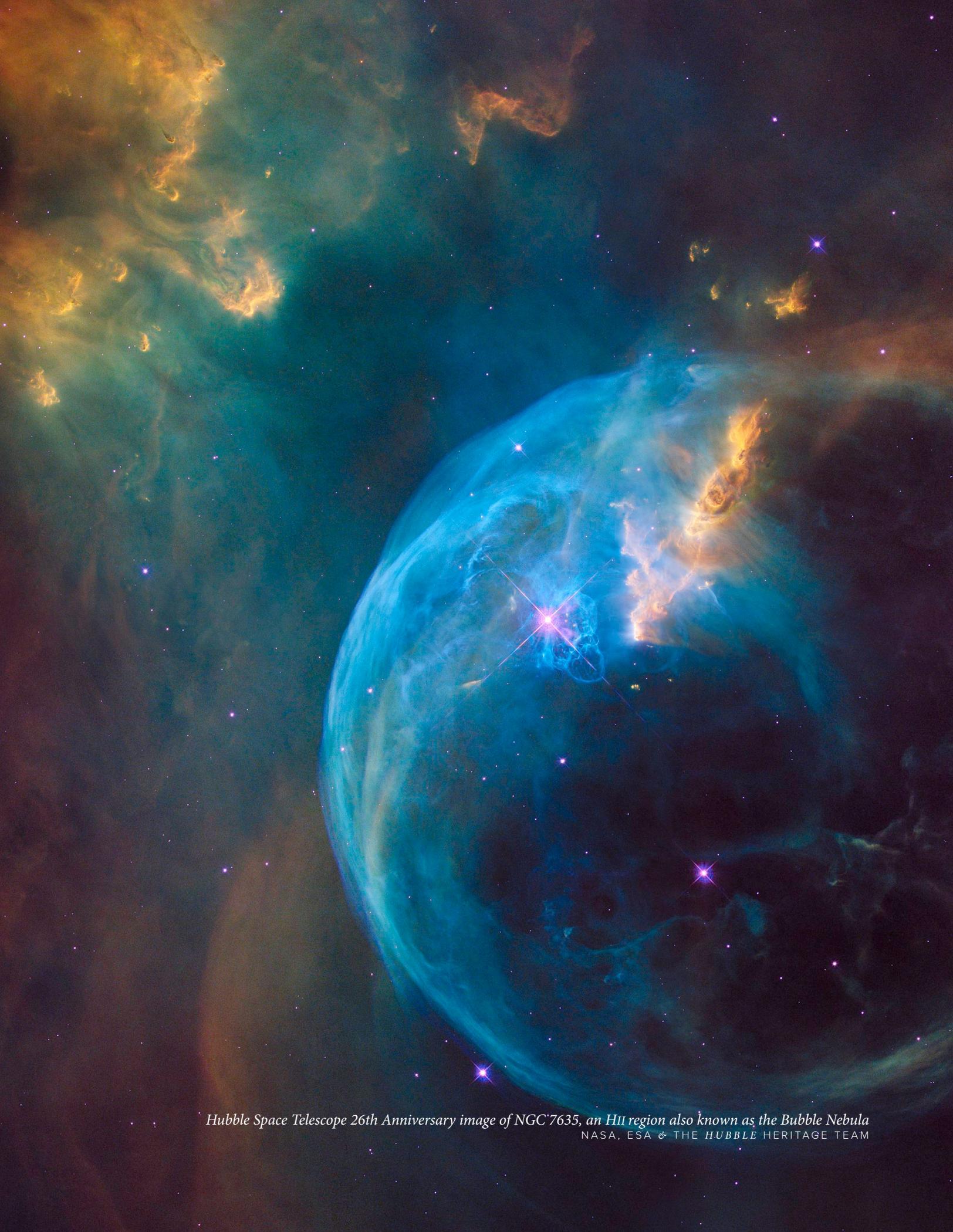

Hubble Space Telescope 26th Anniversary image of NGC 7635, an HII region also known as the Bubble Nebula
NASA, ESA & THE HUBBLE HERITAGE TEAM

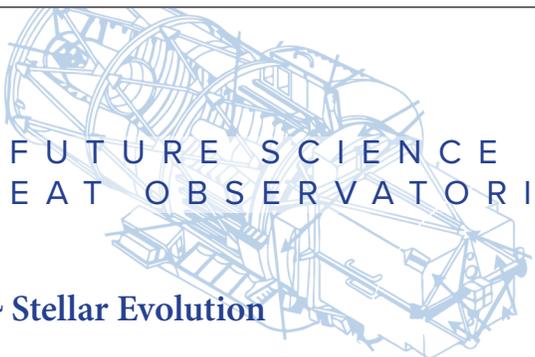

# PAST & FUTURE SCIENCE *with the* GREAT OBSERVATORIES

## 2.1. Galactic Processes *&* Stellar Evolution

Studies of the coupled evolution of stars and ISM in our galaxy and in nearby galaxies supply the detailed physical foundation needed to understand the processes that drive the evolution of galaxies across cosmic time as well as set the initial conditions for the formation of planetary systems. They directly measure the flows of matter and radiation in galaxies and the cycling of baryons between the ISM and stars. Although our understanding of these processes in our own galaxy and those in the local neighborhood is mature compared to our understanding of galaxies in the early universe or the conditions within protoplanetary disks, key problems remain that have broad implications for cosmic evolution. Three guiding principles emerge from these studies. First, environment matters and galactic processes are influenced by whether they occur in a dwarf galaxy, in the outer regions of disk galaxy, or in the center of a large disk galaxy. Second, rapid processes that evolve on human timescales are important, including supernovae, protostellar outbursts, stellar pulsations, and fluctuating X-ray binaries. Finally, as we will explore in this section, multi-wavelength observations are essential for observing the broad range of energetic phenomena found within galaxies.

### 2.1.1 Galactic Processes *&* Stellar Evolution Science enabled *by the* Great Observatories

NASA's Great Observatories have revolutionized our understanding of galactic science, by allowing concurrent observation of stellar processes across the electromagnetic spectrum for the first time. Key to this revolution was the ability to access correlated processes operating over widely different energy ranges, providing a comprehensive view of the physics driving these phenomena. This section presents a few examples illustrating how multi-wavelength observations with the Great Observatories have enabled scientific breakthroughs in galactic science.

#### *SUPERNOVA 1987A: OBSERVING A BLAST WAVE AND THE PRODUCTION OF DUST*

The explosion of SN 1987A in the Large Magellanic Cloud provided astronomers the first opportunity to observe a nearby supernova across the electromagnetic spectrum. Over the first few years, astronomers witnessed the optical evolution of the supernova, X-ray and γ-ray emission from the decay of radioactive materials, UV emission from the material expelled by the supernova's progenitor star (a massive blue supergiant), and rapid dust formation in the supernova ejecta (McCray & Fransson 2016). The detection of neutrinos coinciding with the explosion also makes SN 1987A the earliest example of the multi-messenger, time-domain astrophysics that is sure to mark the next century of astronomy. NASA Great Observatories, coupled with ESA's *Herschel* and *XMM-Newton* observatories, played a leading role in watching the blast wave and emerging ejecta.

***The interaction of previous stellar mass loss with radiation and blast wave*** — *HST* resolved an equatorial ring of previously expelled stellar material, 0.6 ly in radius, and two additional rings, at ±1.3 ly, excited by UV and soft X-rays produced in the supernova explosion. The supernova blast wave collided with the equatorial ring around 5000 days later and continues to produce soft X-rays, optical, and mid-IR emission as it heats clumps of gas and dust. This emission has been observed by *HST*, *Chandra*, *Spitzer* and XMM. The ability to map these structures at sub-arcsec resolution in both optical/UV and X-rays with *HST* and *Chandra* provided the first spatially resolved study of a supernova blast wave evolving in real time (Fig 2.1-1). Without the Great Observatories, our understanding of the physical processes happening in the blast wave would have been missed either for lack of concurrent observations in some crucial energy range not accessible from the ground (UV and X-rays), or for the inability of following the evolution of these processes over the decades-long lifetime of these telescopes.

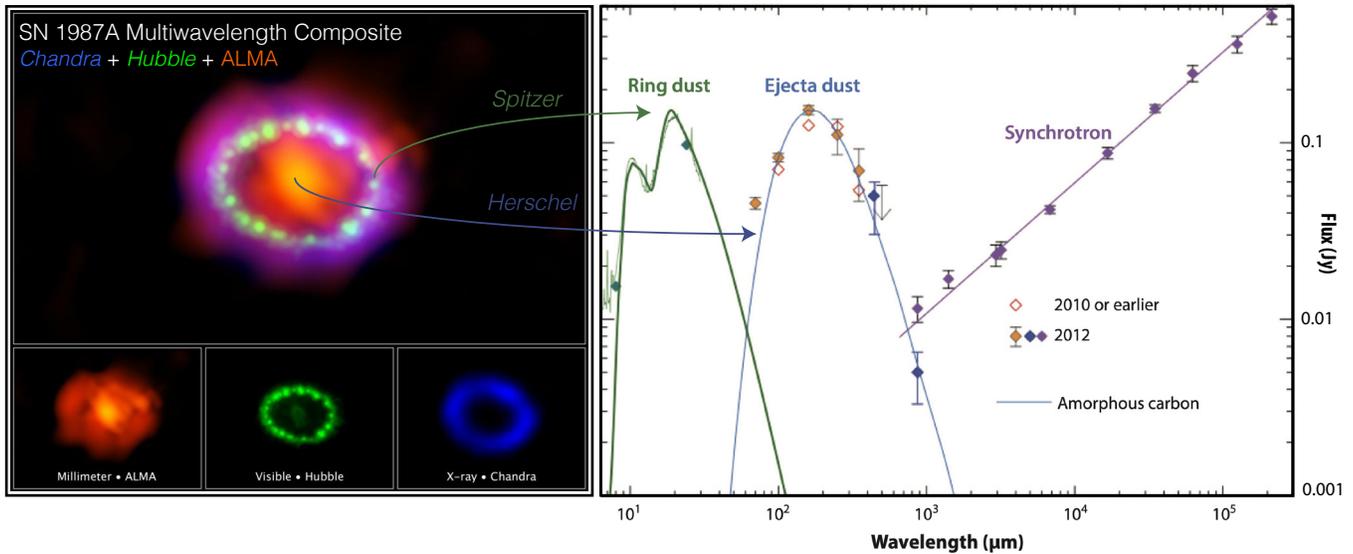

**Fig. 2.1—1. The Great Observatories view of Supernova 1987a.** *Left: 30th anniversary image of SN 1987a (credit: NASA, ESA, and NRAO). This composite image shows the blast wave from the original explosion has moved past the ring of material expelled by the star (Frank et al. 2016). Right: SED of the components from infrared through radio wavelengths (McCray & Fransson 2016). The Spitzer observations provide evidence for the destruction of pre-existing dust in the ring (Dwek et al. 2008, 2010) while the ALMA and Herschel observations detect newly formed dust in the supernova debris (Matsuura et al. 2011, Indebetouw et al. 2014).*

*The destruction and formation of dust by supernovae* — The temporal evolution of the ratio between mid-IR (measured with *Spitzer*) and X-ray luminosity (measured with *Chandra*) has provided evidence that dust from the progenitor is being destroyed by the supernova blast wave (Dwek *et al.*, 2008). In contrast, toward the inner debris from the supernova, the extinction of optical and near infrared emission from the supernova ejecta was the first indication that dust is quickly formed in the debris of the explosion. Ultimately, a large reservoir (~0.5 M) of cold dust (~20 K) at the center of SN 1987A was discovered in the far-infrared (Herschel) and sub-millimeter (ALMA; Matsuura *et al.*, 2011, 2015). *Spitzer* observations revealed an equatorial millimeter (ALMA; Matsuura *et al.*, 2011, 2015). *Spitzer* observations revealed an equatorial ring of warm dust, with order of magnitude fluctuations in density. Measuring the composition and evolution of these two dust populations was crucial for understanding how ISM dust is transformed by supernovae, quantifying the dust destruction rate by the supernova blast vs. the efficiency of the formation of chemically enriched dust in the supernova ejecta.

### UNRAVELING THE ECOSYSTEM OF THE GALACTIC CENTER

The center of our Galaxy is a unique laboratory for highly detailed studies of a supermassive black hole and the complex network of processes in its surroundings. The Great Observatories, aided by *Herschel* and SOFIA, studied the different components of this region: the complex environment within 1 pc of the black hole, Sgr A*, and the massive ($2$–$6 \times 10^7\ M_\odot$), dense Central Molecular Zone (CMZ) within ~200 pc of Sgr A* (Fig. 2.1-2). While Sgr A* hosts our closest example of a supermassive black hole, the CMZ is a laboratory for studying star formation in the centers of galaxies, in starburst galaxies, and in galaxies at the peak of the star formation density, $z \sim 1$-$3$ (Kruijssen & Longmore 2013).

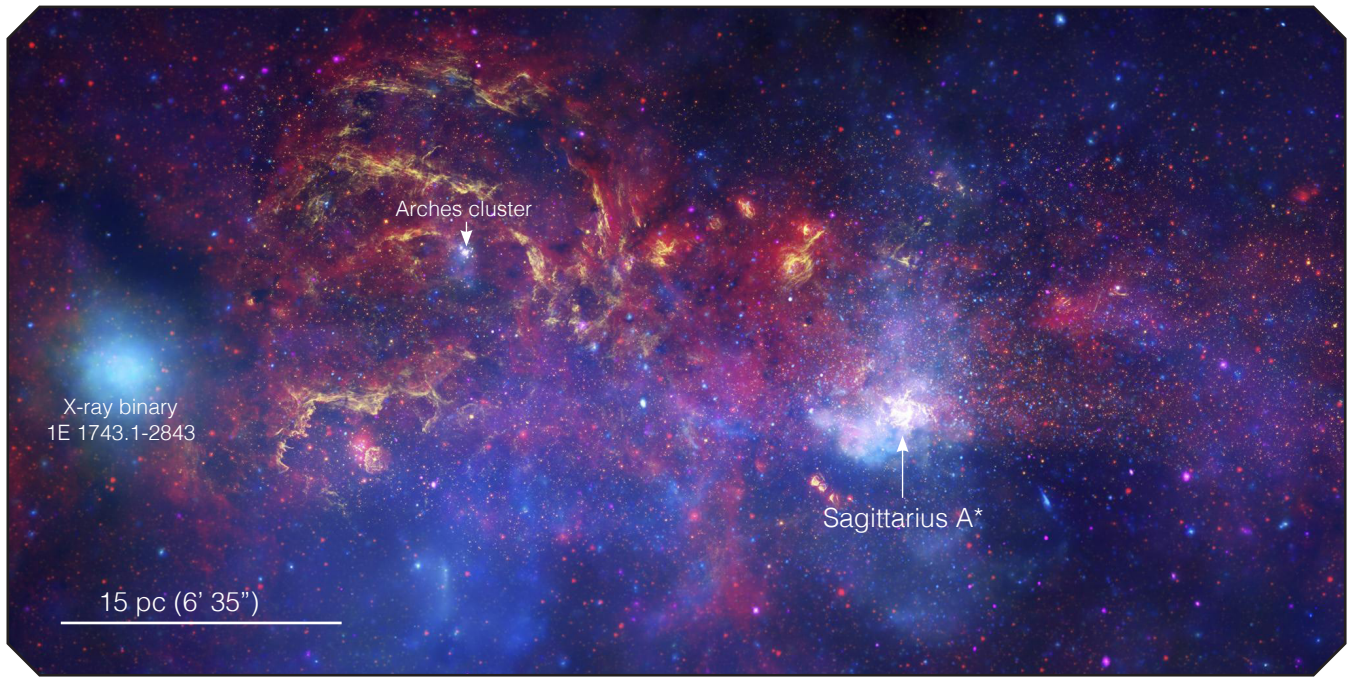

**Fig. 2.1—2. The Great Observatories view of the Galactic Center.** *Multi-wavelength montage of Sgr A and the central molecular disk combining data from Spitzer (red, Stolovy et al. 2006), HST(Pa-α in yellow; Wang et al. 2010) and Chandra (blue; Wang et al. 2002). A few features of interest have been labeled. (Composite credit: NASA, ESA, SSC, CXC, and STScI).*

***Sgr A\* and its immediate surroundings*** — Deep *Chandra* observations spatially resolved the accretion flow fed by colliding winds of massive stars (e.g., Russell *et al.*, 2017). Model fits to this flow provide evidence for an outflow fed by the accretion, which explains why Sgr A* is so faint (Wang *et al.* 2013). Simultaneous observations of Sgr A* at X–ray (*Chandra*, *XMM-Newton*, *Swift*), infrared (*Spitzer* and ground-based telescopes), and submillimeter wavelengths constrained the emission mechanisms and physics responsible for radiation flares from Sgr A* (e.g., Nowak *et al.* 2012; Ponti *et al.* 2015; Yuan *et al.* 2018; Boyce *et al.* 2019). *Chandra* also discovered "echoes" in neutral iron X-ray fluorescence excited by past flares. These evolving echoes trace dense gas structures in the region (Churazov *et al.* 2017), and provide a record of Sgr A* activity over the last 100 years (e.g., Koyama *et al.* 1996).

Star formation and feedback in the central molecular zone (CMZ) – The properties of the two massive star clusters in the CMZ, the Arches and Quintuplet clusters (e.g. Figer *et al.* 1999, Rui *et al.* 2019), were studied with the *HST*, showing evidence for a top-heavy IMF in the Arches (Hosek *et al.* 2019), and a population of high–mass stars via Pa–α observations (Wang *et al.* 2010; Dong *et al.* 2011). *Herschel* mapped the spatial distribution of the cold and dense gas (e.g., Molinari *et al.* 2011), while *Spitzer* and *Herschel* measured the bolometric luminosities and star formation rates in the clouds comprising the CMZ (Barnes *et al.* 2017). These data altered our picture of how the gas distribution in the CMZ is structured and showed that the star formation rate (SFR) is lower than expected based on relationships established for local star forming regions. *Chandra* observations also showed hot and diffuse gas in the CMZ, including SN remnants, as well as candidates of pulsar wind nebulae (e.g., Wang *et al.* 2002; Johnson *et al.* 2013). This material imposes substantial background pressure on all clouds in this region, altering the structure of the molecular clouds.

    While the stars in the immediate vicinity of Sgr A* have been extensively studied in the near-IR with large aperture telescopes equipped with adaptive optics, and the large structures in the CMZ gas have been mapped with ground-based radio interferometers, our view of the Galactic center would be severely incomplete in absence of the Great Observatories. The absence of *Hubble*, *Spitzer* (and *Herschel*) would have resulted in a much less detailed probe of the stellar population and diffuse matter in the CMZ. Similarly, without *Chandra* we would have had no means to study the interactions between Sgr A* and its environment, and the effects of stellar feedback produced by SNR and pulsars on the multi-phase gas in the CMZ.

### *A DETAILED VIEW OF STAR FORMATION & EVOLUTION IN THE MILKY WAY*

Star formation is the conversion of interstellar baryonic matter into the stars that form the backbone of galactic structure, drive the energetics of the ISM, and are the source of elements heavier than Lithium. Studies of star formation in Milky Way and nearby galaxies aim for a detailed physical description of the entire process, from the formation of molecular clouds, to the fragmentation of clouds and ensuing collapse and accretion of gas, and finally to the feedback which disperses the molecular gas. Because young stars are often deeply embedded in dusty clouds, the study of star formation is inherently multi–wavelength, with a strong emphasis on mid/far-IR and X-ray data - wavelengths that would have been missed without the contribution of the Great Observatories and *Herschel* (**Fig. 2.1-3**).

***Star Formation Laws of Individual Molecular Clouds*** — Although star formation laws have been established on galactic scales for decades, IR surveys from space were required to measure these within individual molecular clouds. *Spitzer* and *Herschel* mapped sixteen molecular clouds within 1 kpc - covering multiple square degrees - at wavelengths from 3.6 to 500 microns. *Spitzer* identified hundreds of protostars and thousands of pre-main sequence stars with dusty disks and envelops across the clouds (e.g. Evans *et al.* 2009; Megeath *et al.* 2012, 2016). *Herschel* detected young, deeply embedded protostars that were not identified by *Spitzer* (Stutz *et al.* 2013), measured the far-IR peak of the protostellar SEDs (Furlan *et al.* 2016), and mapped the column densities and temperatures of the parental molecular clouds (Schneider *et al.* 2013, Lombardi *et al.* 2014; Stutz & Kainulainen 2015; Pokhrel *et al.* 2016). These data showed that, in molecular clouds, the star formation rate per area scales as the 2nd power of the gas column density, although the normalization of the power-law can vary from cloud to cloud (Heiderman *et al.* 2010; Gutermuth *et al.* 2011; Lada *et al.* 2013).

***The formation of young clusters and high mass stars*** — Stellar clusters are the sites of high mass stars formation, can be detected in distant galaxies, and are being used to trace star formation over the cosmic time (e.g. Krumholz *et al.* 2018). In the nearest 2 kpc, combined *Chandra* and *Spitzer* observations have mapped the spatial distribution of stars in young, embedded clusters. These data show that young clusters often exhibit hierarchical sub-structure and are typically elongated and aligned with their filamentary, parental gas (Gutermuth *et al.* 2009; Kuhn *et al.* 2014; Megeath *et al.* 2016). When combined with kinematic data from Gaia DR2, the clusters are shown to be often expanding as their natal gas is dispersed, and that individual sub-clusters in hierarchically structured regions are moving apart to form associations or multiple bound clusters (Kuhn *et al.* 2019; Karnath *et al.* 2019). Deeper into the galactic plane, imaging with the Midcourse Space Experiment at 8 μm (MSX; Egan *et al.* 1999), and subsequently with *Spitzer* (e.g., Peretto & Fuller 2009), revealed large, cold, and massive molecular clouds with column densities so large that they appeared as infrared dark clouds (IRDCs) at this wavelength. These clouds were subsequently recognized as the pro-genitors of young clusters (Rathborne *et al.* 2006). *Spitzer* enabled investigations into the density structure of these clouds (Butler & Tan 2009) and provided the first identifications of young high–mass stars in these clouds at infrared wavelengths (Pillai *et al.* 2006). Subsequent far-IR and sub/mm imaging with *Herschel* detected more moderate–luminosity stars in IRDC (Henning *et al.* 2010) and mapped the dust emission from the IRDCs (Molinari & al. 2010). Today, numerous time– intensive projects target IRDCs with ALMA to study high mass star (e.g., Henshaw *et al.* 2017) while X–ray observations have detected intermediate mass pre-main sequence stars in these clouds (e.g., Povich *et al.* 2016).

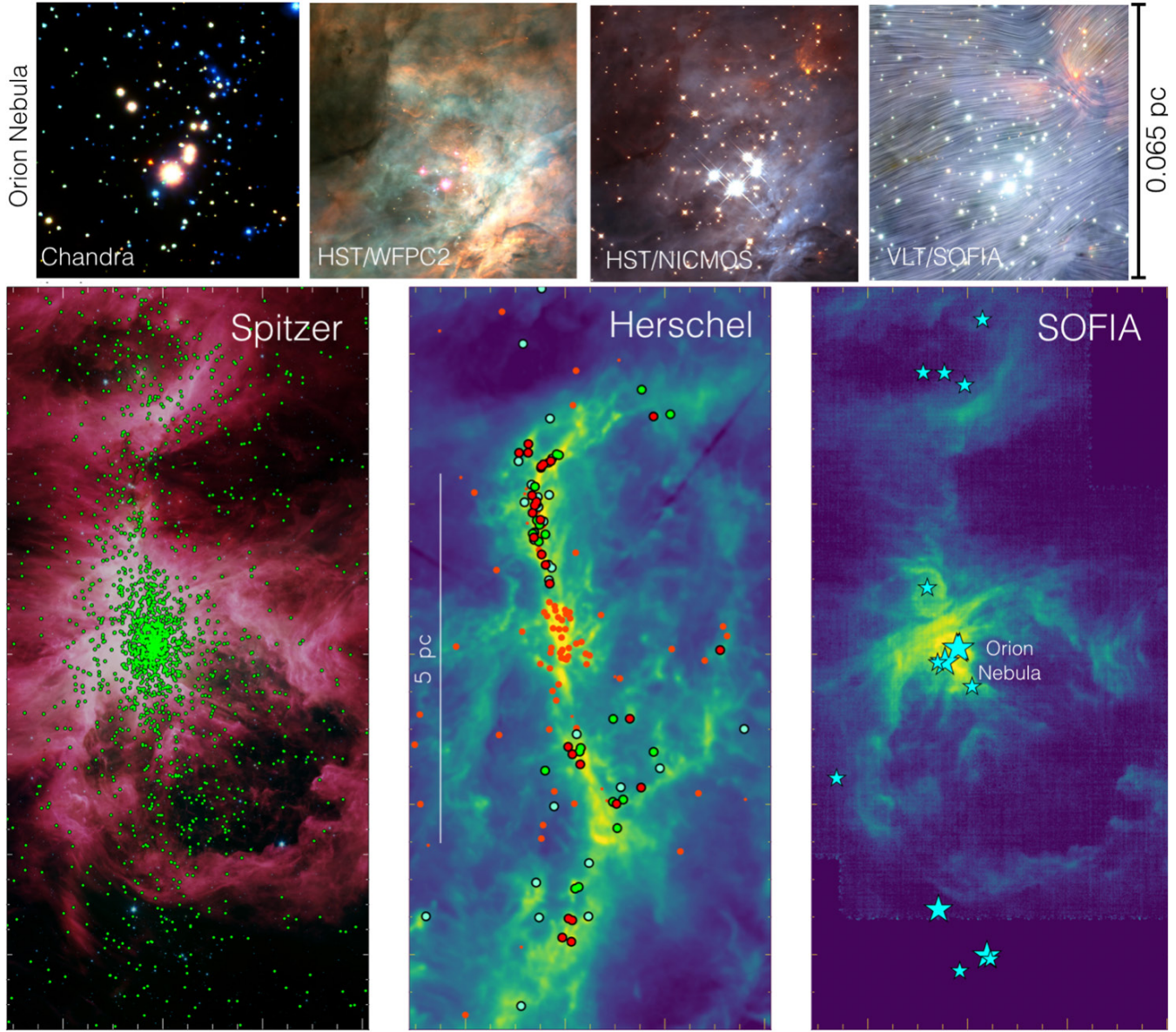

**Fig. 2.1—3. The Great Observatories view of the Orion Nebula.** *Top Row: images of the center of the Orion Nebula from Chandra in X-ray with ACIS, HST in visible light with WFPC2, HST in near-IR with NICMOS, and a composite of a near-IR image from the VLT overlaid with a SOFIA far-IR polarimetry data made with HAWC+. Bottom row: images of the Orion Nebula Cluster and Integral Shaped Filament obtained with IRAC onboard Spitzer with the positions of Spitzer identified pre-main sequence stars with disks overlaid (Megeath et al. 2012; 2016), a column density map made with PACS/SPIRE on Herschel with the positions of protostars and candidate protostars overlaid (Furlan et al. 2016, Stutz & Gould 2016), and the integrated [CII] intensity made with upGREAT on SOFIA (Pabst et al. 2019) with the location of O-B3 stars (Brown et al. 1994).*

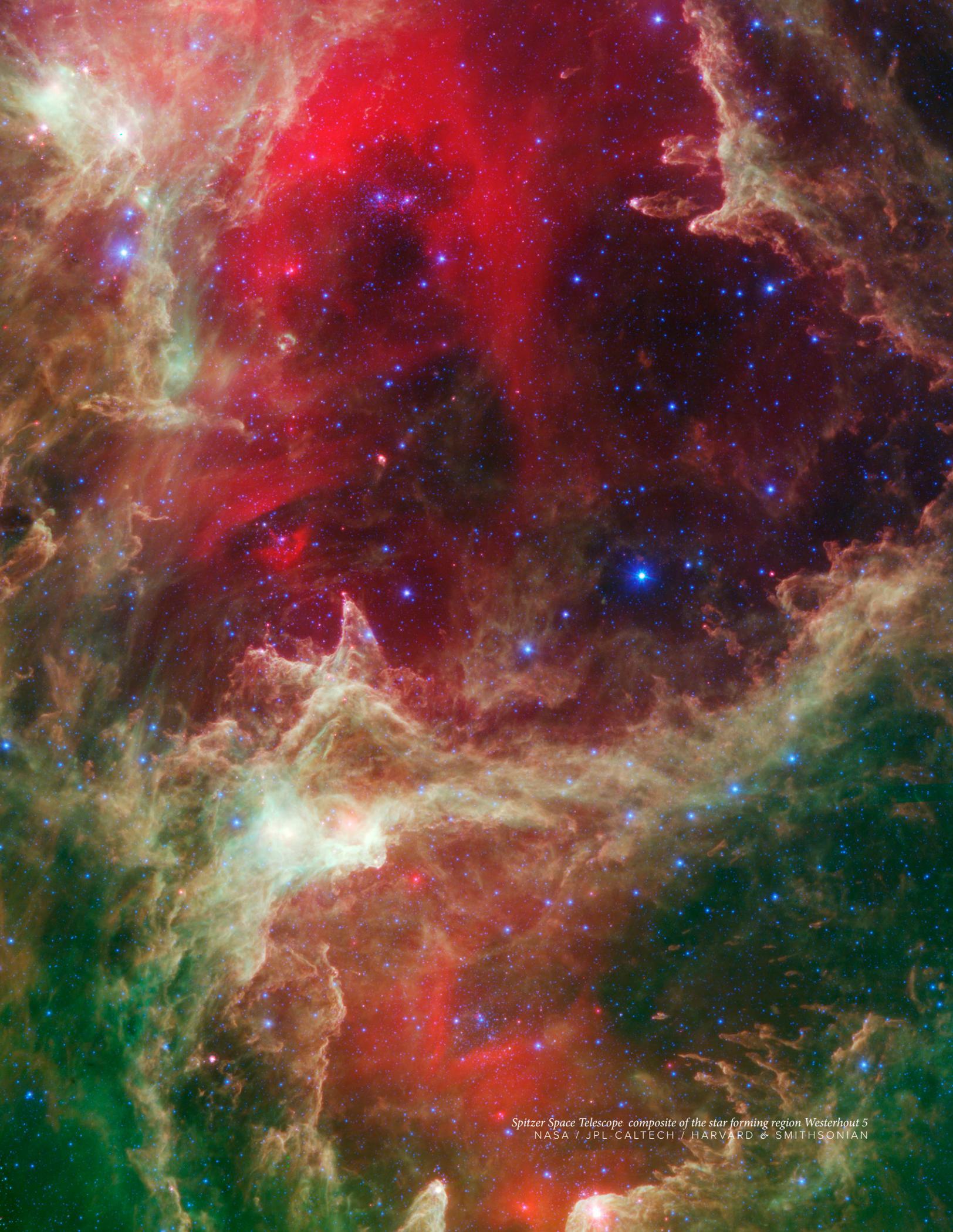
Spitzer Space Telescope composite of the star forming region Westerhout 5
NASA / JPL-CALTECH / HARVARD & SMITHSONIAN

### 2.1.2. Questions *for the* Next Decade

A new generation of multi-wavelength observatories, operating in concert, will be crucial to study of the connections between star formation, stellar evolution and the ISM and the role of the environment within and beyond the Milky Way. As an example, we have identified four key science questions that focus on different parts of the baryonic cycle between the ISM and stars, illustrated in **Fig. 2.1-4**:

- How is star formation influenced by the local environment?
- How can time domain observations advance our understanding of stellar evolution?
- What is the "micro-physics" of stellar feedback, and what is its role in generating cosmic rays, driving turbulence, and quenching/regulating star formation?
- How do the properties and life cycle of dust – from its formation, processing and growth – relate to the processes of stellar evolution and feedback?

*STAR FORMATION IN DIVERSE ENVIRONMENTS*

In the next decade, studies of the Milky Way and nearby galaxies will measure the dependence of star formation on the structure of molecular clouds, the metallicity, the external radiation field, the magnetic field strength, and the galactic tidal field. This will allow us to extrapolate our detailed understanding of star formation near the Sun to galaxies across cosmic time. The need to resolve the low mass end in the IMF (requiring high angular resolution images over large areas), as well as to probe embedded infrared sources and X-ray activity from young stars, make a strong case for a new generation of Great Observatories covering the electromagnetic spectrum.

*How does the star formation rate depend on environment?* — The star formation rate (SFR) quantifies the conversion of interstellar gas into stars, a fundamental step in baryon cycles within galaxies. Previous studies of nearby clouds (Sec 2.1.1) show that the SFR per surface area varies by two orders of magnitude as a function of gas column density (Gutermuth *et al.* 2011). This motivates measurements of the SFR as a function of the natal conditions across the diverse environments found in our galaxy as well as the Large Magellanic Cloud (LMC) and Small Magellanic Cloud (SMC). Synthetic observations of hydrodynamic simulations show that integrated emission cannot trace individual star forming regions (Koepferl *et al.* 2017). This motivates obtaining an observational census of young stars and protostars, extending previous studies of the Gould Belt clouds to more distant regions. For example, source counts from ALMA observations of the CMZ currently show orders of magnitude lower SFRs for a given gas column density compared to nearby star forming regions (Ginsburg *et al.* 2018). *JWST* will have the sensitivity to target extreme environments in the inner galaxy; future wide field X-ray and near- through far-infrared observations will enable the census of young stars and protostars needed to probe star formation laws from the Galactic center to the outer regions of the Milky Way.

*Does the IMF vary?* — The presence of systematic variations in the IMF with environment would have implications for both the physical mechanisms underlying the mass function and the utilization of high mass stars as tracers of star formation in distant galaxies. Near the Sun, the small clusters and groups of young stars populating the southern end of the Orion A cloud are deficient in massive stars compared to the Orion Nebula Cluster at the southern end of the cloud (Hsu *et al.* 2012, 2013). In the Arches cluster near the Galactic center, *HST*/WFC3 data suggest that the cluster IMF down to 1.8 solar masses is unusu-ally shallow and that the IMF is top heavy. Future investigations designed to confirm and characterize such variations include surveys of the IMF in low stellar density regions of molecular clouds within 1.5 kpc of the Sun and the study of dense, extreme clusters across the Milky Way and in nearby dwarf galaxies. The former requires the means to reliably identify young stars and constrain their masses over wide fields. Embedded young stars and sub-stellar objects can be identified by their X-ray emission, presence of mid-IR disks, or proper motions in the near-IR, and characterized by spectrographs deployed on ground-based telescopes. Far-IR telescopes will be required to resolve deeply embedded protostars. These searches can also detect variations in the IMF below the hydrogen-burning limit. Studies of distant clusters require both the identification and characterization of low-mass stars in regions with significant extinction. Here, again, *JWST*, the *Nancy Grace Roman Space Telescope* (formerly *WFIRST*, hereafter *Roman*) and future generation X-ray and far-IR telescopes with arcsecond or better resolution will provide the opportunity to identify pre-main sequence stars, while spectrographs on *JWST*, in concert with 8-30 meter ground-based telescopes will determine their masses.

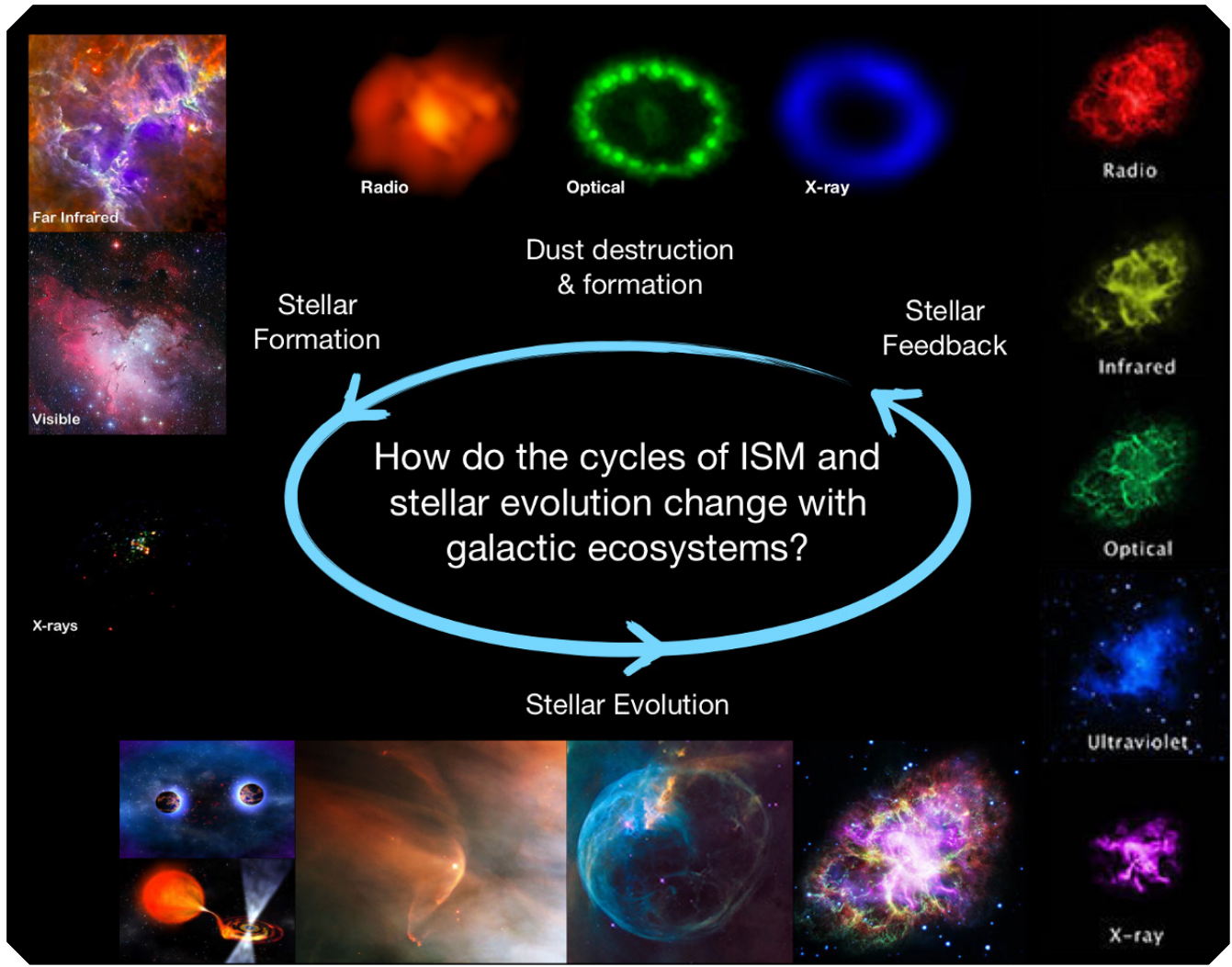

**Fig. 2.1—4. A multi-wavelength view of the baryonic cycle.** *The key to understanding the formation and evolution of the Milky Way and other galaxies is the cycle of baryons between the interstellar medium and stars. This cycle is dominated by a number of processes, including gas accretion, star formation, stellar evolution, and feedback through winds, SNe, and gas heating that can require a multi-wavelength approach.*

## STELLAR EVOLUTION IN REAL TIME

***How do binary interactions affect the evolution of massive stars and determine the properties of compact object binaries?*** — The nature of Wolf-Rayet (WR) stars, evolved stars with large mass loss rates, has been uncertain. Originally thought to be the evolutionary endpoints of the most massive O-stars, UV spectroscopic observations suggested that WR stars needed to go through a luminous blue variable (LBV) phase (e.g., Fullerton *et al.* 2006; Hirschi 2008; Puls *et al.* 2008). However, recent measurements of the binary fraction contradict this idea (see e.g. Sana *et al.* 2012), leaving binary interactions as an explanation for WR phase and winds. Thus, binary interactions may be the dominant effect on the population statistics of massive stars, which are more likely to form in binaries. This has implications for our understanding of double compact object binaries like those responsible for the gravitational wave signals observed by LIGO. Of particular interest are the WR + black hole binaries, IC 10 X-1 and NGC 300 X-1, as these systems are expected to be the direct progenitors of binaries that will be detectable with upgrades to LIGO/VIRGO (e.g., Binder *et al.*, 2015; Laycock *et al.*, 2015). However, the most massive stellar binaries are inaccessible to future gravitational wave facilities (e.g., LISA) due to their short orbital periods (Moore *et al.* 2019). Space-based EM observatories will therefore provide the only method to continue studying gravitational wave progenitors in the coming decades. In particular, time-resolved UV, X-ray and IR imaging and spectroscopy are the primary tools that can be used to probe the interacting stellar winds from massive binaries, with arcsecond-scale angular resolution often required to isolate binaries in crowded fields (e.g., Nicols *et al.* 2015, Gull *et al.* 2016, Lau *et al.* 2019) .

***How important are variations in mass accretion for star formation?*** — The formation of low mass stars is punctuated by bright outbursts driven by episodes of rapid accretion. During these bursts, the luminosity of a young star or protostar increases by factors of 2 to 100, with commensurate rises in the mass accretion rate. It is not known whether most of the stellar mass is accreted during these episodes, implying that these are essential for understanding stellar masses and the origin of the IMF, or whether most mass is accreted in a quiescent mode (Dunham *et al.* 2010, Fischer *et al.* 2017). The phenomenology of these outbursts, namely the distribution of luminosities, their frequency, and their duration, as well as the basic physical mechanisms, are all poorly constrained (Hartmann *et al.* 2016, Fischer *et al.* 2019). Space based observatories are essential for measuring the amount of material accreted during the bursts. For less embedded, more evolved young stars, UV observations can directly measure the variations in accretion (Ingleby *et al.* 2014). For protostars, mid to far-IR observations spanning the time dependent spectral energy distribution are needed to measure fluctuations in the source luminosity (Fischer *et al.* 2012). In addition, a burst in X-ray emission can also be detected (as in the case of the protostar V1647, Kastner *et al.* 2004). Triggers from forthcoming synoptic surveys will provide many new events in the near future, motivating the need for follow-up imaging and spectroscopic capabilities at X-ray, UV, mid-IR, far-IR wavelengths. With these observations, we can determine the fraction of the total stellar mass accreted during these bursts.

## THE "MICRO-PHYSICS" OF STELLAR FEEDBACK

Stellar feedback plays an essential role in the working of galaxies, especially star-forming ones. Theorists have invoked several potential channels for stellar feedback – thermal heating from supernovae, radiation pressure, stellar outflows and winds, cosmic rays, and turbulence – to varying degrees, in order to describe the structure of galaxies and the regulation of star formation. Nearby galaxies and the Milky Way are ideal laboratories for investigating key questions about stellar feedback. We highlight two of these questions for their acute dependence on multi-wavelength facilities needed to probe the complex, multi-phase ISM.

*What accelerates cosmic rays?* — Understanding the origin of cosmic rays requires probing a range of acceleration mechanisms within and outside the Galaxy. While shocks arising in supernova remnants (SNRs) are believed to be the primary source of cosmic rays with energies below 3 PeV (Ackermann *et al.* 2013, see Fig. 2.1-5), interacting stellar winds in massive star associations (Binns *et al.* 2007) and pulsar wind nebulae (see review by Weinstein 2014) may also play an important role in the Galaxy. The signatures left by cosmic ray acceleration processes in the γ-ray secondary photon radiation, as detected by *FERMI* and ground based observatories such as VERITAS, imply that these different sources contribute to the galactic cosmic ray energy spectrum. The limited angular resolution of current γ-ray observatories, however, makes it difficult to disentangle the contribution of the individual sources since they often coexist in the same high mass star forming regions. A well-studied example is the γ-ray source G78.2+21, a 7,000 year old SNR located in proximity of the Cygnus cocoon. This SNR shows a cavity filled with trapped, freshly accelerated cosmic rays of unknown origin. It is still to be determined if these cosmic rays have been originally accelerated by the nearby SNR, or by the wind of pulsars and massive stars within the cavity (Aliu *et al.* 2014). Solving these puzzles requires combining a new generation of more sensitive γ-ray observations, maps of shocked gas from high spatial resolution X-ray data and optical and infrared maps of stellar and diffuse matter in these super-bubbles.

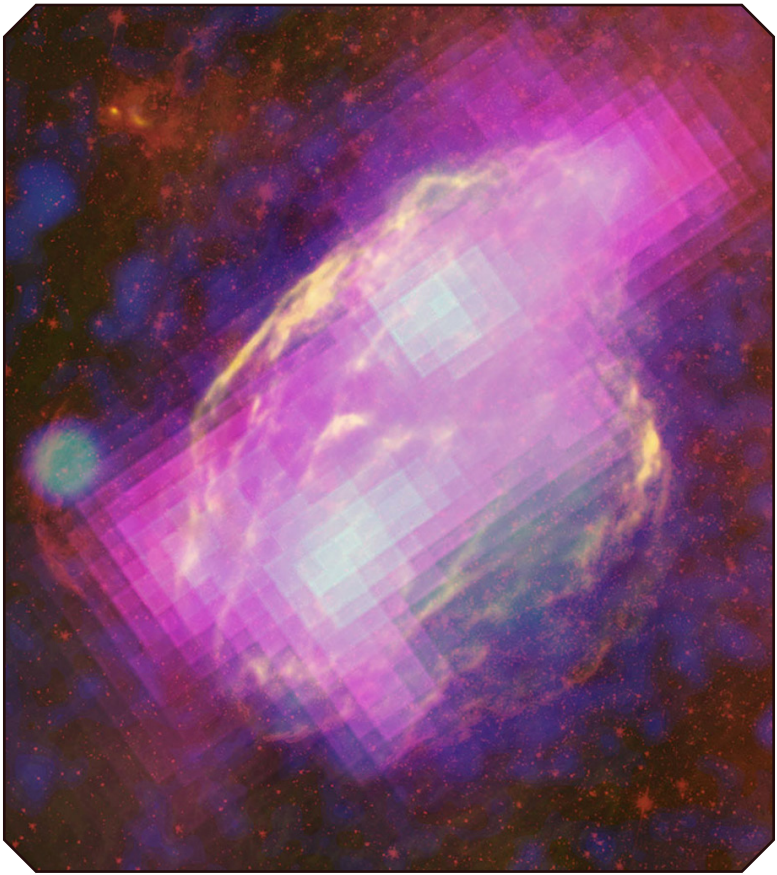

**Fig. 2.1—5. A Multi-wavelength view of the W44 supernova remnant.** *Embedded in the molecular cloud that formed its progenitor, multi-wavelength observations are required to separate individual sources in the remnant and investigate their role in cosmic rays production and acceleration. GeV γ-rays detected by Fermi are shown in magenta. Filamentary structures in the remnant are detected in the radio (VLA, yellow) and infrared (Spitzer, red). Blue shows X-ray emission mapped by ROSAT (Ackermann et al. 2013).*

*How is turbulence driven in a magnetized ISM?* — Turbulence in the ISM couples large and small-scale structures (e.g., molecular clouds), provides support against gravity, and is thought to regulate star formation in galaxies (e.g. Federrath & Klessen 2013). Radio maps of HI in dwarf galaxies and *Herschel* observations of molecular clouds show the ISM exhibits filamentary structures that are signatures of turbulence (Elmegreen & Scalo 2004; Arzoumanian *et al.* 2019). The turbulence can be driven by gravitational, magnetorotational, thermal, and cosmic ray streaming instabilities as well as stellar feedback. The morphology of the magnetic fields in the filamentary structures is an important constraint on the origin of the turbulence. Far-IR polarimetric observations of dust grain alignment by space-based and airborne telescopes can efficiently measure the magnetic field structure in large samples of molecular clouds and the diffuse ISM in the Milky Way and local group galaxies (Andersson *et al.* 2015, Hoang & Lazarian, 2016, Fissel *et al.* 2019, Chuss *et al.* 2019). Far-IR spectroscopy from space can directly measure the energy dissipation of turbulence in molecular clouds, providing constraints on driving mechanisms and the lifetimes of turbulent motions (Larson *et al.* 2015).

Space based observations can also assess the role of stellar feedback in driving turbulence. Due to the large energy released, supernovae are likely an important driver of turbulent motions in galaxies (Mac Low & Klessen 2004). X-ray observations with arcsecond angular resolution and a few eV energy resolution can resolve supernovae remnants (SNR) spatially and kinematically across the local group (Lopez *et al.* 2019). These would also be sufficient to measure proper motions of knots in 102-104 years old SNR out to the distance of the LMC (Patnaude & Fesen 2009; Lopez *et al.* 2019), directly constraining models. Visible, mid-IR and far-IR observations can also be used to estimate the energy input into molecular cloud turbulence from outflows driven by young stars (Maret *et al.* 2009; Neufeld *et al.* 2009; Manoj *et al.* 2016; Hartigan *et al.* 2019). Such studies have been begun with *HST*, *Spitzer* and *Herschel*, and will be expanded to greater distance and a wider range of environments with the next generation of space telescopes.

### THE LIFE CYCLE OF DUST IN GALACTIC ENVIRONMENTS

***Where does dust in the ISM come from?*** — Infrared and optical surveys of Local Group galaxies with resolved stellar populations suggest that there are still sources of interstellar dust unaccounted for (Meixner *et al.*, 2006, 2010; Boyer *et al.*, 2015a,b). Observations of the SMC with *Spitzer* and *Akari* show that ISM dust can come from stellar sources (mainly AGB stars and SNe), if all SNe are net producers of dust (Boyer *et al.*, 2012). Observations of $z > 6$ sub-mm galaxies suggest that large masses of dust ($\geq 10^8\ M_\odot$) may have formed on short timescales (~500 Myr or less), well before AGB stars could matter. Core collapse SNe could be the source of this early dust, provided they produce ~$0.1 - 10\ M_\odot$ of dust per explosion that survives the reverse shock (Dwek *et al.*, 2009; Gall *et al.*, 2011). However, other mechanisms of dust production, including dust growth in the ISM, may be required to explain the dust budget in galaxies (Draine, 2009). Given the essential role of dust for star formation and the physical and chemical evolution of galaxies, these scenarios need to be fully explored over a broad sample of targets covering the Local Group and beyond. This will require a new generation of optical and infrared space telescopes capable of resolving individual stars in more distant galaxies, and with the necessary sensitivity to measure the dust content in star forming regions and the diffuse ISM. Studying the production or destruction of dust in supernovae requires ultraviolet, mid to far-infrared and X-ray telescopes with higher effective areas, and the ability to follow the evolution of dust emission in SNR over decades, as done in the case of SN1987A.

***What is the composition of interstellar dust?*** — The properties of dust grains provide a record of the fundamental processes of growth and destruction as well as the physical underpinnings of the interstellar extinction curves. Knowing the dust grain composition is also fundamentally important for understanding the polarized far infrared and microwave emission that is the main source of confusion for interpreting the cosmic microwave background (Hensley *et al.*, 2018). Dust mineralogy is inferred from gas phase abundances measured in the UV (Jenkins, 2009), and spectroscopic features from the IR (Draine & Lee, 1984) to the X-ray (Lee & Ravel, 2005; Zeegers *et al.*, 2017). Incorporating all wavelength regimes is necessary to provide a complete model of interstellar dust mineralogy and size distributions as they change with the environment. Substantial uncertainties however remain. The abundance and form of carbonaceous dust is still debated, whether graphite, large organic molecules like polycyclic aromatic hydrocarbons (PAHs), or amorphous carbon. Silicate and carbonaceous grains are frequently treated as separate and non-interacting grain types, but models with composite grain species can also explain extinction and emission from the UV to the infrared (Zubko *et al.*, 2004; Jones *et al.*, 2013) as well as far-IR polarization. X-ray spectroscopy can directly test the proposed models for grain growth and processing, and is complementary to spectroscopy in the 1–10 μm regime. This requires similar or better spectroscopic resolution than *Chandra* (R > 1000) and effective areas that are 3 – 100 times larger than *Chandra* in the 0.2 – 2 keV range. UV and mid-IR imaging and spectroscopy of thousands of stars are needed to constrain how the properties of dust change across the Local Group (Gordon *et al.* 2019).

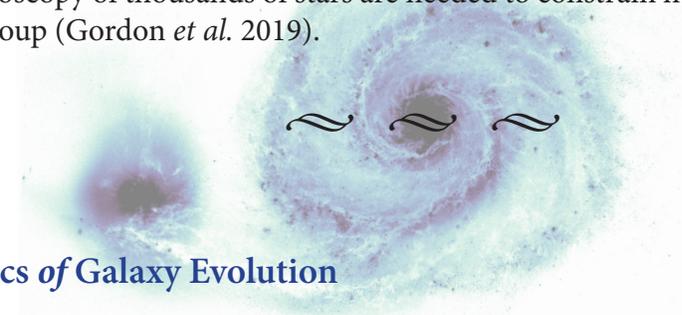

## 2.2. Astrophysics *of* Galaxy Evolution

Galaxies are complex ecosystems. They are the fundamental gravitational structures of the Universe — the gathering sites of the cold gas that condenses to form stars, the most massive of which explode, releasing the heavy elements in their centers into the surrounding interstellar medium (ISM). These elements then condense to form dust grains, which regulate the flow of radiative energy and the balance of heating and cooling in the surrounding gas. Radiation from young stars, active galactic nuclei and cosmic rays heat galaxies on large scales, maintaining gas in physically distinct phases. Exploding stars drive shocks into the ISM, and this energetic feedback can both amplify and inhibit the formation of future generations of stars. A growing supermassive black hole at the heart of a galaxy irradiates and drives fast winds into its surroundings and can launch relativistic jets that inflate large scale radio lobes, injecting mechanical and radiative energy on a wide range of physical scales, and preventing halo gas from cooling and accreting on the galaxy.

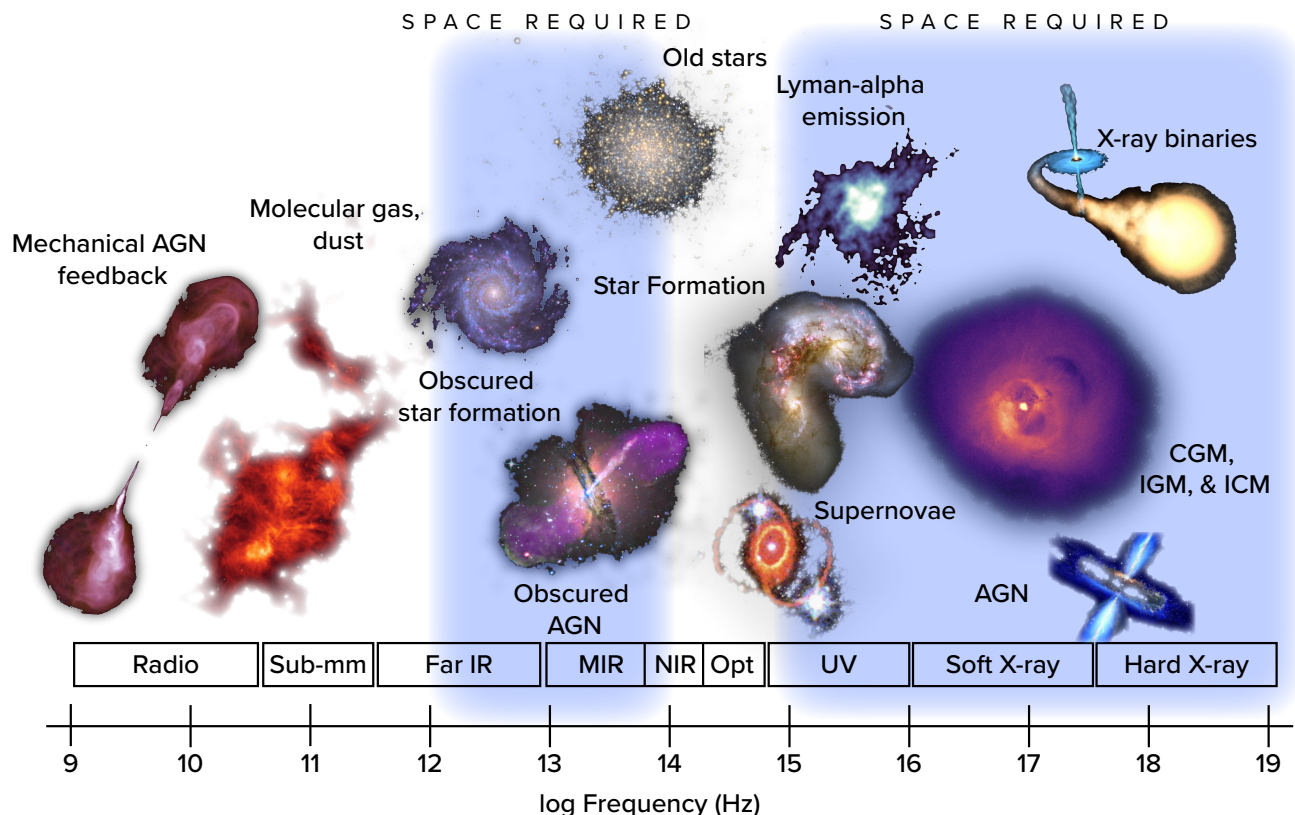

**Fig. 2.2—1. The Panchromatic Nature of Galaxies.** *A rough (and non-exhaustive) sketch of where the primary observable components of galaxies lie in frequency space. From star-forming clouds, to exploding supernovae, to active galactic nuclei, these energetic processes span an extremely wide range of energy and wavelength. Shaded blue regions indicate the frequency regimes that require observations from space.*

Because the processes of star formation, stellar death, black hole accretion, and the heating and cooling of the ISM emit a wide range of wavelengths, galaxies by their very nature require a multi-wavelength approach (Fig. 2.2-1). This need becomes more pressing as we try to understand galaxy evolution over cosmic time, as, for example, the critical diagnostics in the X-ray, UV, optical, and infrared that we use to measure the star formation rate, the age and mass of the stellar population, the star formation history, and the role of active galactic nuclei (AGN), are redshifted to longer wavelengths. As these features stretch across bands and in and out of atmospheric windows, multiple platforms on the ground and in space must be brought to bear to understand the formation conditions, the arc of evolution and the true nature of galaxies at all epochs.

Although incredible progress has been made charting the rise and evolution of galaxies in the Universe in the last decade, such as the joint growth of stellar and central massive black hole mass (e.g., McConnell *et al.* 2013), the bimodal separation of star formation into distinct modes with divergent gas consumption timescales (main sequence and starburst; e.g. Elbaz *et al.* 2011), and the accumulating reservoirs, production pathways, and physical conditions of gas and dust at early epochs (e.g. Michalowski *et al.* 2015, Decarli *et al.* 2016) — there are still many unanswered fundamental questions:

- **How and when do dust and heavy elements build up in a galaxy's ISM, and circulate through the circumgalactic and intergalactic media?**
- **What causes stars to stop forming (often abruptly) in galaxies?**
- **What drives the ten-fold decrease in the average star formation rate density that began nearly 7 Gyr ago?**
- **How does a galaxy's environment and the growth of their central supermassive black holes regulate star formation?**
- **When did the first galaxies and supermassive black holes form?**
- **What sources are responsible for re-ionizing the universe, what are the processes by which AGN and stars control star formation?**

Some of the most significant advances in galaxy evolution made by the original Great Observatories are summarized below, followed by a listing of some of the driving questions in galaxy evolution that will require the next set of sensitive, panchromatic space facilities.

## 2.2.1   Galaxy Evolution Science enabled *by the* Great Observatories

### THE GALACTIC "MAIN SEQUENCE"

Star formation and stellar mass are tightly correlated for galaxies of all sizes, such that most galaxies lie on a "main sequence of star formation". Galaxies on the main sequence undergo secular, long-lived star formation. A minority of galaxies reside above the main sequence and are undergoing rapid bursts of star formation over short timescales (< 1 Gyr), predominantly triggered by galaxy-galaxy interactions. In contrast, galaxies well below the main sequence are quenched, no longer forming stars: they are "red and dead". The area between the main sequence and the quenched regime is sparsely populated, so quenching happens on short timescales. The seminal paper on this topic, Noeske *et al.* (2007), combined *GALEX* and *Spitzer* star formation rates with stellar masses derived in part from *HST* data for over 2000 galaxies out to $z = 1$. Without *Spitzer*, the fact that the main sequence is not merely a local phenomenon would have been missed entirely. Star formation becomes increasingly obscured beyond $z > 0.5$, so that unobscured tracers in the UV/optical account for only 10% of the star formation rate in massive galaxies.

In fact, not only is the main sequence not a local phenomenon, but it is observed to be in place by $z = 4$, when the Universe was less than 2 Gyr old (e.g., Salmon *et al.* 2015). Beyond $z \sim 1.5$, far-IR observations are necessary to measure the star formation rates, as the mid-IR no longer reliably corresponds to the total infrared luminosity, which measures the heating from young stars (e.g., Elbaz *et al.* 2011). The fact that the main sequence is observed even in the very distant Universe suggests that star formation proceeds over billions of years. Without UV/optical star formation rates for local galaxies from *HST* and ground based observatories and mid/far-IR star formation rates for distant galaxies from *Spitzer* and *Herschel*, this broad picture of galactic star formation would have remained unknown.

### THE CO-EVOLUTION OF GALAXIES & SUPERMASSIVE BLACK HOLES

The exquisite spatial resolution afforded by *HST* led to the first well-constrained dynamical masses of central supermassive black holes in nearby galaxies, and the realization that black holes seem to be ubiquitous in galaxy nuclei. Furthermore, investigation into the properties of the black holes and their host galaxies uncovered the remarkable symbiotic relationship between the two (Ferrarese & Merritt 2000, Gebhardt *et al.* 2000, McConnell & Ma 2013; Kormendy & Richstone 1995; Magorrian *et al.* 1998). These observational discoveries were made possible by *HST* and ground-based optical and radio telescopes. Informed by increasingly sophisticated galaxy evolution simulations (e.g., Kauffmann & Haenalt 2000, Granato *et al.* 2004, Di Matteo *et al.* 2005, Croton *et al.* 2006, Hopkins *et al.* 2006), astronomers now think black hole feedback plays an integral part in regulating the growth of massive galaxies (cf. the review by Fabian 2012), although the details of this process are not yet understood.

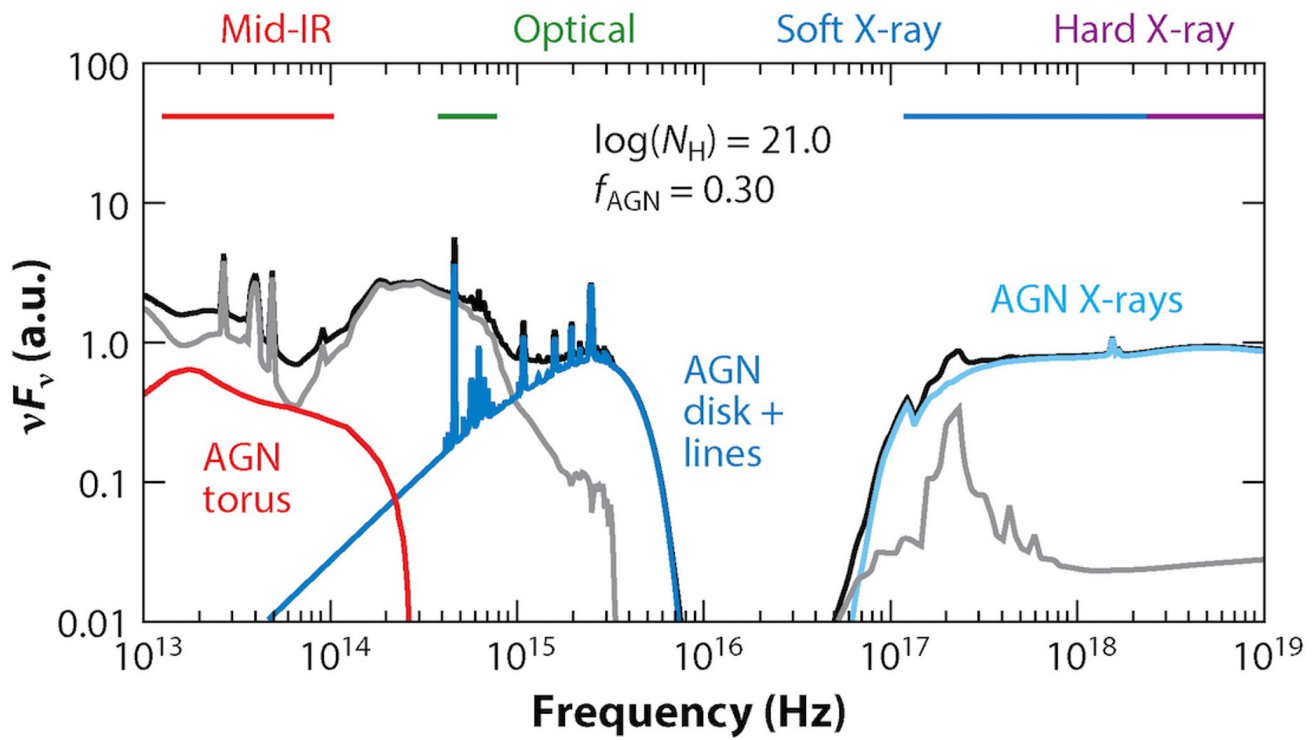

**Fig. 2.2—2. Multiwavelength emission of AGN and galaxies.** *Studying AGN-Galaxy co-evolution requires a multi-wavlength suite of observatories with imaging and spectroscopic capabilities. AGN emit most strongly in the X-ray, UV (unless obscured), and mid-IR, while measurements of the host galaxy light (shown in grey on the left) which provide the mass, redshift, and star formation rate, come from the optical and far-IR. Figure from Hickox & Alexander (2018).*

Mulitwavelength observations with the Great Observatories were fundamental in the discovery that black hole activity and star formation are linked over cosmic time. Galactic nuclear activity can be quantified by the rate at which supermassive black holes accrete material from their surrounding environment. The accretion rate can be measured with X-rays using *Chandra* and *XMM* for all but the most obscured black holes (Shankar *et al.* 2009, Aird *et al.* 2010) and mid/far-IR for obscured black holes (Delvecchio *et al.* 2014). Multiwavelength studies show that the cosmological (i.e. averaged over large areas) black hole accretion rate density peaks at the same epoch ($z \sim 1–3$) as the star formation rate density (Madau & Dickinson 2014, and references therein). The similarities of the star formation rate density and black hole accretion rate density provides a fundamental understanding of how and when mass growth occurs in the Universe. The coincident growth of stellar and black hole mass join with the black hole–bulge mass relationship to provide circumstantial evidence that galaxies co-evolve with their black holes. Without a multiwavelength suite of space telescopes, measurement of both the star formation rate density (requiring UV, optical, and IR) and black hole accretion rate density (requiring X-ray and IR) would not have been possible. This is easy to understand, since the spectrum of a single galaxy harboring an AGN can be decomposed into spectral regions dominated by stars and accretion ([Fig. 2.2-2](#)). Whether the focus is on individual galaxies or the population as a whole across cosmic time, co-evolution studies demand a multiwavelength approach, and this has been a major strength of the Great Observatories.

Moreover, the types of galaxies and AGN contributing to the buildup of mass evolves with time. Today, the majority of stellar mass is formed in galaxies with SFRs two orders of magnitude smaller than at $z \sim 1–3$. There are also fewer luminous quasars today than in the past. Both of these results have led to an understanding of "cosmic downsizing"— the majority of mass buildup shifts to smaller galaxies as the Universe ages because massive galaxies evolve more quickly. Observing downsizing required a multiwavelength approach, as stellar mass is best measured in the optical/near-IR, while star formation rates require both UV/optical and mid/far-IR data. Wide area surveys are critical for identifying luminous, and correspondingly rare, quasars as a function redshift, while deep fields are required for finding less luminous AGN and distant host galaxies.

### *THE GALAXY-HALO CONNECTION*

The idea that galaxies eject gas into the circumgalactic and the intergalactic medium is now well substantiated (Tumlinson *et al.* 2017). Most of the baryons in the Universe lie in the gas between galaxies (Peeples *et al.* 2014). UV studies of absorption lines imprinted on the spectra of distant quasars show that this gas is metal-enriched, and that these metals are more abundant in the gas around galaxies than in the general intergalactic medium (Steidel *et al.*,1994; Adelberger *et al.* 2003; Rudie *et al.* 2012; Turner *et al.* 2014). This metal enriched gas may account for a significant fraction (>25%) of the baryons in the halo of a typical galaxy (Werk *et al.* 2014). Galactic outflows can redistribute and even eject metals formed by stars and supernovae (SNe) out of galactic disks, helping to drive galactic ecosystems. The strong physical impact of metal content on galaxy evolution is clear in the present day Universe, with many robust metal-driven physical processes uncovered, including the tight relations among a galaxy's gas-phase metallicity, stellar mass, luminosity, and star formation rate (e.g. Tremonti *et al.* 2004; Cresci *et al.* 2018), the excitation conditions and structure of star-forming gas clouds (Bolatto *et al.* 2008), the physical properties of the dust (e.g. Sandstrom *et al.* 2010), and the balance of heating and cooling in the ISM (Smith *et al.* 2017).

Feedback can impact global galaxy properties either by ejecting large quantities of gas, stopping or delaying large-scale accretion, or acting over a galaxy's molecular gas reservoir to stabilize it against gravitational collapse. Our understanding of feedback comes from observing ionized (UV/optical) and molecular (sub-mm) outflows, as well as tracing energy injection into the IGM through X-ray and radio data. Distinguishing between AGN and starburst driven feedback, which may each dominate over different galaxy mass regimes, requires the ability to observe galaxies over a wide range of wavelengths from the radio through the gamma-ray bands.

There is a great deal of empirical evidence for the existence of galactic winds in nearby starburst galaxies (e.g., Armus *et al.* 1990; Heckman *et al.* 1990, 2000; Strickland *et al.* 2000; Fischer *et al.* 2010; Sturm *et al.* 2011; Martin 1999; Martin *et al.* 2012; Veilleux *et al.* 2005, 2013 – see Fig. 2.2-3) and in star forming galaxies at $z > 2$ (Kornei *et al.* 2012; Shapley *et al.* 2003; Spilker *et al.* 2018). In fact, in our own galaxy, we have evidence of outflows launched by a powerful central wind through gamma ray discovery of the *Fermi* bubbles. X-ray observations now provide a clear link between the *Fermi* bubbles and the Milky Way (Ponti *et al.* 2019), and UV observations of distant quasars have been used to constrain the kinematics, age and mass of the bubbles (Bordoloi *et al.* 2017).

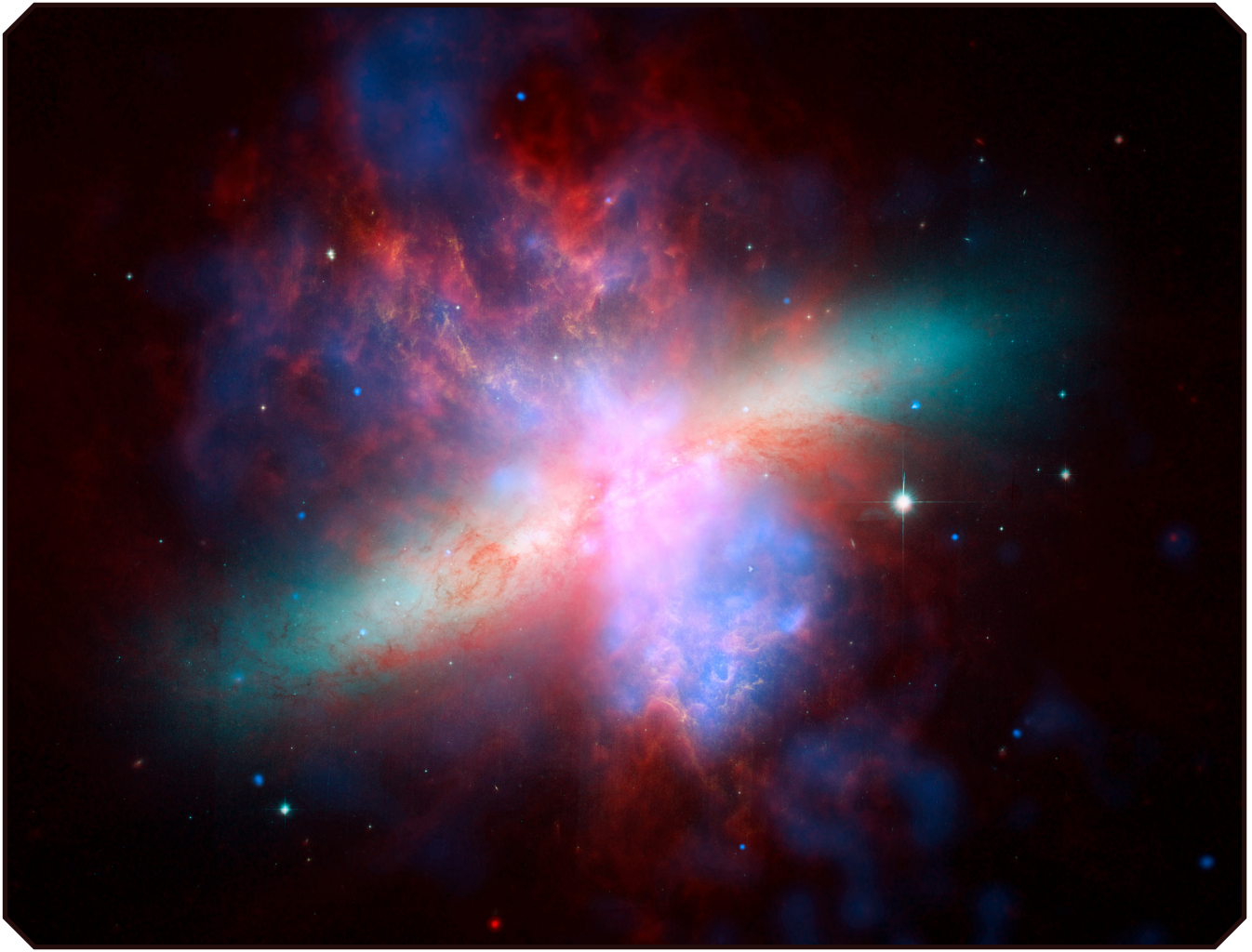

**Fig. 2.2—3.The Superwind in M82.** . *Image of the nearby starburst galaxy, M82, taken with NASA's Great Observatories. M82 is experiencing an intense starburst and driving a galactic outflow along the minor axis. Here, the stars and the warm atomic gas, as seen with HST, are in green and orange, respectively. The dust, as seen with Spitzer, is in red. The hot plasma, as seen with Chandra, is in blue. The complex, multi-phase nature of the bi-polar outflow, driven by the combined effects of young stars and supernovae, is evident, as the gas escapes the galactic disk and interacts with the CGM.*

X-ray evidence for high-velocity outflows (~0.1c – 0.5c) is also prevalent in AGN as evidenced by Warm Absorbers (Laha *et al.* 2016) and Ultra-fast Outflows (e.g. Tombesi *et al.* 2010). Far-IR or molecular outflows and/or extremely turbulent interstellar media have also been seen in some luminous, high-redshift quasars and dusty IR-bright AGN (e.g., Diaz-Santos *et al.* 2017). It is believed that feedback plays a key, but poorly understood, role in establishing the shape of the galaxy mass function at the low and high mass ends (Baldry *et al.* 2008), the mass-metallicity relation (Tremonti *et al.* 2004), and even the existence of a galaxy main sequence (Elbaz *et al.* 2010), and the heating and enrichment of the intergalactic medium.

### *CHARACTERIZING GALAXIES FROM COSMIC DAWN TO THE EPOCH OF REIONIZATION*

The combination of *HST* and *Spitzer* has allowed for a significant improvement in the identification of distant galaxies. In the epoch of re-ionization, at $z > 6$, strong nebular emission lines (Hα and [OIII]) redshift through the *Spitzer* bandpasses. These lines are observed to be much stronger than those seen in star-forming galaxies at lower redshift. Depending upon the precise redshift, galaxies with extremely strong emission show anomalous near-IR colors. Combining IRAC (3.6–4.5 μm) with *HST* colors produces more precise photometric redshifts, allowing astronomers to pin down galaxies at $z \sim 6-8$, which are crucial contributors to the reionization of the universe. Observations of IR colors constrain the strengths of emission lines, allowing measurements of the ionizing photon production efficiencies. So far, photon production efficiencies appear to be higher at higher redshifts and higher for bluer galaxies, with significant impact on models of reionization. However, the dominant source or sources of reionizing photons is still unknown and requires future multiwavelength observations. (Oesch *et al.* 2014, Duncan *et al.* 2015, Song et al 2016, Finkelstein *et al.* 2013, Smit *et al.* 2014, 2015, Salmon *et al.* 2015, Bouwens *et al.* 2016).

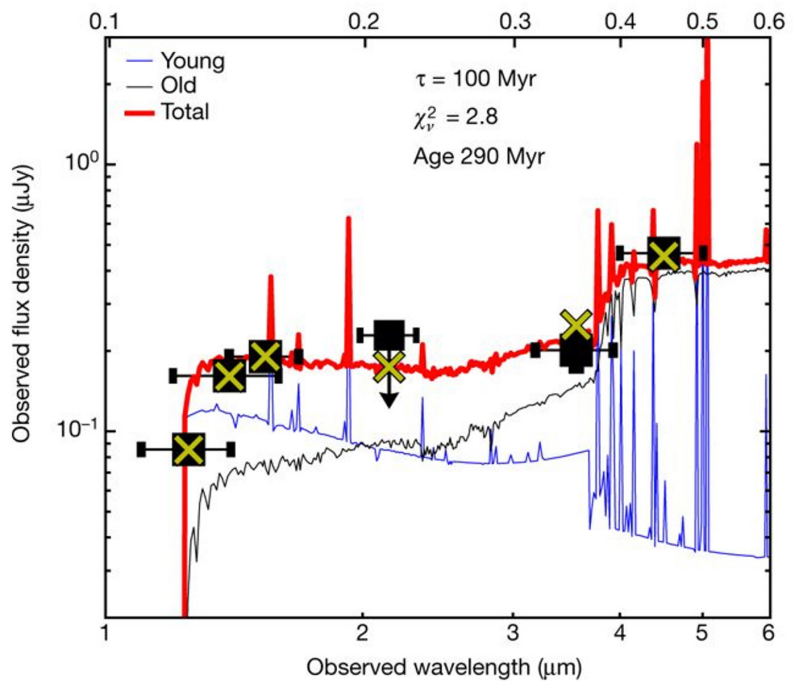

**Fig. 2.2—4. Nature *of the* Earliest Galaxies.** *Best-fit spectral energy distribution for the redshift z=9.11 gravitationally lensed galaxy, MACS1149-JD1, from Hashimoto et al. (2018). The photometric points come from HST, Spitzer, and the VLT. Rest frame wavelengths (in microns) are along the top. Recently a detection of the [OIII] 88-micron emission line with ALMA secured the redshift of this galaxy, and helped establish a size and star formation rate. The photometric and emission line data suggest that the dominant stellar component of MACS1149-JD1 formed about 250 Myr after the Big Bang, at z~15. The combination of gravitational lensing and multiwavelength observations from the ground and space were essential to determine the properties of this early galaxy.*

Massive galaxy clusters act as natural telescopes by magnifying the light of distant, background galaxies. The Frontier Fields were chosen to optimize lensing by massive foreground clusters at z~1, in order to detect and measure young galaxies at $z > 8$, when the Universe was less than a Gyr old. *Hubble* imaging provided the clearest picture of the structure of these distant galaxies and supplied the rest-frame UV data necessary to constrain their star formation. *Spitzer* imaging of the Frontier Fields was critical for estimating the stellar mass of the galaxies, a measure of the integrated star formation since their birth. Together, *Spitzer* and *Hubble* provided constraints on the magnitude and critical scale/spatial locations of star formation, providing insight into the growth histories of the most distant galaxies (Oesch *et al.* 2014; Oesch *et al.* 2016; Zheng *et al.* 2014; Bouwens *et al.* 2015; Zitrin *et al.* 2015; Stark *et al.* 2013; Coe *et al.* 2013, 2015; Infante *et al.* 2015; Hashimoto *et al.* 2018; Song *et al.* 2016; Johnson *et al.* 2017).

Follow-up observations using *HST* and the largest ground-based telescopes have confirmed the redshifts of five galaxies at $z > 8$, and one above $z > 10$, showing that some massive galaxies were already in place by a few hundred Myr after the Big Bang and that massive galaxy buildup was already underway at $z > 10$ (**Fig. 2.2-4**). Working together, the Great Observatories have provided our first glimpse of these early galaxies, resulting in the discovery that the low-mass end slope of the stellar mass function steepens significantly with increasing redshift, implying reduced feedback in these distant galaxies.

### 2.2.2. Questions *for the* Next Decade

#### *HOW DO STARS FORM IN THE EARLY UNIVERSE?*

The star-formation rates (SFRs) of galaxies as a function of redshift are a fundamental prediction of theoretical models, thus robust measurements of total SFRs, especially at high-z where galaxies are rapidly growing, can constrain a variety of physical processes and simulations. Obtaining accurate SFRs for individual galaxies always necessitates a multi-wavelength approach to account for obscuration. When using broad-band surveys to piece together the global star formation rate density (SFRD) as a function of epoch, large and uncertain corrections are often needed to extrapolate to the population as a whole. Current data suggest that from $0 < z < 2.5$, at $\log(M^*/M_\odot) > 9.5$, ~50% of all star-formation is obscured. Beyond $z > 3$, the relative contribution of unobscured and obscured star formation to the global buildup of stellar mass is unconstrained because relatively few detections of galaxies in the IR are available at these epochs (**Fig. 2.2-5**). To understand the role of environment, AGN and feedback in shaping galaxy growth in obscured and dust-free systems, requires measuring thousands of individual galaxies over large areas in the rest frame UV/optical and FIR.

Does star formation in these early galaxies proceed similarly to star formation at $z < 3$? Is it long-lived, regulated mainly by the gas supply, and is that gas

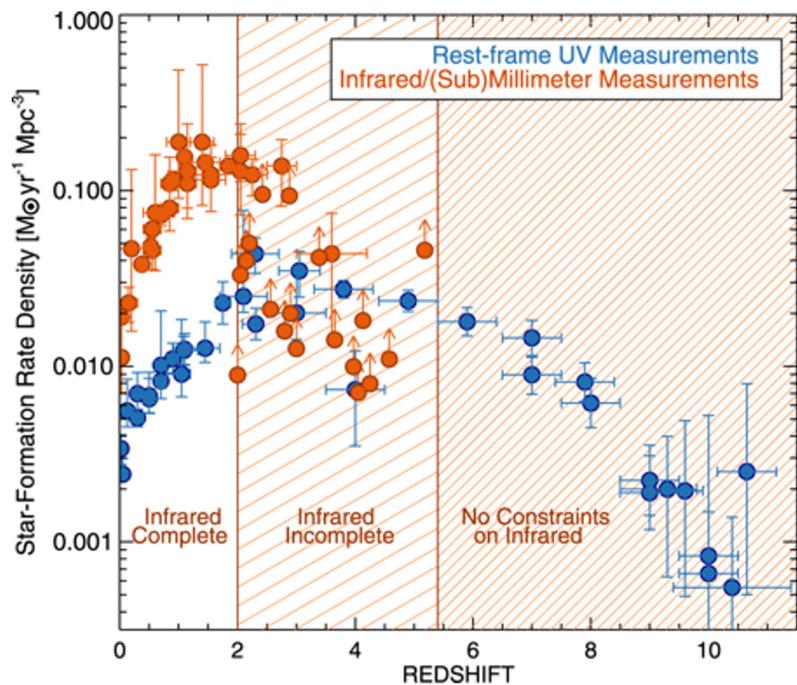

**Fig. 2.2—5. Star Formation over Cosmic Time.** *Star Formation Rate Density (SFRD) as a function of redshift (adapted from Casey et al. 2018). A significant fraction of the light ever emitted by stars is absorbed in the infrared. Currently, we have very little information on the IR-derived star formation rate in galaxies at z > 3. Future multi-wavelength observations of this dusty, high-z population, in the sub-mm, infrared and hard x-rays would allow a complete derivation of the obscured star formation and black hole accretion rates in galaxies as they build up towards cosmic noon.*

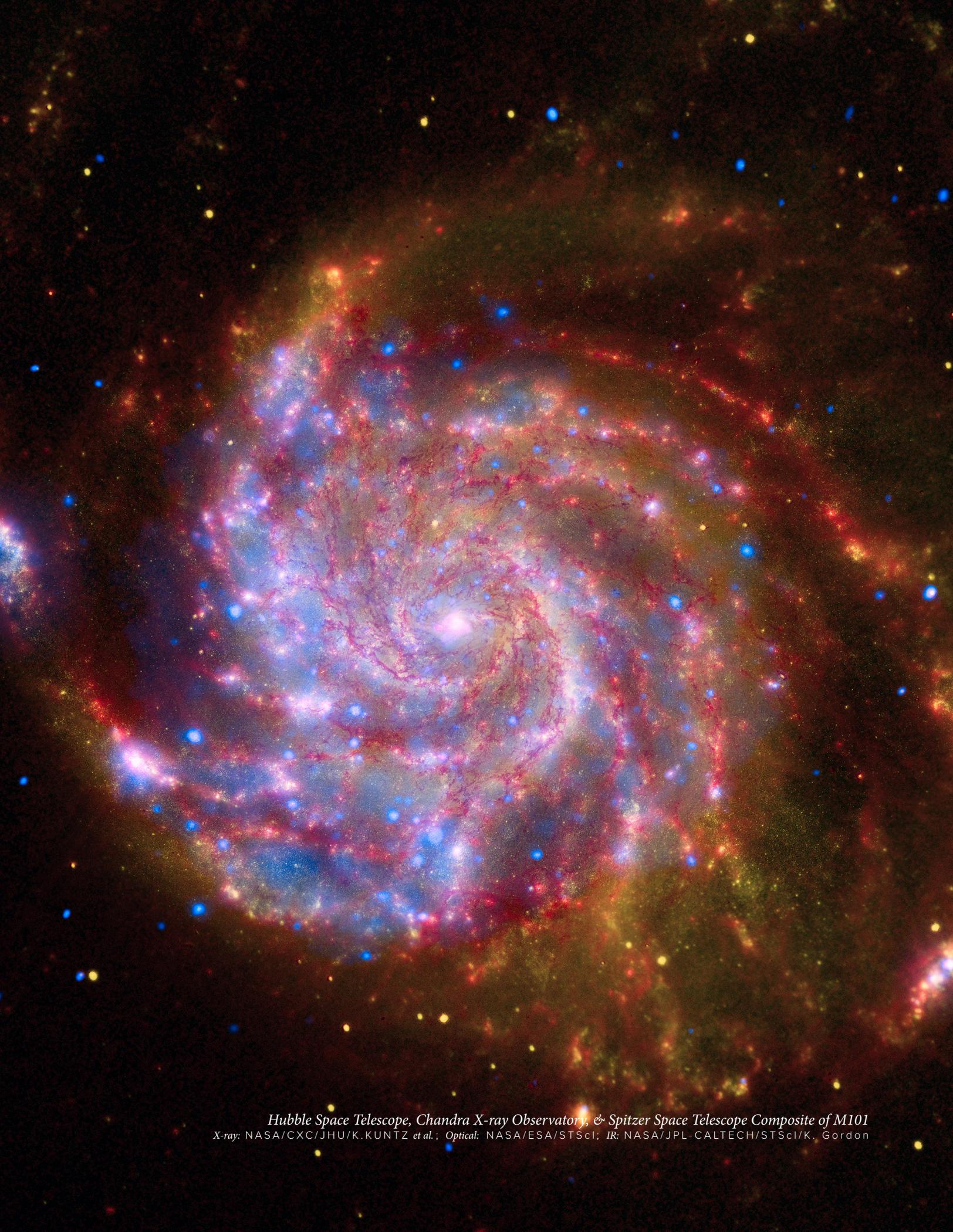

*Hubble Space Telescope, Chandra X-ray Observatory, & Spitzer Space Telescope Composite of M101*
X-ray: NASA/CXC/JHU/K.KUNTZ et al.; Optical: NASA/ESA/STScI; IR: NASA/JPL-CALTECH/STScI/K. Gordon

supply tied directly to galaxy stellar mass? In other words, precisely when was the galaxy main sequence in place? At some point in the evolution of the Universe, the bulk of star formation must be unobscured, as metals will not have had sufficient time to form inside galaxies. Uncovering how early galaxies (z > 4) assemble most of their mass, at what point they arrive on the main sequence, and how quickly they leave it, calls for a multi-wavelength approach over wide areas to build up good population statistics.

The rest-UV colors of galaxies at $z > 6$ are fairly blue (as observed in the NIR), leading many to conclude the obscured star formation is negligible. In fact, recent measurements of sub-mm number counts at $z > 4$ indicate that obscured star formation may be just as ubiquitous as unobscured star formation in the early universe (Zavala *et al.* 2018). Therefore, a full census of star formation at high redshift requires observations at longer wavelengths that are not attenuated by dust. Astronomers need to be able to concretely determine the multi-wavelength counterparts of high redshift galaxies as their key parameters (mass, SFR) are measured in different wavelength regimes. At these redshifts, NIR and MIR observations will be used to measure the stellar mass, while optical observations constrain the un-obscured star formation. Observing the obscured star formation in these systems will also require a large-aperture space telescope optimized for 0.1 - 1mm observations (~20 – 200 microns rest) that is able to cover large areas (tens of sq. degrees) to average out cosmic variance and detect rare objects, with an aperture large enough to manage confusion and reach SFRs of a few $M_\odot$ yr$^{-1}$ at $z > 5$. Furthermore, since it is critical to distinguish star bursts from obscured AGN, and accurately measure the star formation rates and black hole accretion rates in individual sources, FIR spectroscopy and deep X-ray imaging will be required (Yung *et al.* 2019ab, Somerville, Popping and Traeger 2015, Smit *et al.* 2012, Casey *et al.* 2018ab, Whitaker *et al.* 2017). If any of these multi-wavelength observations are absent, or wildly unmatched in sensitivity or resolution, fundamental relationships will be missed, and the true nature of black hole and galaxy coevolution will remain shrouded in mystery.

### *HOW WAS THE UNIVERSE REIONIZED?*

The Reionization of the neutral hydrogen in the diffuse intergalactic medium begun after the formation of the first stars, galaxies and black holes, marked the end of the dark ages. While it is believed that high-energy ultraviolet (UV) photons (> 13.6 eV) from early galaxies were responsible for the most of the ionizing budget, this conclusion depends on a variety of assumptions, including the fraction of ionizing photons that escape the galaxies, the faint end slope of the galaxy luminosity/mass function and the intrinsic production rate of ionizing photons.

The escape fraction is still largely uncertain, though not for a lack of substantial observational efforts. Dozens of nights of ground-based observing, and hundreds of hours of *Hubble* integration have been dedicated to direct detection of escaping ionizing radiation at $z < 4$, where the ionized IGM is transparent enough to make this measurement. While the vast majority of studies have yielded non-detections, a few dwarf galaxies with high escape fractions have recently been directly detected in the nearby universe and at $z \sim 3$, and a few faint detections have also been realized by stacking L* (those at the knee of the luminosity function) galaxies (Steidel *et al.* 2018; Izotov *et al.* 2018). The majority of recent theoretical studies predict that dwarf galaxies dominate reionization not just due to their large numbers, but because they preferentially have higher escape fractions. However, the sparse observational data does not allow us to constrain the nature of the ionizing sources at high-redshift, as these dwarf galaxies (log M* = 6–8) are much smaller than the majority of the targeted (and detected) galaxies. Deep studies in the UV/optical would allow direct detection of Lyman continuum radiation for dwarf star-forming galaxies at $z = 0.1–3$, true low-redshift analogs to the likely dominant sources of reionization (Siana *et al.* 2010, Nestor *et al.* 2011; Vanzella *et al.*

2012, 2016; Izotov *et al.* 2016, Smith *et al.* 2016; McCandliss & O'Meara 2017; Steidel *et al.* 2018). In order to clearly separate emission from hot stars and AGN, as well as detect any underlying older stellar populations, deep X-ray and IR imaging and spectroscopy with moderately high spatial and spectral resolution, will be required.

The numbers and masses of dwarf galaxies in the early Universe is also unconstrained. Although most of the Universe's stellar mass resides in galaxies with $\log(M^*/M_\odot) > 9.5$, the smallest dark matter haloes that are capable of forming stars are still unknown. While the luminosity function of low-redshift galaxies has a fairly flat faint-end slope, this steepens significantly at $z > 4$, such that at the highest redshifts we can currently probe, it is believed that the dominant contributor to the SFRD, and to reionization, are galaxies below the detection limits of even the deepest *HST* surveys. We need to observe much further down the luminosity function, to discover where it deviates from its steep slope.

Theory predicts that star formation in lower-mass halos should begin to be inefficient, both due to lack of atomic line cooling in mini halos, and Jeans filtering in more massive halos, after the onset of reionization. These should combine to cause a turnover in the number densities of galaxies at the very faint end. Studies of the Frontier Fields have found no evidence for a turnover down to M ~ -15 mag, which corresponds to $\log(M_h/M_\odot) < 10$, consistent with this idea. As the slopes are steep, exactly where this function turns over not only has a strong impact on the SFRD, but observing this turnover places key constraints on the physics of gas cooling in galaxies, and also on reionization. Observing to the "end" of the UV luminosity function requires an optical/near-infrared space telescope capable of directly detecting galaxies down to M = -13 mag at $z = 7$, estimated to have $\log(M_h/M_\odot) \sim 9$, beyond the point where many models predict a turnover. While *JWST* will go significantly deeper than *HST*, the deepest, blank-field luminosities reachable by *JWST* will be about -15.5 mag, still brighter than the expected turnover point. Lensing will likely probe a few magnitudes deeper, but results will be limited by systematics. In particular, the uncertainties in the magnification corrections will be high, and thus direct detections are required. (Yung *et al.* 2019ab, Somerville, Popping and Traeger 2015, Finkelstein et al 2012, 2015, Bouwens *et al.* 2014, 2015, 2017, Livermore *et al.* 2017).

### *HOW DO SUPERNOVAE & AGN REGULATE THE GROWTH OF GALAXIES?*

From large scales to small, star formation is an extremely inefficient process. In galaxies, only a few percent of gas is converted into stars in a free fall time. The stellar mass–to–halo mass ratio peaks at about 1 part in 30 for a halo mass of $10^{12} M_\odot$, and then falls steeply above and below this halo mass (e.g., Moster *et al.* 2010). To avoid over-producing stars in numerical simulations, it is often necessary to invoke some form of strong negative feedback (e.g., Somerville & Davé 2015; Hopkins *et al.* 2014). Powerful feedback from AGN is often invoked to explain the fall off in star formation efficiency at the high mass end (e.g., Springel *et al.* 2005; Henriques *et al.* 2015), which is truly remarkable given the vastly different physical scale of the AGN and its host galaxy. At the low mass end, supernovae and stellar feedback are thought to explain the drop, as starbursts overcome their relatively feeble gravity and eject gas into the intergalactic medium. However, this appealingly simple picture has challenges (e.g. Smith, Sijacki & Sijing, 2019; Henden *et al.*, 2019). The importance of different feedback mechanisms on all scales is a key open question in astrophysics, precisely because the sub-grid physics is only crudely modeled, even in the most advanced physical simulations.

Directly or indirectly measuring feedback in galaxies, and determining the driver for feedback as a function of galaxy type, is extremely important for understanding galaxy growth as a function of redshift. However, not all observations fit in with the basic paradigm of AGN playing a central role in quenching star formation in the most massive galaxies. For example, recently, Spilker *et al.* (2018) found molecular mass outflow rates a factor of two higher than the SFR (the ratio of the outflow to star formation rate is typically referred to as the mass loading factor) in a star forming galaxy at $z = 5$ that, to all appearances, lacks an AGN. This is surprising because large mass loading factors at low-redshift are usually seen in galaxies with luminous AGN. Post-starburst galaxies, without AGN signatures, are seen to retain large molecular gas reservoirs, although they have a deficit of dense gas. How these results fit into the simple model for galaxy quenching at $z < 2$ remains unclear. Probing stellar and black hole mass assembly within similar populations requires wide-area FIR and X-ray surveys to find luminous AGN, and deep IR and X-ray surveys, including spectrsocopy, to measure faint star formation and black hole accretion in moderate mass galaxies. Co-spatial, multi-tiered X-ray and FIR surveys over substantial areas that include a range of environmental density and densely sample the luminosity function and redshift are needed to make progress.

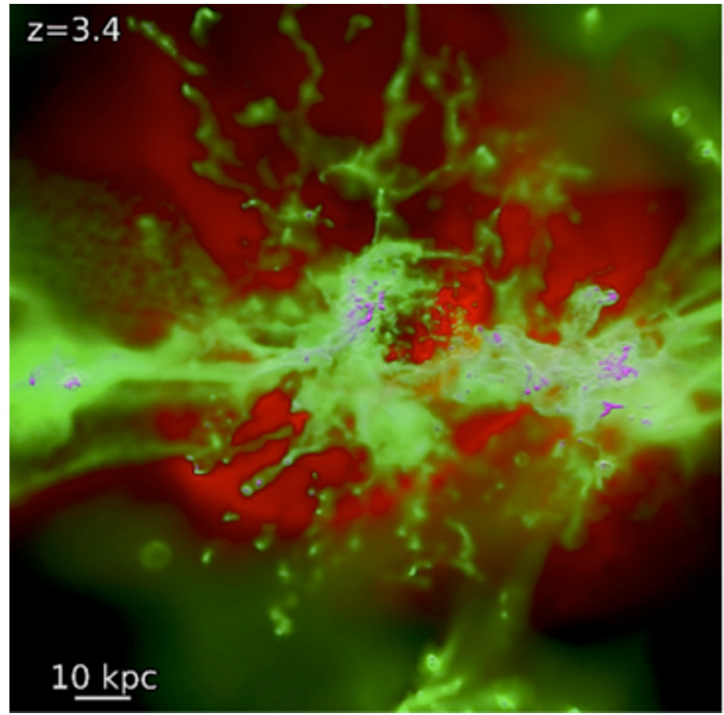

**Fig. 2.2—6. Galactic Feedback at High Redshift.** *Simulation of a Milky Way mass progenitor galaxy at z=3.4 from Hopkins et al. (2014) showing the multi-phase nature of outflows and feedback driven structure in in the cold (magenta), warm (green) and hot (red) gas. Detecting and analyzing these components at cosmic noon and earlier will require the next generation of multiwavelength space observatories.*

We are currently limited to studying only bright, actively accreting AGN, particularly as we move back in time. The most luminous of these may follow a different evolutionary scenario than more moderate mass galaxies. The production of luminous quasars requires a major merger, fueling a burst of star formation and triggering rapid growth of a supermassive black hole. In this merger-driven scenario, the AGN goes through an obscured growth phase, where the accretion disk is hidden by a dust torus. This phase ends when the AGN launches winds powerful enough to blow away some of the obscuring gas and dust, revealing the central source (a quasar) in the optical, UV and soft X-rays. The hard X-rays and the far-infrared are best at penetrating the dust to reveal the growing supermassive black hole during these phases. As the winds clear away the circumnuclear dust, the galaxy's star formation is quenched by galaxy-scale, AGN-driven outflows. Less powerful AGN may not follow this path. Morphological studies show no enhancement of the merging fraction in AGN samples (e.g., Cisternas *et al.* 2011).

Since galactic outflows are by their very nature multi-phase, the signatures of feedback also need to be studied across many wavelengths. By probing the atomic and molecular gas as it responds to radiation and shocks, and directly measuring the mass and outflow rates of warm and cold molecular gas, the next generation of great observatories will be able to directly constrain models of SNe and AGN feedback (**Fig. 2.2-6**).

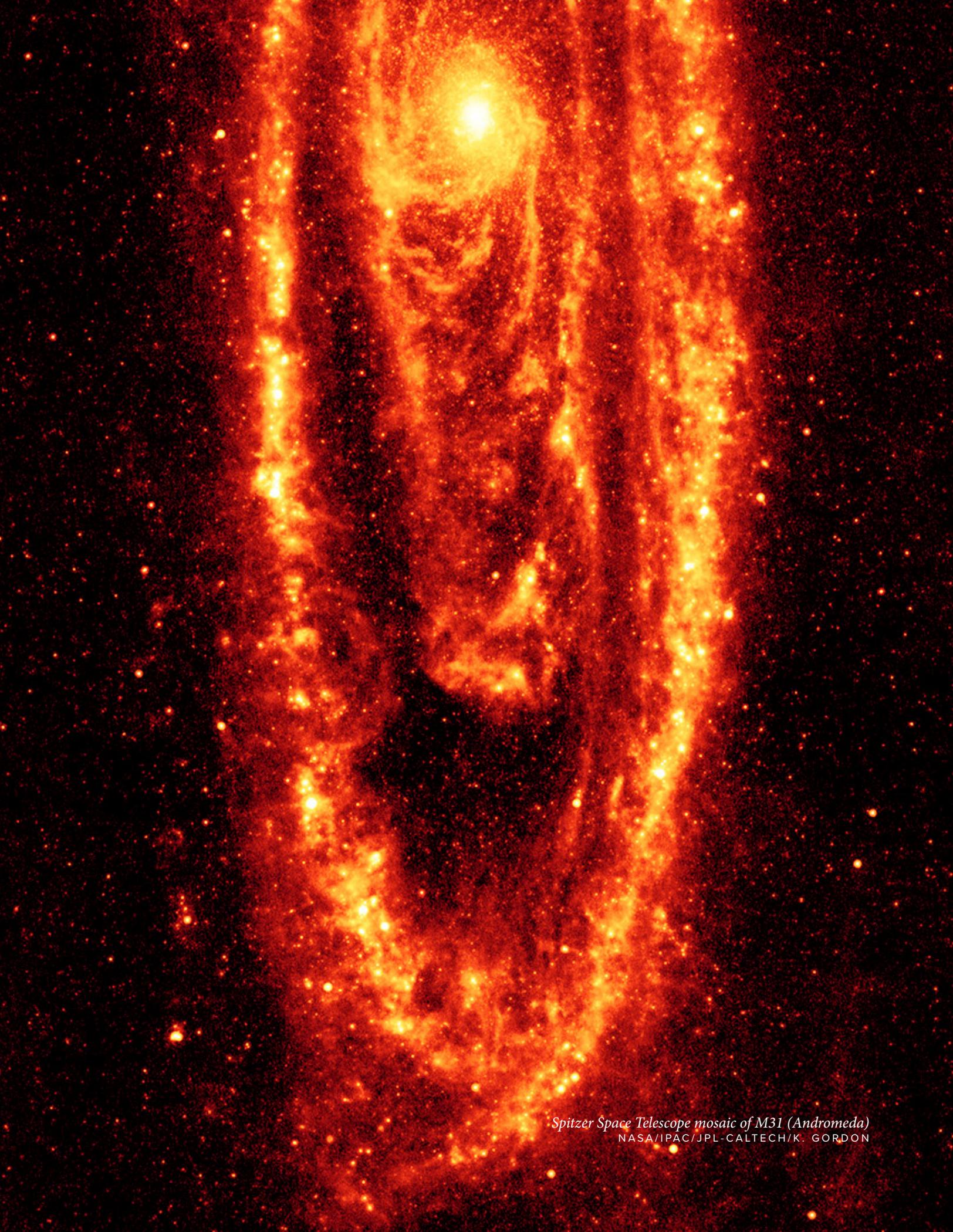
*Spitzer Space Telescope mosaic of M31 (Andromeda)*
NASA/IPAC/JPL-CALTECH/K. GORDON

The evolving gas fractions of galaxies and the presence of dust also play non-negligible roles in the ability to detect and measure AGN over cosmic time, as the number of obscured AGN increases with redshift. In the GOODS-S field, which has some of the deepest *HST* and *Chandra* imaging, only a few hundred X-ray detected galaxies have measured multiwavelength properties (Xue *et al.* 2011), and these are limited to the most massive galaxies (M* >$10^{10.5}$ $M_\odot$). The COSMOS survey delivered many obscured AGNs using stacking techniques, over a large range of stellar masses ($10^7$–$10^{11}$ $M_\odot$) and redshifts (Paggi *et al.* 2015, Mezcua *et al.* 2016, Fornasini *et al.* 2018). Teasing out the exact relationship between black holes and their hosts has not yet been possible, due to the limited sensitivity and resolution of telescopes operating at the requisite wavelengths.

Finding obscured AGN, an important stage of black hole growth, requires FIR and X-ray telescopes matched in sensitivity and resolution to *JWST*. The match in resolution is essential for counterpart identification, since many galaxies are contained in one *Spitzer* beam size. At $z \sim$ 1-3, *JWST* will make strides in our understanding of black hole--galaxy coevolution by resolving the centers of galaxies and picking out obscured AGN (Kirkpatrick *et al.* 2017), but beyond z~3-4, finding obscured AGN will require the next generation FIR and X-ray observatories. For example, line sensitivities below ~$1 \times 10$-20 Wm-2 in the far-IR, well beyond the reach of *JWST* and 100-1000x fainter than what was achievable with *Herschel*, will enable measurements of key diagnostics of accretion and star formation in typical galaxies at these epochs.

### *HOW DO GALAXIES ACCRETE GAS, MAKE METALS, & INTERACT WITH THE CGM & IGM?*

Galaxies grow in an evolving equilibrium between accretion from circumgalactic gas, star formation, and powerful galactic outflows. Understanding this "baryon cycle" is a key challenge for galaxy formation models, connecting small scale processes such as local feedback and enrichment of the ISM from supernovae and stellar outflows, to large scale flows of gas through the halos of galaxies. The gas involved spans temperatures from tens of degrees to millions of degrees and so necessitates a multi-wavelength approach from the FIR to X-rays. Galaxies are not completely isolated and can be strongly affected by their intergalactic environments. Besides feedback from AGN and starbursts, ram-pressure stripping of galaxies in cluster or group environments, for example, can play an important role in quenching star formation.

The metal content of gas in galaxies, both locally where it is produced in the ISM, and on large scales in the circum- and inter-galactic medium, constitutes a powerful and unique probe of baryon cycling and the galactic ecosystems (see **Section 2.4.2**). Despite the rapid pace of new observational and theoretical insights, the absolute chemical enrichment history of the gas in galaxies remains elusive. In part, this is purely an observational limitation — the typically-employed rest-frame optical indicators become challenging to observe from ground, as they redshift into infrared passbands. More significantly, the strong emission lines employed by current and planned abundance surveys retain the same decades-old systematic uncertainties impacting their conversion to underlying metal abundances. These uncertainties are principally impacted by unknown temperature structure in the ionized gas of galaxies, leading to the remarkable result that we do not know if galaxies in the local universe have, on average, supersolar or subsolar metal abundance (e.g., Kewley *&* Ellingson 2008). *JWST* will employ rest frame optical methods to measure abundance in moderate mass, low-attenuation galaxies out to $z \sim$ 3. And although the faint "auroral" lines can be two orders of magnitude weaker than strong abundance-sensitive transitions, *JWST* will build on recent ground-based success in detecting these lines in bright galaxies to yield temperature-unbiased metallicities (Sanders *et al.*, 2015).

While the strong gas temperature dependence of classical abundance measures is an important hurdle to overcome, another significant challenge relates to their sensitivity to dust extinction. Most of the star formation in the Universe has occurred in highly obscured regions (e.g., Whitaker *et al.*, 2017), making it inaccessible to UV and optical abundance tools. This has presented only limited difficulties locally, where highly obscured galaxies are rare. But future efforts to chart the history of metal enrichment of gas in galaxies through the peak of cosmic star formation will require the use of tracers in the infrared and X-ray regime that can penetrate high dust obscuration.

New approaches under development to measure the metal content of gas can resolve these local uncertainties and chart, for the first time, the full chemical enrichment history of the Universe, from the nearly pristine proto-galaxies driving the epoch of reionization to the massive and metal rich galaxies where most stars reside today. To make progress on this fundamental goal, we must leverage a powerful multi-wavelength suite of tools, coupling absorption and emission line studies in the UV and X-rays in galaxies and their haloes, recalibrated traditional optical emission line metallicity techniques, new FIR abundance tools that are insensitive to temperature and dust obscuration, dust emission as a secondary metal abundance indicator, and even radio-continuum free-free emission as a promising new metal abundance normalization (Croxall, *et al.* 2013; Ferkinhoff, *et al.* 2015; Fernandez-Ontiveros *et al.* 2016; Smith *et al.* 2019).

Multi-wavelength observations are essential to the study of both the hot intra-cluster medium and the stripped cooler gas from galaxies, including resolved and integrated dust emission (FIR), and interstellar gas in emission and absorption from coolest (in the sub-mm/mm) to hottest (in the UV and X-ray). The spatial, thermal, chemical, and kinematic state of the IGM from which galaxies and clusters grow, can provide important constraints on early, and late galactic feedback. To achieve these goals, new capabilities are required to conduct sensitive observations of distant galaxies and the faint gas that surrounds low-redshift galaxies. Ultimately, we want to better understand the intimate connection of the IGM, the CGM and the gas in galaxies, and how this interdependence evolves with galaxy mass, environment, star formation history, and the growth and activity of supermassive central black holes over cosmic time.

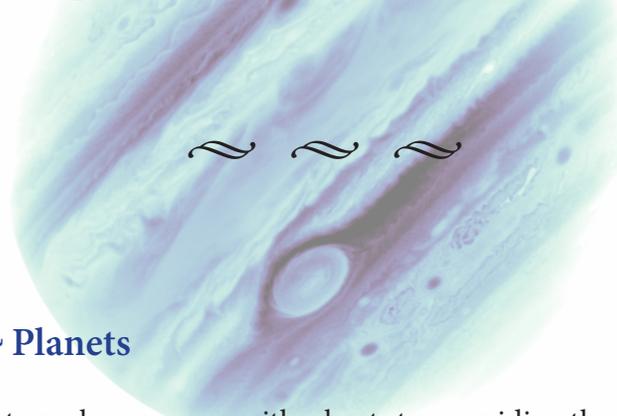

## 2.3. Origin *of* Life *&* Planets

Life is thought to be a planetary phenomenon with a host star providing the necessary energy. Tidal pools along shorelines of a water-bearing (but not too wet) world are good places to solve the twin miracles that are the biochemical origins of life: metabolism (chemical disequilibrium can be tapped as a local energy source) and reproduction (polar clays can "spontaneously" give rise to structures capable of collecting and separating genetic material which are permeable to water). The collapse of a molecular cloud core into a protostar plus planet-forming circumstellar disk is the first step. The star-forming environment (e.g. the richness and density of the host star cluster as well as the ambient ionizing radiation) may also play a key role in dictating the initial conditions of planet formation, driving organic chemistry, and ultimately the frequency of potentially habitable worlds (Lichtenberg *et al.* 2019). Interactions with the host star play a fundamental role in both the evolution of the proto-planetary disks and the long-term evolution of emergent planetary systems. Physical and chemical processes spanning many orders of magnitude in density, temperature, and size require observation over a wide range of wavelength from the far-infrared and sub-mm (tracing key molecular species and atomic fine structure lines) to gamma-rays (tracing the distribution of 26Al, whose decay is a critical heat source in forming proto-planets).

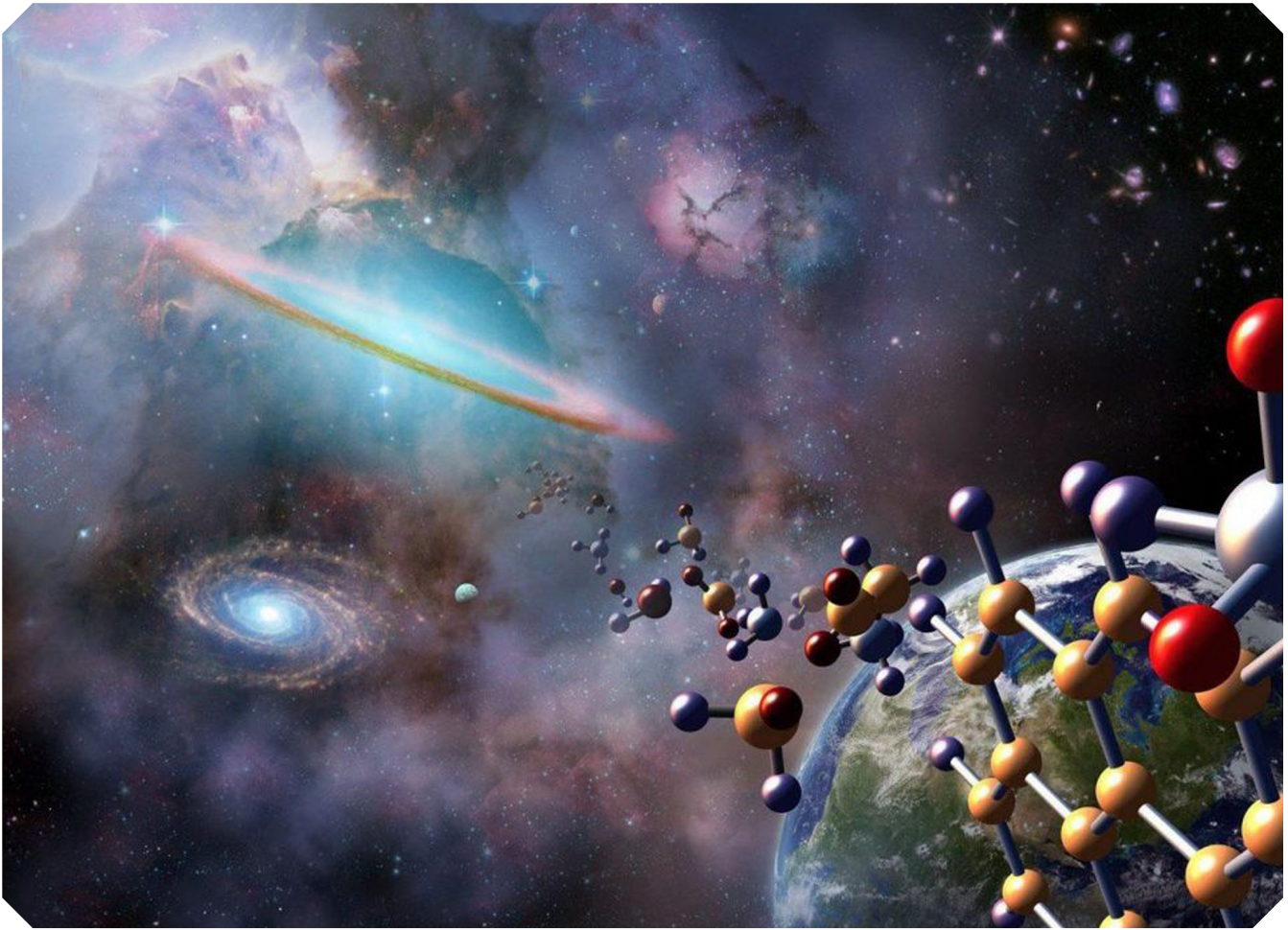

**Fig. 2.3—1. From the ISM to Life.** *Molecules traced from the ISM to collapsing cloud cores, to circumstellar disks, to planets, and the biochemical origins of life (Jenny Mottar).*

### 2.3.1   Planet Formation *&* Origin *of* Life Science enabled *by the* Great Observatories

#### CIRCUMSTELLAR DISKS *&* PLANET FORMATION

Planets form in circumstellar disks that are a nearly universal outcome of the star formation process itself. Which environmental aspects of star formation impact planet formation and prospects for habitability? Planet formation is complex, likely involving several physical processes including collisional growth of solids, particle aggregation in pressure bumps, non-linear solid growth regimes such as streaming instability and pebble accretion, rapid gas accretion onto critical cores (or prompt local gravitational instability), and subsequent orbital migration. A tremendous amount of information has been gathered over the past 30 years concerning the structure and evolution of circumstellar disks from IRAS, *HST*, ISO, *Spitzer*, *Herschel*,

WISE, and many ground-based observatories including ALMA. This information characterizes the initial conditions and timescales of planet formation.

Theory predicts that planets clear some of the surrounding material in the disk dynamically as they form (Zhu *et al.* 2012). By modeling spectral energy distributions spanning the optical to mm wavelengths, including *Spitzer* spectra, large disk holes and gaps were studied in detail (e.g., Espaillat *et al.* 2014). SED-inferred clearings in disks were quite large, encompassing orbits beyond Neptune, or about 30 - 50 AU. These large clearings were confirmed with ground-based high contrast imaging of scattered light tracing small grains (e.g. Avenhaus *et al.* 2018) and millimeter-wave imaging in thermal emission tracing larger grains (e.g. Andrews *et al.* 2011). More recently, the high-resolution of ALMA imaging has found much smaller AU gaps in the large dust grain distribution of disks (ALMA Partnership *et al.* 2015, Andrews *et al.* 2018), pointing to the ubiquity of gaps in disks.

In addition to disk gaps, there are other indirect signatures of planets in disks that have also been revealed with a multiwavelength approach. Using *Spitzer* and *Herschel*, we have seen that even young 1 Myr old stars host disks that have undergone significant dust grain growth and settling, the first steps of planet formation (e.g. Grant *et al.* 2018). A diversity of disk structures has also been seen in ground-based near-IR and sub-mm data, such as spiral arms, warps, and vortices that have been linked to forming planets (Benisty *et al.* 2018). We have started to see evidence for planets interacting with the gas in disks and distorting the profiles of IR lines (e.g., Brittain *et al.* 2014) as well as in the mm with ALMA (Teague *et al.* 2018).

How does the disk structure and composition dictate planetary system architectures as well as composition? Surveys have identified thousands of exoplanets (e.g., Fischer *et al.* 2014). However, the vast majority are located less than 10 AU from their host star. Current facilities can probe disks down to radii of 10 AU (Bae *et al.* 2018). Therefore, we do not currently have much information on the detailed structure in the inner parts of disks for comparison to exoplanet statistics. However, we do have some information on the composition of proto-planetary disks in both the inner and outer disk. Connecting *Spitzer*, *Herschel*, and ground-based mm, SED modeling of disks has revealed their dust composition and water ice content (McClure *et al.* 2013). FUV emission lines have probed the gas in the inner most (less than 1 AU) disk (France *et al.* 2017). *Spitzer* spectroscopy, along with ground-based studies, has revealed water and molecules within 10 AU (Pontoppidan *et al.* 2010). FIR lines seen with *Herschel* and SOFIA probe farther out in the disk and also deeper in the disk atmosphere (Bergin *et al.* 2013), and ALMA has extended this work into the mm (e.g., Miotello *et al.* 2017). Clearly, a multi-wavelength approach is necessary to fully map the chemical complexity of planet-forming disks and link this to potential habitability in resulting planetary systems.

Finally, debris disks around main sequence stars were an unexpected surprise during the calibration of IRAS (Neugebauer *et al.* 1984). Statistical as well as individual object studies with ISO, *Spitzer*, *Herschel*, and WISE have attempted to place these discoveries, and the properties of our own Solar System debris belts in context (Meyer *et al.* 2007; Wyatt *et al.* 2008). Perhaps the best-studied target, highlighting the value of multi-wavelength space-based imaging and spectroscopy, is Fomalhaut (see **Fig. 2.3-2**). In this system, we observe multi-belt debris, not unlike our own asteroid and Kuiper belts, as well as perhaps planets, spanning a wide range of temperatures and orbital radii.

### *DISCOVERY & CHARACTERIZATION OF PLANETS*

Which formation processes dictate the properties of emergent planet populations that impact planet hab-itability? A crucial step towards answering this question is to assess the outcomes of planet formation by measuring the planet mass function, orbital separation distribution, and composition of planets, all as a function of each other, host star mass, and system architecture. A wide range of techniques are needed to obtain the complete census of exoplanets and search for critical dependencies.

*Direct Imaging*. *Hubble* has made fundamental contributions, obtaining some of the first images of planetary mass objects as companions to nearby stars. Early work focused on the detection of brown dwarf companions to normal stars, particularly in star forming regions where sub-stellar companions are much brighter than at late times (e.g. Lowrance *et al.* 1999). Further work focused on the search for planetary mass companions to stars with debris disks (e.g. Schneider *et al.* 2014). Finally, Kalas *et al.* (2008) reported detection of Fomalhaut b, at the same time as the announcement of the first detections of gas giant planets around HR 8799 (Marois *et al.* 2008). Follow-up of the magnitude of the scattered light signal, variability, and subsequent astrometry, have left the original interpretation in question. However, the importance of achieving these high-contrast imaging results with *HST* should not be underestimated. Regardless of the ultimate interpretation, optical and near-IR *HST* observations of Fomalhaut b provide a powerful characterization of planets (or swarm of planetesimals in collision) dynamically connected to the long-studied debris disk. Characterization observations with *HST* of HR 8799 soon followed their discovery. This multi-planet system, in a debris disk system surrounding an intermediate mass star, will remain a touchstone for many years to come. *HST* has also recovered Beta Pic b, detected with ground-based telescopes (Lagrange *et al.* 2009). *Spitzer* has also played a critical role in direct imaging, by providing powerful infrared upper limits for planets that have not been detected in thermal emission (e.g. Janson *et al.* 2012).

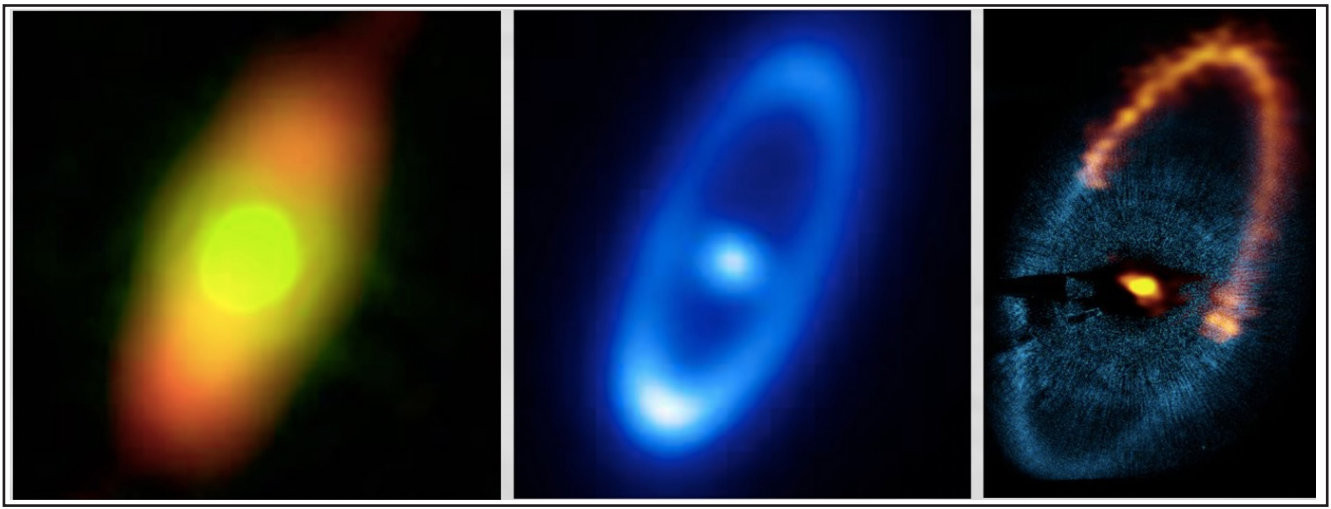

**Fig. 2.3—2. Multiwavelength observations of the Debris Disk surrounding Fomalhaut.** *These data, made with Hubble, Spitzer, Herschel and ALMA, trace dust grains created by collisions within a planetesimal belt extending from 13 to 19 AU around the 2 solar mass star. This belt – although not detected directly - replenishes the grains lost to Poynting-Robertson drag and radiation pressure, as traced by the IR and sub-mm measurements (Stapelfeldt et al. 2004, Acke et al. 2012). The narrow width of the belt may be the result of shepherding by planets (Boley et al. 2012). The IR and visible light measurements with Spitzer and Hubble put tight limits on the masses of the planets, and the Hubble data has detected a planet candidate (Kalas et al. 2005, Marengo et al. 2009, Janson et al. 2012). These data illustrate how multi-wavelength observations map the structure of planetesimal belts, place constraints on the rate of dust production and the properties of the grains, and directly constrain the properties of planets.*

*Characterization*. *Hubble* and *Spitzer* have many firsts in the realm of exoplanet characterization. While many of the targets were initially discovered via ground-based radial velocity or transit observations, space-based transmission spectroscopy, and more recently, space-based thermal infrared secondary eclipse observations are now primary characterization techniques.

The first direct chemical analysis of the atmosphere of a planet orbiting another star was done with *HST*/STIS (Hd 209458b; Charbonneau *et al.* 2002). This opened up an exciting new phase of extrasolar planet exploration, where astronomers can compare and contrast the atmospheres of planets around other stars, and potentially search for chemical biomarkers of life beyond Earth. Its atmospheric composition was probed when the planet passed in front of its parent star, allowing astronomers for the first time ever to see light from the star filtered through the planet's atmosphere. Scientists detected the presence of sodium in the planet's atmosphere, at levels less than predicted, leading to one interpretation that high-altitude clouds in the alien atmosphere can be inferred.

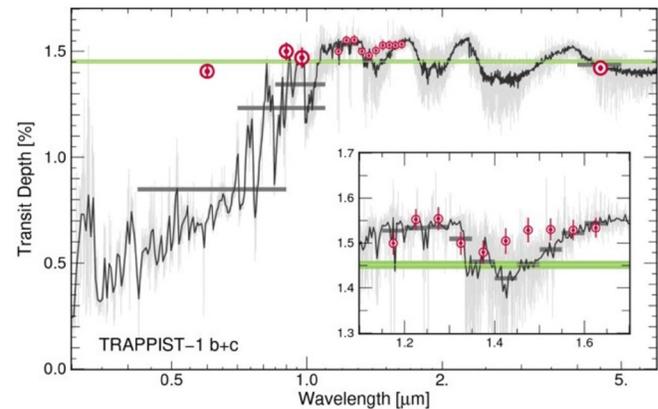

Fig. 2.3—3. **Habitable Zone Rocky Exoplanets.** *Transit depth of Trappist-1 b+c as a function of wavelength (Zhang et al. 2018). These small rocky bodies in the habitable zone appear to have bulk densities consistent with about 10 % water. Multi-wavelength characterization of the atmospheres, including HST/WFC3 (inset), is still ongoing.*

*Spitzer* pioneered the technique of secondary eclipse observations for exoplanets. Firsts included the direct detection of thermal emission from an extra-solar planet (Deming *et al.* 2005, Charbonneau *et al.* 2005), and the detection of atmospheric water vapor (Tinetti *et al.* 2007, Knutson 2007). The ability to characterize atmospheres in both reflected light (with *Hubble*) and emission (with *Spitzer*) provides a whole that is more than the sum of the parts (e.g. Sing *et al.* 2016). The characterization of these atmospheres has steadily pushed to smaller sized planets, around smaller host stars, such that the limit in relative photometric precision with *HST* and *Spitzer* (30-50 ppm) can be used to study mini-Neptune and super-Earth planets (e.g. GJ 1214 b; Kriedberg *et al.* 2014). A landmark achievement has been the discovery, and characterization, of rocky planets in the habitable zone around a nearby late M dwarf at the hydrogen-burning limit, Trappist-1, using *Spitzer*, *HST*, Kepler and ground-based data (Zhang *et al.* 2018 and Fig. 2.3-3).

Comparison of phase resolved observations enable us to understand whether energy redistribution is efficient or hampered, yielding hot spots offset from the sub-solar point. Putting together a global energy budget where reflected light studies are combined with thermal emission phase curves provides a powerful technique to study the structure of the atmospheres as well as prevailing wind patterns linked to energy redistribution. Spectral retrieval techniques are also providing initial estimates of atmospheric composition. Ultimately these inventories of key molecular species will enable confrontation of new data with models of formation, migration, and evolution of the atmospheres.

*Transits*. The exoplanet revolution began with radial velocity reconnaissance on modest aperture telescopes dedicated to long-term monitoring programs. It continued with miraculous discoveries of their nature and diversity. Even the null results from *Hubble* on transits in the 47 Tuc cluster remain noteworthy. However demographic studies enabled by the Kepler and K2 missions have accelerated the pace of discovery to an unimaginable pace. Kepler has found more exoplanets than any other facility, enabled occurrence rate estimates over a wide range of planet to host star size, and orbital period, provided enhanced samples of multi-planet systems, including the possibility of using transit timing variations to provide independent mass estimates for some planets. There is now, for the first time, an estimate of the number of Earth-like planets in the habitable zone. We also understand that gas giant planets are rare around lower mass stars, while lower mass planets (like super-earths) are more common. There is a local minimum in the occurrence rates of planets between 1.5-2 Earth radii (Fulton *et al.* 2017).

### *THE STAR DISK/PLANET CONNECTION*

The connection between proto-planetary disks and their host stars is still largely unexplored and is best probed with a multi-wavelength approach. The lifetime of the disk places an upper limit on the timescale for giant planet formation given that gas must be present for gas giant planets to form. The presence of gas in the disk is dictated by the rate at which gas is eroded by photo-evaporative winds. However, different mass loss rates are predicted by models of X-ray, EUV, and FUV photo-evaporation (e.g., Alexander *et al.* 2014). Observational work attempts to link high-energy X-ray radiation from the star (measured with *Chandra*) to MIR emission lines (measured with *Spitzer*) thought to be diagnostic of high-energy radiation fields (Espaillat *et al.* 2013, Pascucci *et al.* 2007). These works have shown that disk photo-evaporation is likely dominated by X-ray photons, providing constraints to disk clearing models and hence disk lifetimes.

Radiation from young stars surrounded by planet-forming disks is characterized by large amplitude periodic, as well as aperiodic, variability from x-rays through the infrared. Geometric structures in the disks also appear to vary as diagnosed by observations of the MIR continuum with *Spitzer* (Muzerolle *et al.* 2009; Espaillat *et al.* 2011). Coordinated observations have been undertaken to search for empirical connections between star and disk emission. Work with *Chandra* and *Spitzer* found that the X-rays due to accretion and coronal emission are not responsible for changes in the IR emission (Flaherty *et al.* 2014), yet may well drive changes in millimeter emission (Cleeves *et al.* 2017). Observations utilizing *Hubble* and the IRTF demonstrate a connection between FUV emission lines, NUV-derived accretion rates, and NIR emission, linking gas and dust in the inner disk to accretion onto the star (Ingleby *et al.* 2013). Space-based X-ray and time-domain studies coordinated with ground-based Hα measurements also find connections between accretion onto the star and the properties of the inner disk (e.g., Guarcello *et al.* 2017).

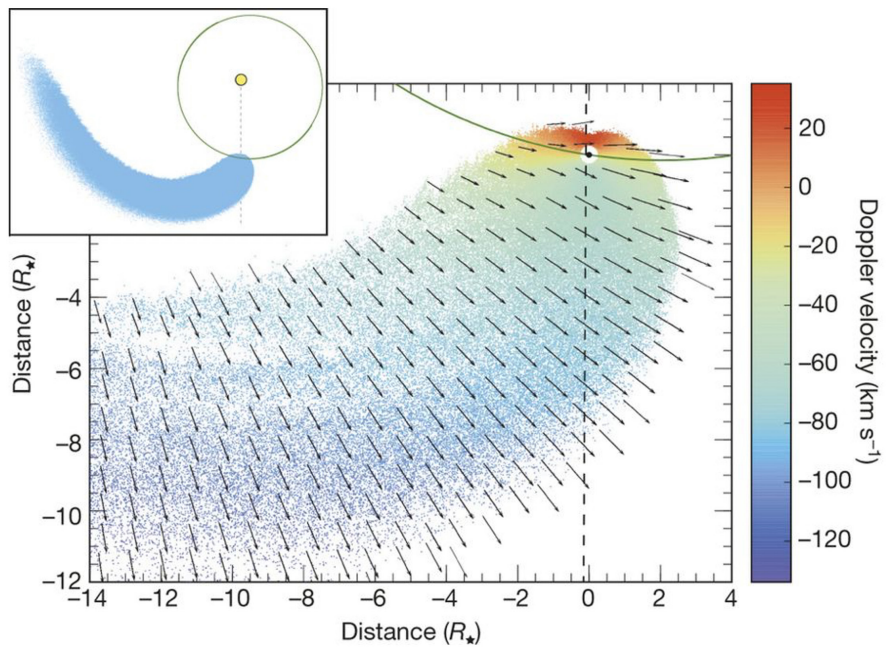

Fig. 2.3—4. **Star-Planet Interactions** *Models of a photo-evaporating cometary cloud of hydrogen in the vicinity of GL436b (Ehrenreich et al. 2015). EUV and X-ray emission from the central star are thought to drive this flow, detected with HST.*

Over the course of a planet's lifetime x-ray and extreme UV photons will cause significant mass loss via photo-dissociation. This process is directly observable today in our own Solar System. Dennerl (2006) noted charge exchange emission emanating from regions around Mars, extending for more than 2 planetary

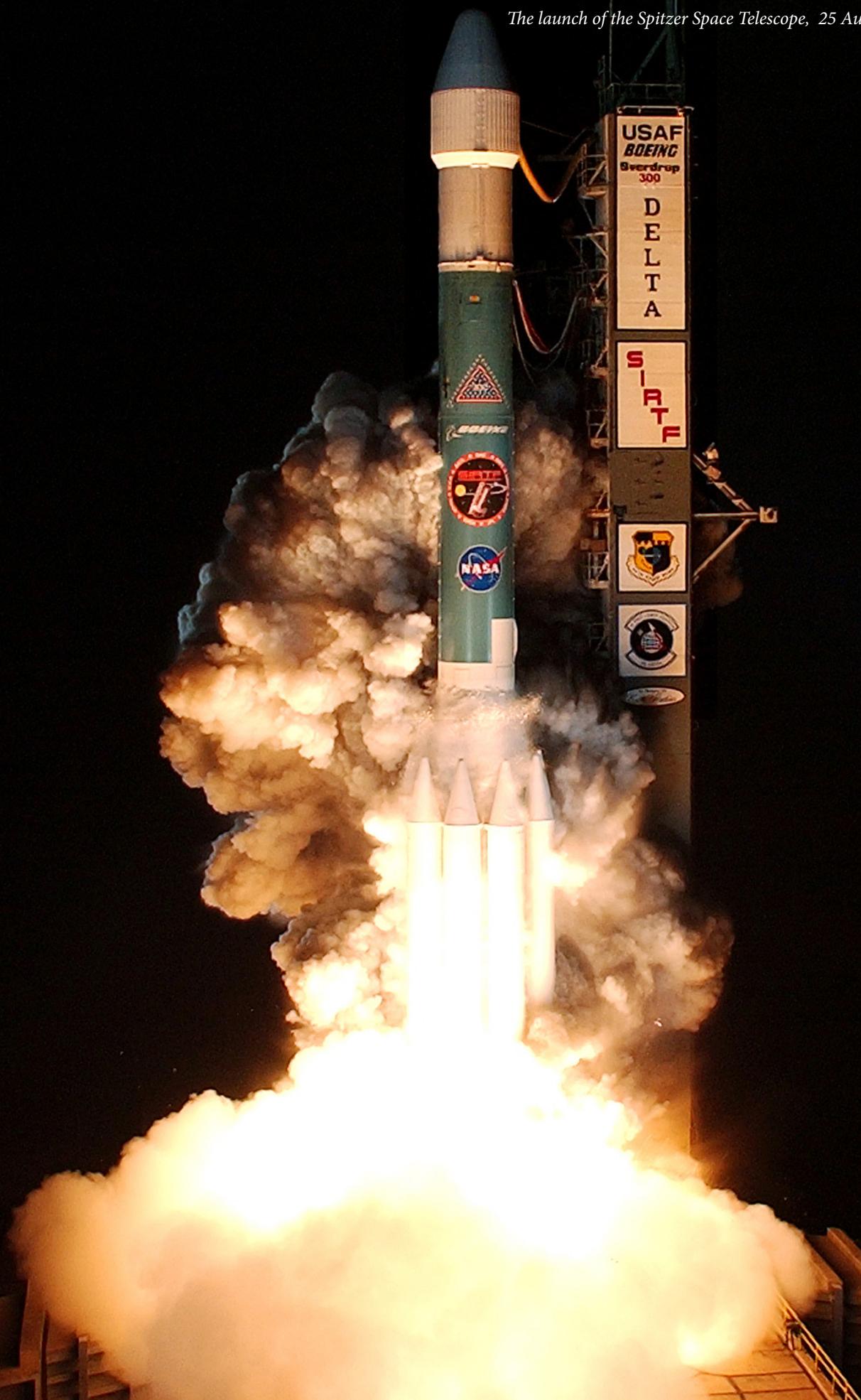

*The launch of the Spitzer Space Telescope, 25 August 2003*
NASA

radii. This is clear evidence not only of an extended thin atmosphere, but that the exosphere is interacting with the solar wind and actively being ablated (Fig. 2.3-4). In the case of exoplanets, X-ray transit measurements of HD 189733b directly measured the exosphere extending to 1.75 planetary radii above the optical cloud tops (Poppenhaeger et. al. 2013). The data indicate a symmetric low-density (1011 cm3) exosphere of about 20,000 K. Much work remains to be done in understanding star-planet interactions, particularly in understanding the role this plays in creating the local minimum in planet radii between 1.5–2 $R_\oplus$ close to their parent stars.

After the formation of a planetary system, heating by stellar UV and X-ray radiation can dramatically the influence the evolution of planetary atmospheres. Stars and planets can have bi-directional interaction through gravity, tides, and magnetic fields. To first order, these interactions in cool stars are unavoidable. A hot Jupiter orbits its host star in about 4 days, meanwhile, the stellar rotation period of a middle-aged star is typically 2-4 weeks. In solar type stars with a convective zone, this must result in tidal stress within the stellar convective envelope which will tend to increase the rotation period of the star and hence dynamo activity. The increased dynamo leads to increased magnetic field strength. Observationally this will make the star appear younger in terms of geochronology. This was most obviously observed in the binary system HD 109733, wherein the exoplanet host has an inferred age of about 1 Gyr, the isolated star without a detected planet has an inferred age of about 5 Gyr (Poppenhaeger *et al.* 2015).

In systems with planets at separations of few stellar radii from their parent stars, the magnetic fields of planet and star can lead to magnetic reconnections and flaring activity. The magnetic field can drag ionized material evaporating from the outer atmosphere of the planet and form a cometary tail of plasma, a magnetized stream of gas accreting onto the star or even Roche-lobe overflow in the most extreme systems. The magnetic field of the planet can be as strong as to shield the planet itself from the blast of violent coronal mass ejections (CME, Cohen *et al.*, 2011) or be the ultimate energy source for a flare capable of ripping off a lunar mass of material from its atmosphere. Beyond this, the enhanced magnetic field near the stellar surface can form very active regions on the star increasing the activity of the star in the X-ray band. The detectability and the significance of star-planet interaction (SPI) at high energies (X-rays and FUV) is demonstrated by studies such as Maggio *et al.* (2015).

### *THE EVOLUTION OF COMPLEX ORGANICS NEEDED FOR LIFE*

What initial conditions in the galaxy are needed for life (e.g. chemical evolution)? Carbon, hydrogen, oxygen, nitrogen, sulfur, and phosphorus (CHONSP) are the elements most commonly found in biomolecules on Earth. Understanding their evolution from the diffuse interstellar medium (ISM), to molecular clouds, to forming stars and planet, and ultimately through the biochemical origins of life is the major scientific question of the next century. Can complex organics form in the ISM, driven by ion-molecule reactions or reactions on dust grains? Must all key steps take place on planetary surfaces? How are the major elements above, and key molecules comprised of them, stored in the ISM and transmuted in circumstellar disks? In order to trace the major reservoirs of volatiles species (particularly carbon, nitrogen, and oxygen needed for water and life), a multi-wavelength approach is essential: from the ionizing radiation in the X-ray to the ultraviolet that drives chemistry, to electronic transitions in the UV/blue, to the vibrational bands of the infrared, to the rotational transitions in the millimeter.

Knowing the abundance and form of oxygen and carbon in the ISM is central to any predictive model of the evolution of complex organics in the galaxy. Space-based characterization of the interstellar medium began with Copernicus in the UV, taking inventory of the gas phase elements depleted in the diffuse ISM, presumably incorporated into dust grains. This legacy was continued with IUE, *Hubble*, and FUSE continued this legacy, together with ground-based millimeter wave astronomy, driving our understanding of the composition and phase of the stuff between the stars. In what form is oxygen in the bulk ISM, relative to solar abundances in the neighborhood of the Sun? Small explorer class missions (such as *ODIN* and *SWAS*) failed to resolve the puzzle, but the mystery deepened in important ways. *Herschel* has also contributed to our understanding, as have x-ray spectroscopy in the ISM. However we still do not fully understand in what form most oxygen in the Universe exists.

IRAS first took census of galactic dust in the mid to far-infrared, identifying the importance of PAHs as an important form of carbon in the ISM. ISO characterized carbonaceous and silicate dust in the ISM. This dust is produced in a range of astrophysical environments, from supernova remnants to AGB stars. Yet it is still debated whether carbon-rich grains dominate silicates, not to mention in what form those carbon grains exist (graphite, HACs, PAHs, or another form). Understanding the carriers and evolution of nitrogen carriers in the solid ISM is just beginning. We know that complex organics exist in the ISM as long-chain and cyclic molecules are observed in molecular clouds (as well as within meteorite samples). The evolution to complex organic molecules is driven by ion chemistry so the radiation environments of molecular clouds matters a great deal. In addition to this connection with X-ray observations, X-ray spectroscopy is another tool yet to be fully exploited to help track the evolution of C, N, and O in the ISM.

Ultimately these materials are processed through the star and planet formation process. Shocks due to infalling material from cloud cores into protostars and circumstellar disks can sublimate a significant fraction of solids (Neufeld & Hollenbach, 1994) putting volatiles back into the gas phase. As planet-forming disks form and evolve, the interplay between solids and gas continues, locking volatiles into solids beyond species-specific ice-lines as a function of orbital radius (with colder material farther from the young star, as well as deeper into the disk midplane). Even refractory species can be transmuted, releasing elements into the gas phase. Oxidation of carbon-rich grains, partially due to radicals tracing high-energy radiation, releases carbon into an oxygen rich nebula resulting in enhanced CO (Gail *et al.* 2002). These carbon grains, formed in the unique environments of carbon-rich evolved stars, cannot reform as solids once broken down. This chemical complexity, which depends on the radiation field from the star, and impacts the observed elemental abundances as a function of orbital radius, contributes to the composition of forming planets along with the dynamics of migration. As we characterize worlds increasingly like our own, through detailed spectroscopic observations, we cannot help but speculate on how to infer the presence of life on other worlds with future observations.

### 2.3.2. Questions *for the* Next Decade

#### *CIRCUMSTELLAR DISKS & PLANET FORMATION*

Several proto-planet candidates have been directly imaged within circum-stellar disks using high contrast imaging in the visible and infrared (e.g. PDS 70b, Hd 100546b; Keppler *et al.* 2018, Quanz *et al.* 2015). It is difficult to distinguish proto-planets in formation from "disk features" which could be transient density perturbations, a structure that is a precursor to a forming planet, or a compact object actively growing in mass. Most models predict formation of a circum-planetary disk surrounding a critical core mass object (e.g. 3-30 MEARTH in a gas–rich disk). An unresolved compact proto-planet surrounded by a tiny circum-planetary disk may be fed by an accretion shock from a vertical flow from the active layers of the circum-stellar disk. Disentangling these three possible emission mechanisms requires a multi-wavelength approach from as broad a wavelength range as possible. Ratios of emission lines can indicate ionization state and composition. Future work, e.g. with *JWST* and other facilities, will include direct empirical constraints on the dominant physical processes at play in gas giant planet formation. The best way to test predictive theories of planet formation is to observe planets in formation, and characterize them in terms of i) temperature, luminosity, orbital location, possibly accretion rates, ii) nature of any attendant circum-planetary disks, and iii) local physical conditions of the circum-stellar disk. To make further progress in identifying forming proto-planets and how/if planets inherit the properties of their proto-planetary disks, we need a multi-wavelength space-based approach to: 1) image more gas giant proto-planets; 2) decipher their effect on line profiles; 3) and resolve the composition and structure of the innermost, terrestrial planet-forming region of the disk which will allow us to directly compare with the parameter space probed by most exoplanet studies. High spatial resolution optical and near-IR spectroscopy will allow us to search for accreting proto-planets via hydrogen emission-lines within disks located in nearby star-forming regions. High spectral resolution will enable us to scrutinize line profiles for signatures of distortion from proto-planets.

There are also great advancements to be made with mapping the location of gas in the inner disk. Electronic transitions from molecules in the EUV/FUV such as H2, CO and other species trace gas closest to the star while infrared vibrational emission bands probe the warm molecular content (including the dominant constituent H2) of the terrestrial planet zone, as well as high altitude layers of the disk. The far-IR and sub-mm can access the cooler outer as well as interior parts of the disk, utilizing emission from optically-thin fine structure lines which are extremely sensitive tracers of remnant gas as well as reservoirs of volatile species. This multi-wavelength approach is necessary to map molecular abundances as a function of orbital radius and stellar properties, as well as isolate snowlines in disks, in order to map a planet's location in the disk to its composition. High-spatial resolution space-based optical and infrared imaging (requiring large apertures) can trace scattered light in the innermost disk, allowing us to search for gaps and structural distortions due to low-mass proto-planets. Finally, multi-wavelength studies of debris disks in scattered light, constraining the albedo as a function of wavelength, searching for spectral features in reflected light and emission, and measuring the emissive radiation as a function of orbital distance, will provide fundamental constraints on dust particle size and composition, including the location of ice lines thought to be critical to planet formation.

### *DISCOVERY & CHARACTERIZATION OF PLANETS*

In the near term, multi-wavelength studies of atmospheric characterization in transmission, and emission, are planned through GTO, ERS, and GO programs on *JWST* (e.g. Beichmann *et al.* 2014; Bean *et al.* 2018) as well as continuing programs on *HST*. It is expected that *JWST* will be used to characterize dozens (if not hundreds) of planets, charting out the diversity of atmospheres for close-in planets. Although challenging, transit observations with *JWST* can, in principle characterize Earth-sized planets in the habitable zone of very late type stars (similar to the Trappist-1 system), providing constraints on important volatile species and cloud properties. Imaging with *JWST* will also have extraordinary sensitivity to low mass companions, particularly when exploited to achieve background-limited performance orders of magnitude below ground-based limits. Nearby, young, faint primaries can be searched for planets < 15 $M_\oplus$ in the thermal IR with *JWST*. *HST* still has a valuable role to play until replaced with more capable UV/visible facilities. It is essential to measure exoplanet transits in scattered light: if a Rayleigh slope can be identified then an absolute reference of the altitude-pressure relationship can be derived. This is crucial to tie relative absorption of molecules from infrared spectra to absolute abundances. Future missions in the mid-IR may measure thermal emission during secondary transits providing additional information on chemistry of atmospheres by detecting key species in absorption.

With results from Kepler, we have estimates of the frequency of planets from < 1 $R_\oplus$ to > 4 $R_\oplus$, as a function of orbital separation out to periods 250–500 days and host star property. The *Roman* microlensing survey will unveil the full demographics of exoplanets down to unprecedented masses (within the Einstein radius of about 1-10 AU). *Roman* will also make major contributions towards imaging planets in reflected light, taking the next steps in proving the feasibility of coronagraphic imaging and spectroscopy in space, possibly with a formation flying star-shade. Future missions could provide revolutionary capabilities for direct imaging. For example, an aperture larger than four meters could be used to image an Earth-sized planet in the habitable zone in reflected visible light. Similarly, a large aperture UV optical capability could provide the first census of such objects, characterizing their atmospheric diversity. A large aperture infrared telescope, with stable mid-infrared detectors, would permit photon-limited characterization of potentially habitable worlds and the search for atmospheric biosignatures. Ultimately, a mid-infrared inteferometer with baselines of 300 meters or more, and sensitivity exceeding *JWST*, will be required to fully characterize a large sample of diverse, potentially habitable worlds. A combination of studies in reflected light in the visible/near-IR, and thermal emission in the mid-IR will be needed to complete the search for, and characterization of Earth-like planets in the habitable zones of nearby Sun-like stars.

### *THE STAR DISK/PLANET CONNECTION*

To make further progress in understanding the connection between stars and disks, future work needs to take a multi-wavelength and multi-epoch, coordinated approach. Future higher resolution (and more sensitive) X-ray spectroscopy can trace the composition of accretion streams to detect any differences in material that builds up the star versus material that is sequestered in the disk which ultimately forms planets. Crucial measurements of the x-ray, EUV, and FUV line and continuum fluxes are still needed for young stars between 3-30 Myr to help explain the "last gasp" of primordial gas rich disks which could still form super-Earths to sub-Saturns with a range of gas to dust ratios. Detecting and characterizing young stellar populations in x-ray emission requires high spatial resolution, and wide-field imaging of star clusters. The ability to resolve multiple systems enables disentangling correlations between x-ray and infrared properties, as well as direct comparisons between components of systems. Directly studying the impact of variable high energy radiation on the chemistry, structure, and evolution of gas rich circumstellar disks will require coordinated x-ray, EUV/FUV, and infrared studies. Similarly, studying the impact of high energy radiation on planetary atmospheres requires sustained access to the x-ray and UV. Constraining levels of x-ray/EUV/FUV emission and monitoring variability as a function of host star mass, and stellar age provides vital information into the long-term evolution of planets, potentially explaining population dichotomies such as the observed gap in transiting planet radii at small separations.

### *THE EVOLUTION OF COMPLEX ORGANICS NEEDED FOR LIFE*

Tracing the evolution of volatile species such as C, N, and O into potentially life-bearing worlds is also an inherently multi-wavelength enterprise. Fundamental work in the ultraviolet, characterizing gas phase depletions along diverse, diffuse, lines of sight remains. Near-term work with *JWST* will enable infrared spectroscopy to be deployed with the spectral and spatial resolution needed, along with the required sensitivity to make major progress. One example is the potential to solve the long-standing puzzle of the ratio of carbon to silicate grains through absorption measurements (3.3/3.4 micron carbon complex versus 10 micron silicates) along the same lines of sight. We still have no idea what are the dominant refractory carriers of nitrogen in the ISM, even though they could dominate the cosmic abundance. Wavelengths longward of 30 microns, beyond the reach of *JWST*, provide access to a rich array of complex organics, nitrogen-bearing compounds, and water. Large complex organics, including amino acids, have been detected in the interstellar medium. It is as yet unclear whether delivering such compounds to planet-forming disks is a key step in the biochemical origin of life, or whether the needed ingredients are manufactured in situ on the surfaces of planets. Tracing the transmutation of C, N, and O in abundance and form, throughout the life cycle of the ISM, and incorporation into forming planets remains a priority. Future x-ray studies will also make important contributions. X-ray absorption is sensitive to heavy element abundance in both the gas and dust phase while studies of x-ray scattering halos can help constrain the maximum size of dust in the ISM.

Ultimately the understanding the emergence of life in the Universe will require its detection on other worlds. Non-equilibrium chemistry is not enough. The study of biomarkers is complicated and no single spectroscopic detection of a molecule can unambiguously identify life on another planet (Doloman-Goodman *et al.* 2018). However, with careful study, understanding both the diversity of planetary atmospheres as well as the context of the planetary system as a whole (stellar environment, composition, evolutionary state), it possible that multiple, consistent signs of life could be detected. For now, we can only focus on biosignatures based on the one example of Earth. This will likely require the confirmation of species like O2 in both the visible (reflected light) and infrared (thermal emission) for a potentially habitable world. Surveys to detect these will require large aperture UV/optical and infrared space-telescopes, as well as long-baseline mid-infrared interferometers. Yet we may be surprised as the first signs of life on other worlds could come in forms we cannot yet anticipate. The scope of these experiments will be daunting, but the significance of the results, positive or negative, may justify the undertaking.

$$\sim \sim \sim$$

## 2.4.  Fundamental Physics

Astrophysics offers two unique and complementary channels by which scientists can explore the fundamental physics of our Universe. First, the universe provides us with extreme and irreproducible experiments in strong-field gravity, high-energy interactions, and high energy-density physics via assorted natural phenomena including black holes, neutron stars, accreting galactic nuclei, exploding stars, and ultra-high energy cosmic rays. Second, the Universe serves as the ultimate backdrop for studies of its dominant and enigmatic diffuse components, the dark matter ($\Omega_c$ = 26%) and dark energy ($\Omega_\Lambda$ = 68%).

### 2.4.1   Fundamental Physics enabled *by the* Great Observatories

NASA's Great Observatories, and related multi-wavelength missions, have over the past two decades contributed to multiple important findings in fundamental physics.

#### *DARK MATTER, DARK ENERGY, & COSMOLOGY WITH THE GREAT OBSERVATORIES*

*Hubble* provided crucial precision photometry of high-redshift supernovae for both teams that contributed to the Nobel prize-winning discovery of Dark Energy in 1998 (Riess *et al.* 1998; Perlmutter *et al.* 1999). Since then *HST* has provided high-precision type Ia SNe observations out to redshifts $z > 1$ (Riess *et al.* 2004) and refined distances to the nearest SNe Ia via Cepheid variable stars (Riess *et al.* 2019), the tip of the red giant branch (e.g. Jang & Lee 2017; Hatt *et al.* 2018), and other calibrators (e.g. Huang *et al.* 2018). These measurements have enabled increasingly precise constraints on the Dark Energy equation of state, and led to surprising recent tension with the *Hubble* constant value inferred from Planck observations of the cosmic microwave background (Riess *et al.* 2019).

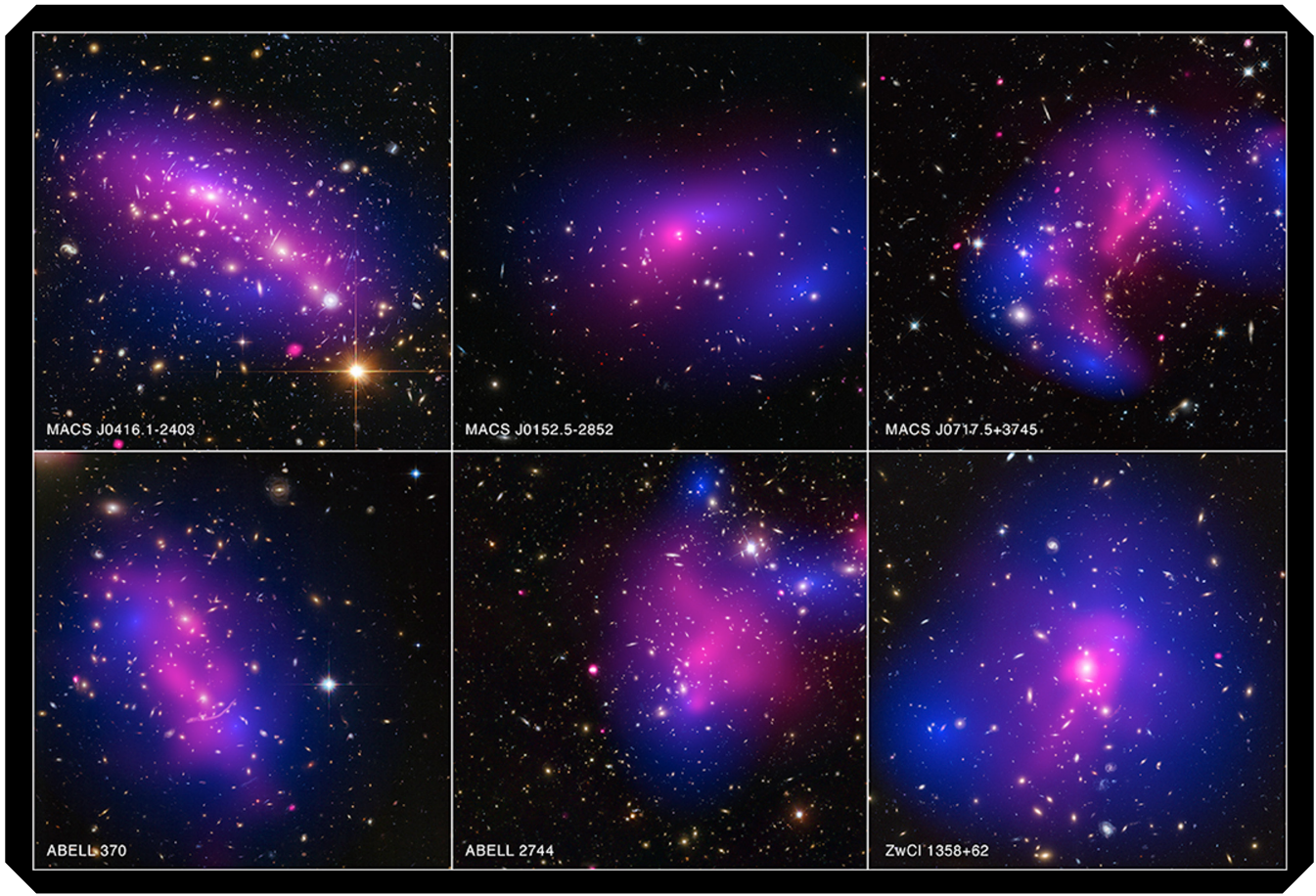

**Fig. 2.4—1. Dark Matter in Merging Clusters.** *Superposed images of galaxies (yellow / multicolor, from ground-based observatories), X-ray gas (magenta, from the Chandra X-ray Observatory), and dark matter (blue, from Hubble Space Telescope weak lensing observations) associated with six actively merging galaxy clusters. These coordinated observations by two Great Observatories have set world-leading constraints on dark matter self-interactions (Harvey et al. 2016). Image from Chandra X-ray Center.*

    The combination of *HST* and *Chandra* high angular-resolution observations of galaxy clusters proved to be a powerful tool for advancing our understanding of the dark matter and probing the cosmology of the universe (Fig. 2.4-1). Gravitational "weak lensing" of background galaxies, observed with *HST*, allows the construction of cluster "mass maps" that trace the dark matter in galaxy clusters, while *Chandra* X-ray imaging reveals the hot gas that accounts for ~90% of clusters' baryonic matter. Together these observations demonstrate, in colliding galaxy clusters like the "Bullet Cluster" (1E 0657–558), that the dark matter must be nearly collisionless, setting strong constraints on any self-interactions (Clowe *et al.* 2006; Harvey *et al.* 2015). *HST* observations of lensing effects in galaxy clusters and across the COSMOS field have probed the matter power spectrum within clusters (Umetsu *et al.* 2014) and in the universe at large (Massey *et al.* 2007). And *Chandra* and X-ray Multimirror Mission-Newton (*XMM-Newton*) observations of galaxy clusters have served to calibrate proxy measurements of cluster mass that are subsequently used to set cosmological constraints on the mean matter density, amplitude of the matter power spectrum, neutrino masses, and the properties of dark energy (Allen, Evrard *&* Mantz 2011). The SNe (primarily *HST*) and cluster cosmology

(primarily *Chandra*) measurements of key cosmological parameters provide critical independent checks as the systematics and biases of the two techniques are very different.

*Spitzer* has proven to be essential in the construction of our current census of massive clusters at low and high-redshift. The IRAC data have been extremely useful for identifying and characterizing high-z clusters since (1) *Spitzer* has been able to map large areas of the sky quickly and identify extremely faint red galaxies, and (2) stellar masses for the majority of cluster galaxies (well below L*) can be accurately measured. *Spitzer* has also allowed for rapid confirmation of cluster candidates selected using different methods and using different observatories, such as X-rays from hot gas trapped in the gravitational potential of the cluster, or the distortion of the background radiation via the Sunyaev-Zel'dovitch (SZ) Effect (Benson *et al.* 2010; Song *et al.* 2012), or via infrared, all-sky surveys like the Massive and Distant Clusters of WISE Survey (MaDCoWS, Gonzalez *et al.* 2018).

### *REVEALING THE NATURE OF EXTREME COSMIC TRANSIENTS*

*Compton*, *Chandra*, *HST*, *Fermi*, and smaller missions including the Neil Gehrels Swift Observatory (Swift) and the Galaxy Evolution Explorer (GALEX) have worked together to reveal the nature of previously mysterious varieties of extreme high-energy transients and transient phenomena. These findings include: discovery of the afterglows and host galaxies of long-duration (Galama *et al.* 1998) and short-duration (Fox *et al.* 2005) gamma-ray bursts (GRBs); studies of GRB counterparts and host galaxies revealing their likely origin in the deaths of massive stars (long bursts; e.g. Bloom, Kulkarni & Djorgovski 2002) and binary neutron star mergers (short bursts; e.g. Fong, Berger & Fox 2010); revealing the nature of soft gamma-ray repeaters as magnetars and measuring their extreme magnetic fields (Kouveliotou *et al.* 1998); discovery and characterization of the tidal disruption of ordinary stars by supermassive black holes in the centers of galaxies (e.g. Burrows *et al.* 2011; Gezari *et al.* 2012); the first X-ray discovery of a supernova shock breakout event (Soderberg *et al.* 2008); and establishing the nature of low-luminosity GRBs as relativistic supernova shock breakout events (Campana *et al.* 2006). Most recently, the binary neutron star merger scenario for short GRBs was validated in dramatic fashion with the discovery of the *Fermi* prompt counterpart, *Chandra* and *Swift* X-ray afterglow, and *Swift* and ground-based UV, optical, and near-infrared kilonova counterparts to the gravitational wave event GW 170817 (Abbott *et al.* 2017a, Fig. 2.4-2).

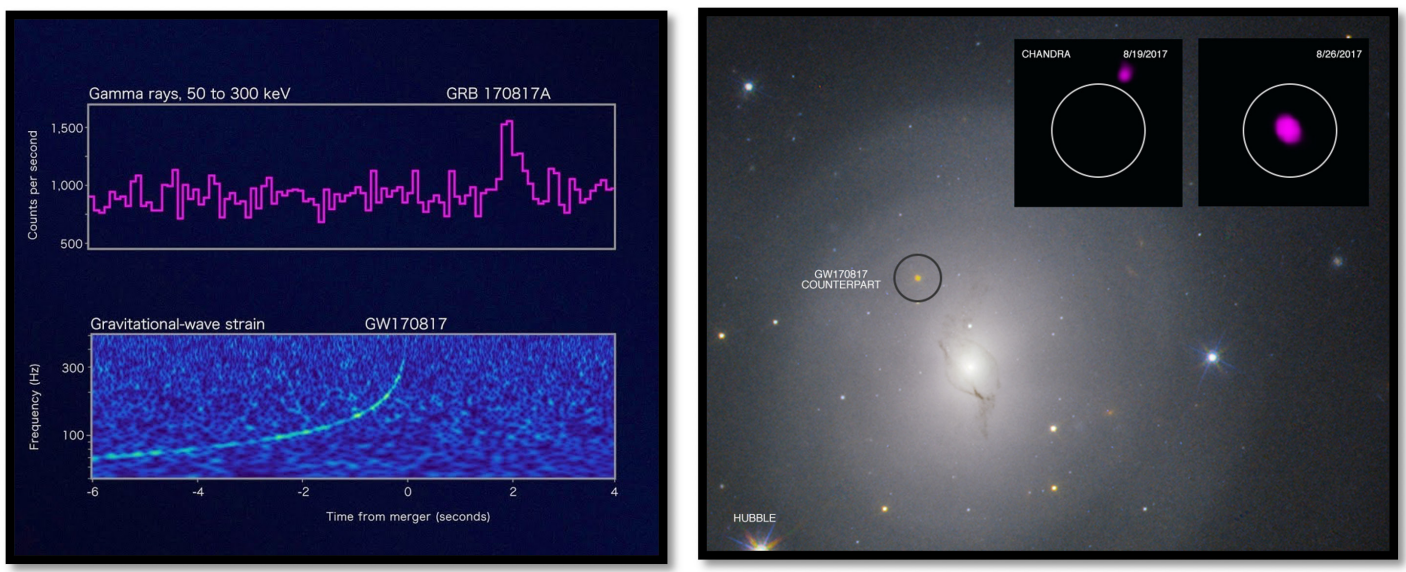

**Fig. 2.4—2. Multi-messenger Astronomy.** *Joint gravitational wave + electromagnetic detection of the binary neutron star merger GW 170817 / GRB 170817A by LIGO + Virgo and Fermi (left), the first multimessenger gravitational wave source and definitive proof that (some or all) short gamma-ray bursts arise from compact object mergers, along with images (right) of its ensuing kilonova (optical / HST) and afterglow (X-ray / Chandra) emission. Images from the LIGO Laboratory and Chandra X-ray Center.*

### SEEKING INDIRECT SIGNATURES OF DARK MATTER

*Chandra*, *Fermi*, and *XMM-Newton* have variously sought indirect evidence of dark matter accumulation and annihilation in the cores of our own Milky Way Galaxy (Ackermann *et al.* 2017), nearby dwarf galaxies (Jeltema & Profumo 2008; Ackermann *et al.* 2015), and rich galaxy clusters (Bulbul *et al.* 2014; Boyarsky *et al.* 2014). While multiple tantalizing claims have been brought forward (e.g. Hooper & Goodenough 2011), none has yet been confirmed with the necessary degree of confidence to make a discovery claim.

### EXPLORING ASTROPARTICLE PHYSICS WITH COSMIC SOURCES

*Fermi* gamma-ray and *Swift* X-ray observations helped to establish the blazar TXS 0506+056 as the first identified cosmic source of high-energy neutrinos (IceCube *et al.* 2018). Joint observations of electromagnetic and neutrino emissions from this source have shown for the first time shown that its relativistic jet – and presumably, the jets of most other active galaxies – accelerate hadrons (protons or other nuclei) as well as leptons (electrons and positrons), and have provided novel constraints on the physics of these jets and their acceleration processes (e.g. Keivani *et al.* 2018). Multi-messenger observation of cosmic sources have also enabled leading constraints on hypothetical fundamental physics phenomena including Lorentz violation at extreme energies and variations in the speeds of light (Abdo *et al.* 2009), neutrinos (IceCube *et al.* 2018), and gravitational waves (Abbott *et al.* 2017).

Discovery and characterization of the *Fermi* Bubbles has been another triumph of multiwavelength astronomy. Originally discovered as a faint haze – and then as a sharp-edged excess – of gamma-ray surface brightness in all-sky maps from the *Fermi* Gamma Ray Observatory, these bipolar plumes of high-energy emission (Fig. 2.4-3) extend 10 kpc above and below the plane of the Milky Way (Dobler *et al.* 2010; Su *et al.* 2010). They have been taken as evidence of an enormous outflow that once emanated from the heart of our galaxy, driven either by nuclear star formation (e.g. Crocker & Aharonian 2011; Lacki 2014) or a past epoch of AGN activity by Sgr A* (e.g. Guo *et al.* 2012). Exploring the nature and origins of this structure has required multiwavelength microwave (Finkbeiner 2004; Planck Collaboration 2015), X-ray (Ponti *et al.*

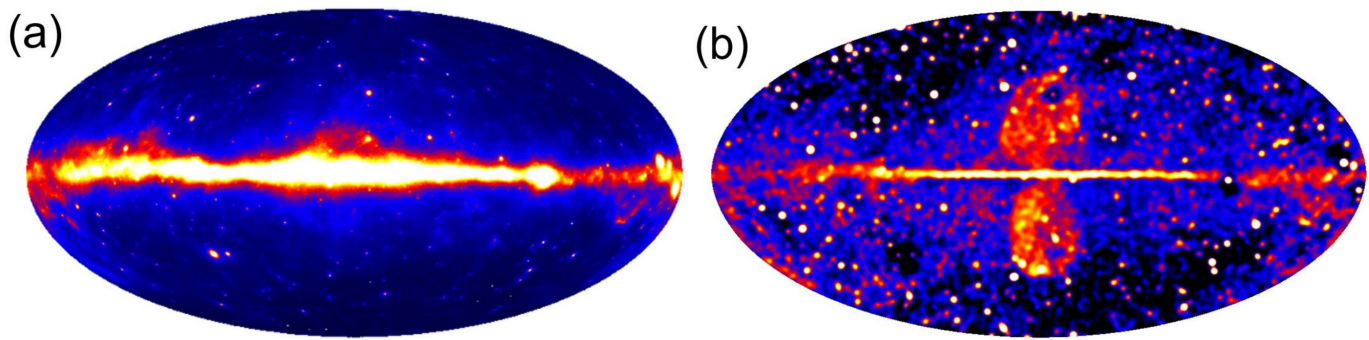

**Fig. 2.4—3. The Milky Way Bubbles.** *All-sky images of the gamma-ray sky as observed with the Fermi Gamma-Ray Space Telescope. (a) An integrated image reveals intense emission from the Galactic plane along with the sky's brightest individual gamma-ray sources. (b) A smoothed image of the hardness ratio of the data (high-energy flux divided by low-energy flux) reveals the "Fermi Bubbles" extending thousands of light years above and below the Galactic Center. Images from Kataoka et al. (2018)*

2019), and TeV observations (Abeysekara *et al.* 2017), and even suggestive analyses of IceCube high-energy neutrino data (Razzaque 2013).

### 2.4.2. Questions *for the* Next Decade

The next generation of Great Observatories, active over the two decades ahead, would advance our understanding of fundamental physics in multiple key regards and answer some of the most pressing questions in our field. Envisioning a program of New Great Observatories that would work in conjunction with next-generation ground-based astronomical and multi-messenger facilities thus offers rich opportunities to explore important questions of fundamental physics.

#### *HOW HAVE BLACK HOLES FORMED & CO-EVOLVED WITH BARYONIC MATTER?*

One of the most important results in modern astrophysics has been the realization that virtually every L* and larger galaxy contains a central supermassive black hole, and that the mass of this SMBH is tightly correlated (Magorrian *et al.* 1998) with the bulge mass of the host galaxy (see **Section 2.2** for a discussion of galaxy evolution). Quasars are seen abundantly up to redshifts of 6, less than 1 Gyr from the Big Bang, and a handful have been found at $z > 7$, barely half a Gyr from the Big Bang. These quasars are so powerful that even at the Eddington limit, their minimum black hole mass is over $10^9\ M_\odot$. Interestingly, there is now growing evidence that the SMBHs in early galaxies are overmassive relative to their bulges compared with galaxies in the local Universe. The huge black hole masses of distant quasars provide a quandary: how were these black holes able to grow so large in such a short amount of time?

It is now recognized that the formation, growth, and evolution of SMBHs and their host galaxies are closely linked, and the appearance of cosmic structures in the present day Universe is intimately tied to this link. The growth and fueling of SMBHs is strongly coupled to the stellar evolution lifecycle, and period activity of the central SMBH regulates star formation and the gas entropy of the larger scale halo.

While a great deal is known about galaxy evolution and black hole growth from the present epoch back to the peak of nuclear activity at $z \sim 2$-3, fundamental questions remain about the earliest stages of black hole growth in galaxies. ALMA is the only facility that will have the capability to observe high redshift black hole hosts in a band outside the mid-IR, due to limitations of imaging resolution and sensitivity. However, virtually all of the key results from galaxy surveys made with the Great Observatories, such as CDF-S (**Fig. 2.4-4**), COSMOS, and Bootes, have come from a multi-wavelength view of galaxies and AGN. The ability to tie stellar populations to black hole growth and periods of activity is key to understanding how galaxies and black holes form and co-evolve.

It is generally thought that black holes form along one of two broad paths (Rees 1984). In the 'light seeds' scenario, central SMBHs initially form from long-term Eddington-rate accretion onto an initially stellar mass black hole formed in a population III stellar cluster (Barkana & Loeb 2001). In the 'heavy seeds' scenarios, radiatively-cooling gas (via collisionally-excited Ly emission) in atomic cooling haloes collapses onto a black hole (Shang *et al.* 2010, Oh & Haiman 2002). *JWST* may be able to glean some insights into the early progenitors of SMBHs, but will always be limited by systematics, such as variable absorption and cleanly disentangling star formation from nuclear activity. The most direct way to probe SMBH seeds at high redshifts is in the soft X-ray (Civano *et al.* 2019). Emission in the hard (>10 keV) X-ray band is unaffected by absorption, and redshift moves this into the soft X-ray band. A mission with a much larger area

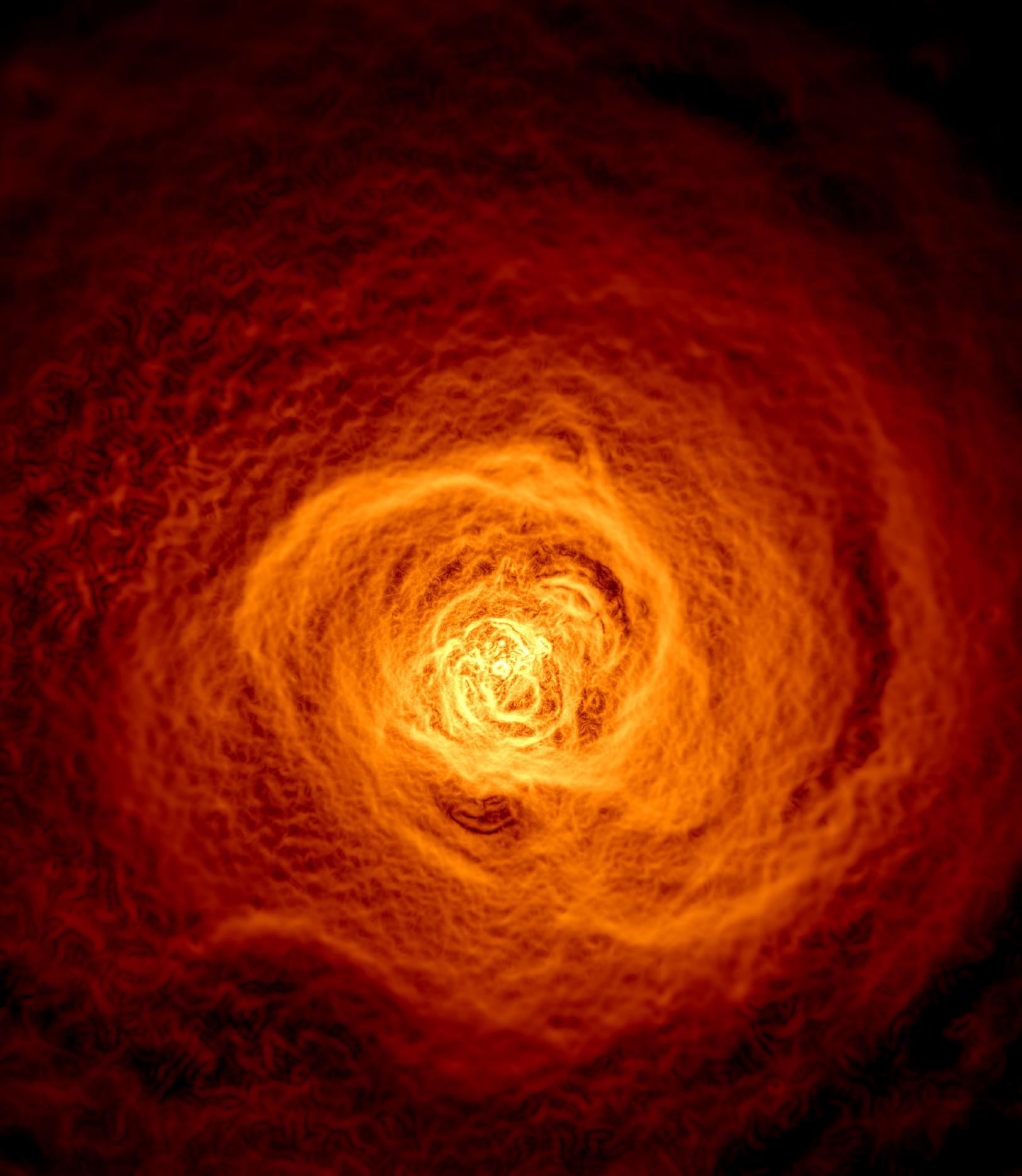

Chandra X-ray Observatory map of the Perseus Cluster of Galaxies. An edge-detection filter has been applied.
NASA / CXC / GSFC / S. WALKER / A. FABIAN *et al.*

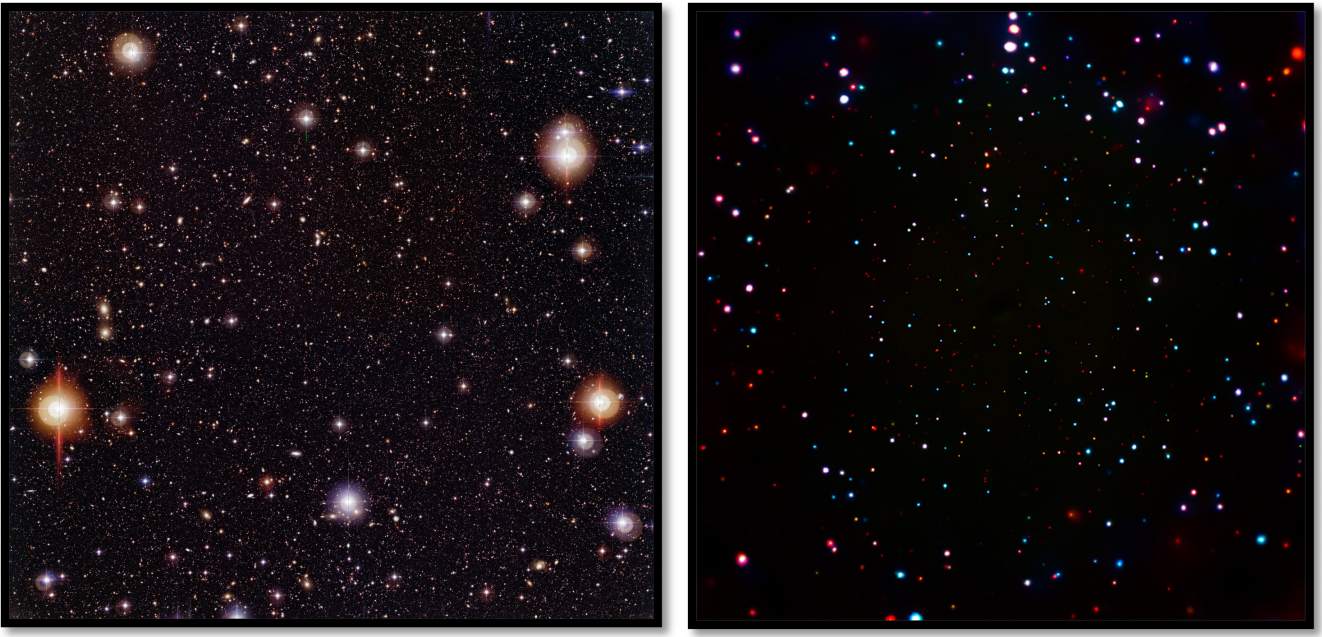

**Fig. 2.4—4. AGN and Galaxy co-evolution.** *ESO optical (left) and Chandra X-ray (right) images of the Chandra Deep Field South. The great majority of the X-ray sources seen in this deepest-ever X-ray exposure are active galactic nuclei, with distances ranging from the local Universe out to z~7.*

than *Chandra*, but with similar spatial resolution, would enormously enhance the science return of *JWST* by being able to map the BH growth over this same high-redshift epoch. Similarly, future large, cooled far-IR missions would be able to detect the gas, dust and young stars in the host galaxies of these growing black hole seeds, revealing the environment necessary to sustain this rapid growth and directly mapping the earliest links between SMBH and stellar mass.

### THE FORMATION OF SUBSTRUCTURE

Understanding the distribution of baryonic and dark matter and the thermodynamic evolution of baryons across all halo mass scales throughout cosmic time are key unsolved problems of astrophysics. On galaxy scales, the complex interplay between the stellar population, the central SMBH, the extended hot gas corona, and the local environment is under investigation by a variety of observational and numerical techniques (see Section 2.2). On larger mass scales, using the power of the current generation of SZ instruments (e.g. SPT and ACT) to detect z~1 clusters, deep *Chandra* and *XMM-Newton* observations are just now starting to characterize their thermodynamic state. These clusters are the progenitors of the most massive clusters, such as the Coma cluster, in the local Universe, and the observation times required with the current generation of X-ray observatories to make basic measurements of abundance and entropy are long (typically hundreds of ks per target). Instruments such as *Roman*, *Euclid*, and the Vera C. Rubin Observatory (formerly LSST) will map out the stellar populations and largely resolve the issue of how dark matter structures form and evolve. The baryons are subjected to forces other than gravity, though, and can evolve through interactions with the stellar populations and via feedback from the central SMBHs. Disentangling the various process will require an instrument with significantly larger sensitivity and imaging spectral resolution than the current generation of instruments. The temperature of the virialized gas in these halos of galaxies to clusters ranges from $10^6$ K to $10^8$ K, so that measuring the temperature, entropy, and elemental abundance of the baryons in these halos will require new capabilities in the X-ray regime.

Using targets identified in the SZ and optical surveys, sensitive X-ray observations are required to make three key measurements. First, the density and temperature profiles of group and cluster mass halos to z~2 need to be measured out to R500. This will require a large area imaging instrument with low background and at least 1 m2 effective area over the soft (0.5-2.0 keV) band pass. Second, a large area calorimeter is required to make imaging spectroscopic measurements at high spatial and spectral (~few eV) resolution. This will allow precise measurements of elemental abundances of both prominent lines but also detection of weaker lines that are not resolvable by Si-resolution instruments. Mapping the profiles of the elemental abundance as a function of radius in these hot gas haloes will give key information about how the gas is enriched from stellar winds and supernovae, and by what processes these elements are distributed through the gas. Additionally, precise measurement of line shapes and centroid will provide critical information about the role of turbulence and non-thermal pressure support in the gas. Finally, deep imaging observations of the hot gas through cosmic time are required to understand the role of AGN feedback in the haloes through cosmic time. Sensitive X-ray imaging is required to detect and characterize the shocks and sound waves in these systems to understand the heating/dissipation mechanisms and how they offset radiative losses.

An alternative method for understanding the distribution at high redshift comes from studying the spatially unresolved IR and X-ray backgrounds. Recent cross-correlation of the unresolved IR and X-ray backgrounds in the AEGIS fields after the subtraction of known point sources shows significant power of scales of ~20" (Cappelluti *et al.* 2013). The origin of the sources responsible for this correlated emission is at present unknown, but is likely to originate in a population of active nuclei at z>4 with significantly higher occupation fraction than for local AGN (Kashlinsky *et al.* 2018). Wide field infrared and X-ray surveys, such as will be made by *Roman* and *eROSITA*, are likely to provide important clues about the population of sources responsible for this emission, but resolution of this question can only be made with the identification of the sources in the IR and X-ray. This will require deep observations by IR and X-ray instruments with sufficient sensitivity and angular resolution to uniquely correlate individual sources. Webb will provide the necessary IR measurements, but there is no comparable X-ray instrument capable of making the necessary observations.

A powerful emerging technique in long-wavelength astronomy to study large-scale structure is Line Intensity Mapping (LIM). Introduced over 20 years ago, initially for studies of 21 cm radiation (Madau *et al.* 1997; Shaver *et al.* 1999), it was subsequently suggested for the far-IR fine-structure lines (Suginohara *et al.* 1999). In LIM, the clustering of line-emitting galaxies is detected as fluctuations in a 3-D spatial-spectral dataset in which the line-of-sight dimension is encoded as wavelength. The technique provides 3D measurements of galaxy clustering and moments of the galaxy luminosity function. For steep luminosity functions, intensity mapping can be an effective way of measuring average intensity and thus constraining the bulk of the luminosity function, as well as the optimal method of measuring the clustering power spectrum. Since individual galaxies are not detected, much of the luminosity function can be below the nominal detection threshold. Surface brightness sensitivity, detector stability, and the ability to map large areas of the sky (tens - hundreds of sq. degrees) are important, but high spatial resolution is not required. A key feature of LIM is the ability to measure cross-correlations among multiple datasets, for example comparing far-IR fine-structure transitions with one another and with HI 21 cm. Prospects for LIM in the far-IR/submillimeter has been examined in several studies (Gong *et al.* 2011; Uzgil *et al.* 2014; Silva *et al.* 2015; Cheng *et al.* 2016; Lidz and Taylor 2016; Serra *et al.* 2016). Ground-based experiments are currently making measurements of CO (Cleary *et al.* 2016; Bower *et al.* 2015), and [CII] in the 1-mm atmospheric window (Crites *et al.* 2014; Lagache 2017), and balloon experiments will target the [CII] 158 micron line in the 240 to 420 μm band (Hailey-Dunsheath *et al.* 2018).

## WHAT IS THE NATURE OF THE BRIGHTEST MULTI-MESSENGER SOURCES?

A new era of multi-messenger astrophysics exploded into prominence in late 2017 with the real-time localizations of the gravitational wave binary neutron star event GW170817 and the high-energy neutrino IceCube-170922A. Ensuing global follow-up campaigns for these two events led to the first electromagnetic counterparts to a gravitational wave (GW) transient (Abbott *et al.* 2017a) and a high-energy neutrino (HEN, IceCube *et al.* 2018), respectively.

Given the novel nature of the discipline, we can anticipate many exciting multi-messenger results over the course of the next 20 years. During this period, capabilities of the global network of GW and HEN observatories will improve substantially, ultimately by an order of magnitude or more (Sathyaprakash *et al.* 2012, LIGO Lab 2019, IceCube Gen-2 Collaboration 2014, KM3NeT Collaboration 2018). The chief challenge for EM observations will be to keep pace with these improvements and continue to discover and characterize EM counterparts to these sources, even as they are revealed in greater and greater numbers, at increasing rates, and to increasingly greater distances and lower fluxes. Only via continuing relentless pursuit of EM counterparts can we hope to maximize the science yield of these sources and discover new multi-messenger source populations.

As one example, rapid-response multiwavelength counterpart searches triggered by binary neutron star merger events will continue to be required to discover and characterize their afterglows and kilonovae. These searches will have to be carried out over relatively large GW-based localizations, repeatedly, even as the distances to these events increase by an order of magnitude or more. In parallel, and to exploit the recent discovery of likely gamma-ray (*Fermi*, MAGIC) and X-ray (*Swift*, NuSTAR) flare-associated neutrino emission from the BL Lac-type blazar TXS 0506+056 (IceCube Collaboration *et al.* 2018, IceCube Collaboration 2018), multiwavelength facilities should seek to provide near-continuous monitoring of the most prominent blazars, AGN, and other likely sources of high-energy neutrinos in order to cross-correlate against the potentially time-variable neutrino emissions of these sources.

Wide-sky monitoring across the EM spectrum will also be required to distinguish different varieties of multimessenger source by their EM emissions. For example, we can anticipate detection of "orphan" afterglows and kilonovae from far off-axis binary neutron star mergers. For these events, the absence of prompt emission to deep limits will serve as the key distinguishing characteristic. Finally, we anticipate a first detection of the next Galactic or M31 supernova, within seconds of core collapse, via its MeV neutrino emissions (Antonioli *et al.* 2004). Whenever this next "nearest supernova" occurs, a comprehensive suite of multiwavelength EM facilities should be ready and prepared to characterize the "once in a lifetime" event, from its earliest moments, in as much detail as possible.

## WHAT CAN WE LEARN ABOUT THE UNIVERSE FROM TRANSIENT OBSERVATIONS?

Many varieties of energetic transients – including neutron star mergers, supernovae, gamma-ray bursts, and tidal disruption events – display UV, X-ray, or gamma-ray radiation at early times. These high-energy signals can be used to localize the transients, pinpoint their times of explosion, and provide valuable insights into the nature of the progenitor and the physics of the transient. In this fashion, multiwavelength observations can serve a crucial role in exploiting transients to learn about the fundamental physics of the universe.

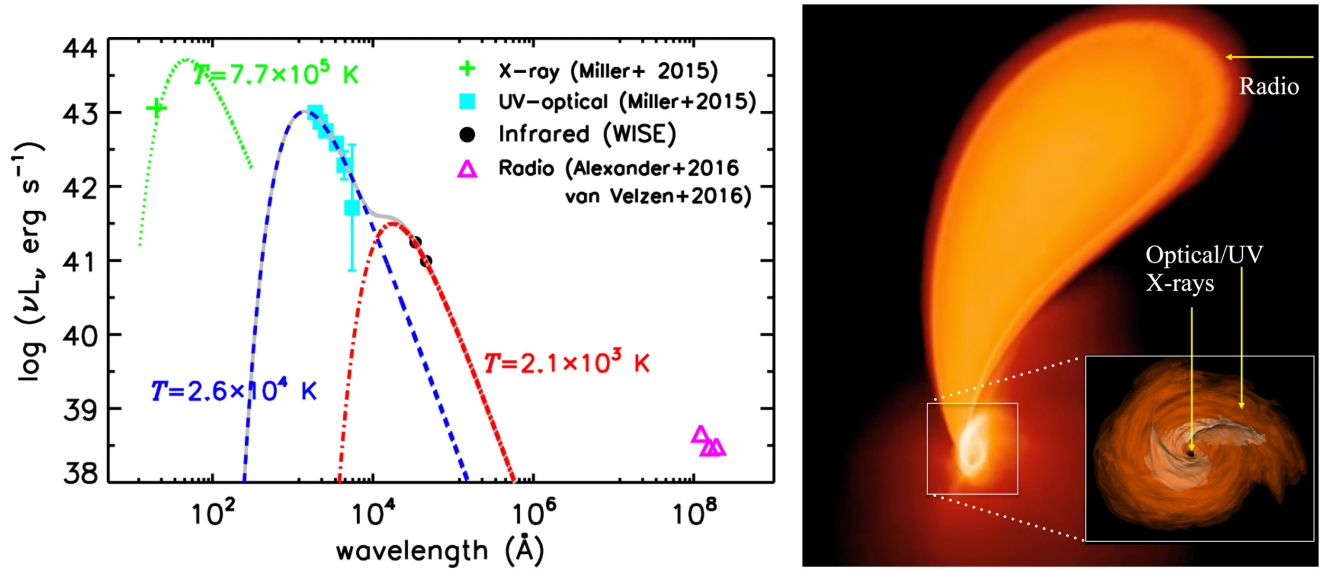

**Fig. 2.4—5. Tidal Disruption Events.** *Left: Multi-wavelength components of emission from the TDE ASASSN-14li. Figure from Jiang et al. (2016). Right: Theoretical model for the origin of these components from Krolik et al. (2016).*

The 2020s in particular promise to be a golden decade of time domain astronomy, with wide-field optical surveys, next-generation radio surveys, and multi-messenger (non-EM) observatories operating at dramatically enhanced sensitivities. There will be a critical need both for wide-field discovery surveys of electromagnetic sources, and for rapid-response follow-up imaging and spectroscopy across the electromagnetic spectrum. Loss of the space capabilities that currently enable astronomy's panchromatic transient discovery and follow-up efforts, just as this epoch dawns, would tragically hobble efforts just at the moment when these new ground-based capabilities promise an extraordinary leap in our capacity to understand and exploit these transients.

Multiwavelength observations of multi-messenger transients have a unique capacity to put our understanding of the physical universe to various high-precision tests. Observations of GW-detected compact object mergers will push utilization of the GW "standard siren" technique (Abbott *et al.* 2017b) to the point where it can provide competitive (< 2%) constraints on the value of the *Hubble* constant, helping to constrain models for the evolution of Dark Energy. Observations of kilonova counterparts to these events will constrain the properties of their mass outflows, with multimessenger modeling leading to new constraints on the dense matter equation of state (e.g., Radice *et al.* 2018).

Multiwavelength observation and modeling of tidal disruption events (TDEs) will continue to probe the dynamics of transient disk and jet formation in the near vicinity of supermassive black holes (**Fig. 2.4-5**), potentially reflecting General Relativistic effects (Jiang *et al.* 2016, Krolik *et al.* 2016). Multiwavelength characterization of the TDE population at large will explore the demographics of massive black holes, potentially revealing new populations expected from studies of black hole formation and galaxy coevolution, including: intermediate-mass black holes, binary massive black holes, and recoiling massive black holes after a binary merger.

Panchromatic studies of gamma-ray burst afterglows have proven their utility in optical and near-infrared studies of their host galaxies (e.g. Cucchiara *et al.* 2015) and the intergalactic medium (e.g. Totani *et al.* 2006, Chornock *et al.* 2013). Due to the extreme luminosity and simple (synchrotron) spectrum of young afterglows, they are capable of providing unique insights into element abundances, clustering, and properties (e.g. density, temperature, ionization state) of absorbing gas along the line of sight, nearly independent of redshift. At high redshift, the insights offer unique cosmological constraints, including quantifying the evolving ionization of the intergalactic medium (Totani *et al.* 2006, Chornock *et al.* 2014). In the UV, infrared, and X-ray bands these insights can only be gathered using space-based observatories; within the optical and NIR bands, the rapid response and low backgrounds achievable with space-based platforms can allow competitive results even with relatively small apertures.

On the rare occasions when cosmologically distant transients are subject to strong lensing by intervening galaxies or galaxy clusters (e.g. Kelly *et al.* 2015), multiwavelength characterization of the event's distinct manifestations and the lensing system can yield a "single step" determination of the *Hubble* constant. Lensed transients can also provide insights into matter clustering in the lensing systems via microlensing variability and image flux ratios (e.g. Mao *&* Schneider 1998, Metcalf *&* Madau 2001), including constraints on dark matter substructure.

### *WHAT IS THE NATURE OF DARK MATTER?*

The abundance of ubiquitous dark matter is extremely well-quantified by observations (e.g. Planck Collaboration 2018), yet its nature remains unknown. Despite thorough searches, direct detection experiments targeting weakly interacting massive particles (WIMPs) in the underground detectors and at LHC at CERN have as yet yielded no evidence of detection. Indirect detection searches that seeks observable signals of dark matter interactions (e.g. decay, annihilation, and scattering) offer a complementary arena for the characterization of dark matter. Indirect detection probes parameter spaces and sensitivities which are unreachable with current and future ground-based dark matter detectors.

One of the most promising avenues to search for the by-products of the decay of dark matter is in the gamma-ray band. Such searches typically are dominated by observations of the galactic center and nearby dwarf spheroidal galaxies as these are the nearest regions with the largest dark matter densities. An excess of gamma-rays in the 3-5 GeV band was observed by the *Fermi* LAT (Morselli *et al.* 2011), although it is unclear if this is due to dark matter decay, a large population of milli-second pulsars (Lee *et al.* 2016), or related to the *Fermi* bubble (Petrovic *et al.* 2014). This degeneracy is only likely to be resolved by a mission with somewhat better angular resolution, and working in a lower energy band is probably optimal. A mission such as the e-ASTROGAM concept (proposed to ESA's M5 call) could provide both the sensitivity and angular resolution required to resolve the emission components and determine what fraction of the gamma-ray emission from the galactic center could originate from decaying dark matter and thus put strong constraints on dark matter cross-sections and particle masses.

Lighter alternative dark matter candidates including Axion Like Particles (ALPs, Marsh 2016) and sterile neutrinos appear naturally in extensions to the Standard Model (Dodelson *&* Widrow 1994, Peccei *&* Quinn 1977). Axion-like particles through their decay can produce observable quasi-sinusoidal features in the X-ray spectra of centers of clusters of galaxies. Deep X-ray observations of active galactic nuclei in cluster centers can constrain ALP mass and photon-ALP coupling. These observations naturally complement direct searches for light ALP dark matter (e.g. Conlon *et al.* 2018). Another viable warm dark matter candidate is sterile neutrinos (Dodelson *&* Widrow 1994). Warm decaying dark matter would produce X-ray photons via its decay process that would appear as an emission line in the X-ray spectra of dark matter dominated objects, e.g. dwarf spheroidals, galaxies, and clusters of galaxies (Abajian *et al.* 2001). Deep and high spectral-resolution X-ray observations of these objects can thus constrain decaying dark matter particle masses and decay rates (Bulbul *et al.* 2014). Heavier dark matter particles such as Weakly Interacting

Massive Particles (WIMPs) can generate gamma-rays through direct annihilation or via production of a decaying secondary particle (Bertone *et al.* 2005). In this connection, we note again a suggestive excess of gamma-rays observed toward the center of the Milky Way (e.g. Hooper & Goodenough 2010).

Given the potentially vast parameter space of notional dark matter particles, space-based investigation of the nature of dark matter is inherently a multiwavelength exploration. Broad band observations of galaxy clusters and strongly-lensed quasars, supernovae and other transients, the Galactic center, and nearby dwarf spheroidal galaxies will continue to yield insights into the nature of dark matter and its interactions (or lack thereof; e.g. Harvey *et al.* 2016) and complement the variety of ground-based investigations.

### *HOW HAS DARK ENERGY EVOLVED WITH COSMIC TIME?*

Cosmological tests based the observed number of galaxy clusters as a function of mass and redshift provide one of the cornerstones of modern cosmology (for a review, see Allen *et al.* 2011). Such measurements have been used to place competitive constraints on a broad range of
cosmological parameters including the mean matter and dark energy densities, the amplitude of the matter power spectrum, the dark energy equation of state, departures from General Relativity on cosmological scales, and total (species-summed) neutrino mass (e.g. Vikhlinin *et al.* 2009, Mantz *et al.* 2010, 2015, Rozo *et al.* 2010, Cataneo *et al.* 2015, de Haan *et al.* 2016; Planck Collaboration 2018).

The next decade will see the launch of *Euclid* and *Roman*, two missions optimized to study the nature of dark energy. *Roman* will make three key investigations that will facilitate an unprecedented understanding of dark energy: a High Latitude Spectroscopy Survey to accurately measure the positions and distances of a large number of galaxies, a Type 1a Supernova Survey to use these objects as standard candles, and a High Latitude Imaging Survey to measure the shapes and distance of a large number of galaxies and galaxies clusters. Cosmological studies are often dominated by systematic uncertainties, and one of the simplest ways to reduce these systematics is to use investigations in other wavebands that are dominated by independent systematic assumptions. For example, precision cluster cosmology provides a "growth of structure" constraint on dark energy, as opposed to the standard candle SN-based approach (Vikhlinin 2017, Oukbir & Blanchard 1992). Additionally, cross-correlation of low-frequency radio and optical/IR survey can remove residual systematics to extremely low levels, provide an independent means of detecting experiment systematics, and extend surveys to higher redshifts than optical/IR-only investigations, giving results that are less influenced by non-linear perturbations in the matter distributions (Camera *et al.* 2017).

The coming decade will see a vast expansion in the size of available cluster catalogs through new surveys at X-ray, optical and mm-wavelengths. The role of dedicated X-ray observations in exploiting these catalogs will remain vital. For example, while the *eROSITA* X-ray survey will find tens of thousands of clusters down to fluxes approximately a hundred times fainter than the previous *ROSAT* All-Sky Survey, it will not have the angular resolution to cleanly discriminate the cluster gas X-ray emission from that of any AGN within these systems. For optical and mm-wavelength surveys, the availability of low-scatter X-ray mass proxies will continue to provide an important boost in both the statistical power and robustness of the cosmological constraints: while these surveys will provide exquisite constraints on the mean masses of clusters using galaxy- and CMB-weak lensing methods, only intensive characterization of individual clusters can provide the clean, low-scatter mass estimates needed to pin down the mass-observable scaling relations and their scatter, as a function of mass and redshift.

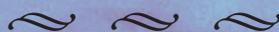

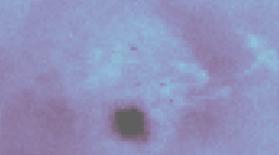

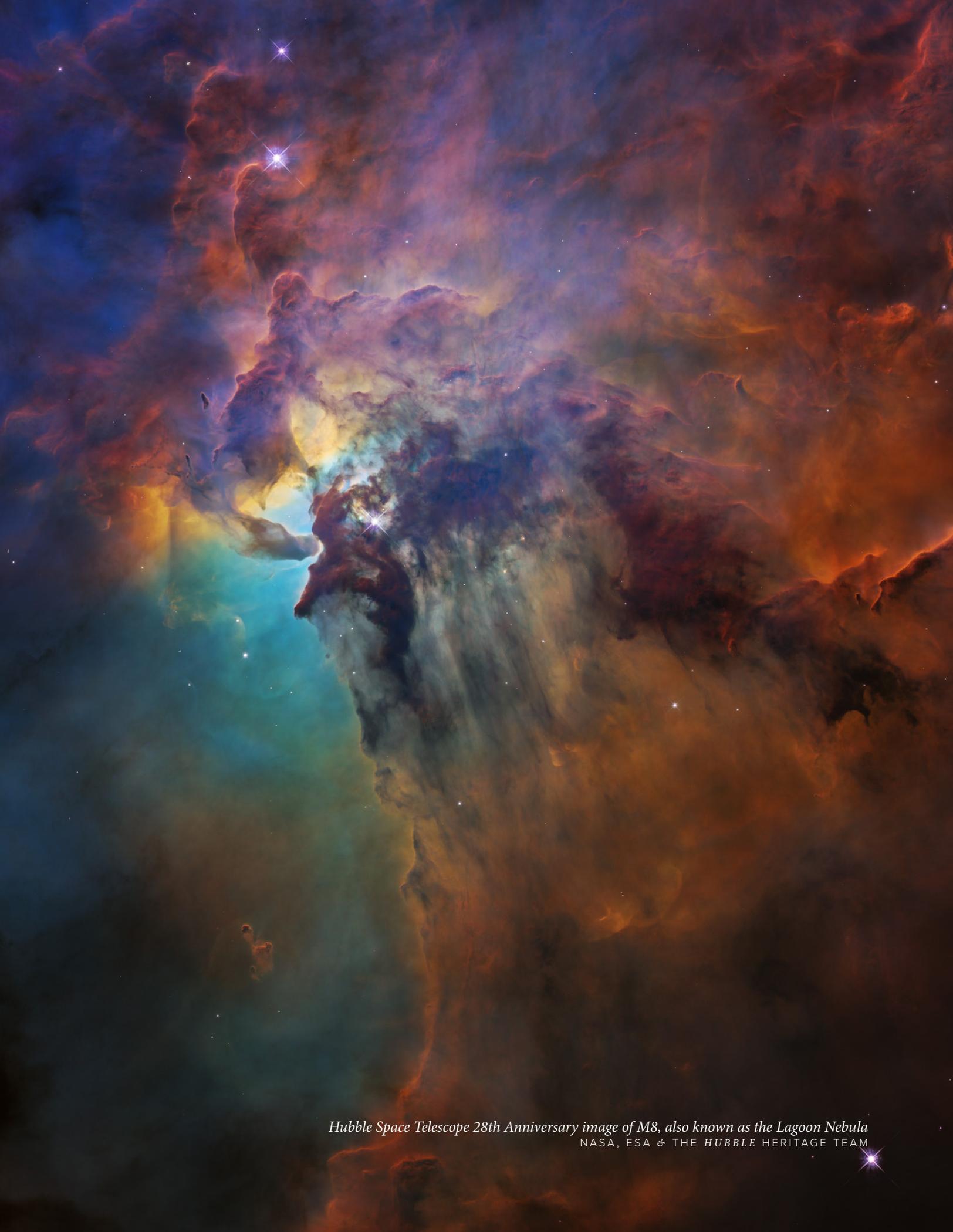

*Hubble Space Telescope 28th Anniversary image of M8, also known as the Lagoon Nebula*
NASA, ESA & THE *HUBBLE* HERITAGE TEAM

# CAPABILITIES, FACILITIES & OPTIONS

## 3.1. *The* Current Landscape *&* Impending Gaps

The astronomical community currently enjoys widespread access to an unprecedented panchromatic capability in astronomy, extending from the very low frequency radio regime to TeV gamma-rays. This multi-wavelength access to space has led to the current Golden Age of astronomy. Our wide-ranging view of the Universe, through a suite of observatories, greatly expands our ability to discover, and then understand, new phenomena, and to test our theoretical constructs.

To a large extent, our existing space-based panchromatic capability derives from NASA's Great Observatories: *Hubble*, *Compton*, *Chandra*, and *Spitzer*, and the community access provided through General Observer (GO) and Guest Investigator (GI) archival research programs. The GO programs are invariably highly competitive, with oversubscription rates of more than 4:1 or 5:1 being common, and with multiple, joint GO proposals being awarded every cycle. In addition, smaller scale missions have extended both the wavelength and sky coverage of the Great Observatories. Examples of past synergistic use of the Great Observatories abound, and are documented in the previous sections, along with a sampling of future science directions that require this synergy. These needs are broad, based in part on our universal quest to understand how the emergence of radiation, light elements and large-scale structures, evolve into galaxies, stars, planets, and life.

A major legacy of the Great Observatories is the wealth of archival data that covers large areas of sky at multiple wavelengths. Combined, the MAST, IRSA and HEASARC archives contain 2.3 PB of data from NASA missions, with 0.3 PB downloaded in 2019 alone. The use of these data in the literature is large and growing. These data enable unique science, provide baselines for investigating time variable phenomena, and establish foundations for future investigations with more targeted missions. Support for maintaining and enhancing archives is an important component of panchromatic astronomy. The archives, however, are not a substitute for maintaining panchromatic capabilities into the future. They cannot provide capabilities commensurate to those of newly developed observatories. They also do not necessarily include newly discovered phenomena; observations by the Great Observatories only cover a fraction of the sky while all-sky surveys are, by their very nature, typically shallow and carried out over broad photometric bands. Nor can archival data provide concurrent capabilities for studying time dependent phenomena.

Unfortunately, the Great Observatories are aging. One, *Compton*, was decommissioned in 2000, although it was replaced by *Fermi* in 2008, which itself is 11 years old. Another, *Spitzer*, was decommissioned in January 2020, although some of its capabilities in the near and mid-infrared will be superseded and expanded upon by *JWST*. The remaining two, *Chandra* and *Hubble*, are 21 and 31 years old, respectively. A Roadmap for NASA Astrophysics in the next three decades entitled "*Enduring Quests, Daring Visions*" has outlined a set of notional future missions necessary to understand emergent astrophysical phenomena. Their vision, taken as a whole, is of a panchromatic mission suite that might function after the Great Observatories cease operating. In this section we identify the wavelength gaps that make this planning necessary, and the capabilities required to fulfill the science goals outlined in Chapter 2.

In **Fig. 3.1** we illustrate how pan-spectral coverage will diminish from now into the 2030s in the absence of new facilities. Without a concerted effort to maintain the type of panchromatic coverage enabled by the Great Observatories, the future of space astrophysics will suffer from major gaps appearing in our electromagnetic coverage. New ground and space-based facilities that are planned to become operational in the next decade will only partially fill these wavelength gaps. In particular, the far-infrared, UV and X-ray regimes are sparsely covered, and the capabilities of existing or planned facilities are not well-matched across the spectrum to answer the pressing and inherently multi-wavelength questions outlined in **Chapter 2**. At the same time, these gaps present an opportunity to develop new missions and strategies for maintaining panchromatic coverage that draw on the lessons of the Great Observatories. Below, we first explore some of the consequences of the impending wavelength coverage gaps. We then summarize some of the lessons from the Great Observatories, and overview potential strategies and opportunities for maintaining and enhancing panchromatic access.

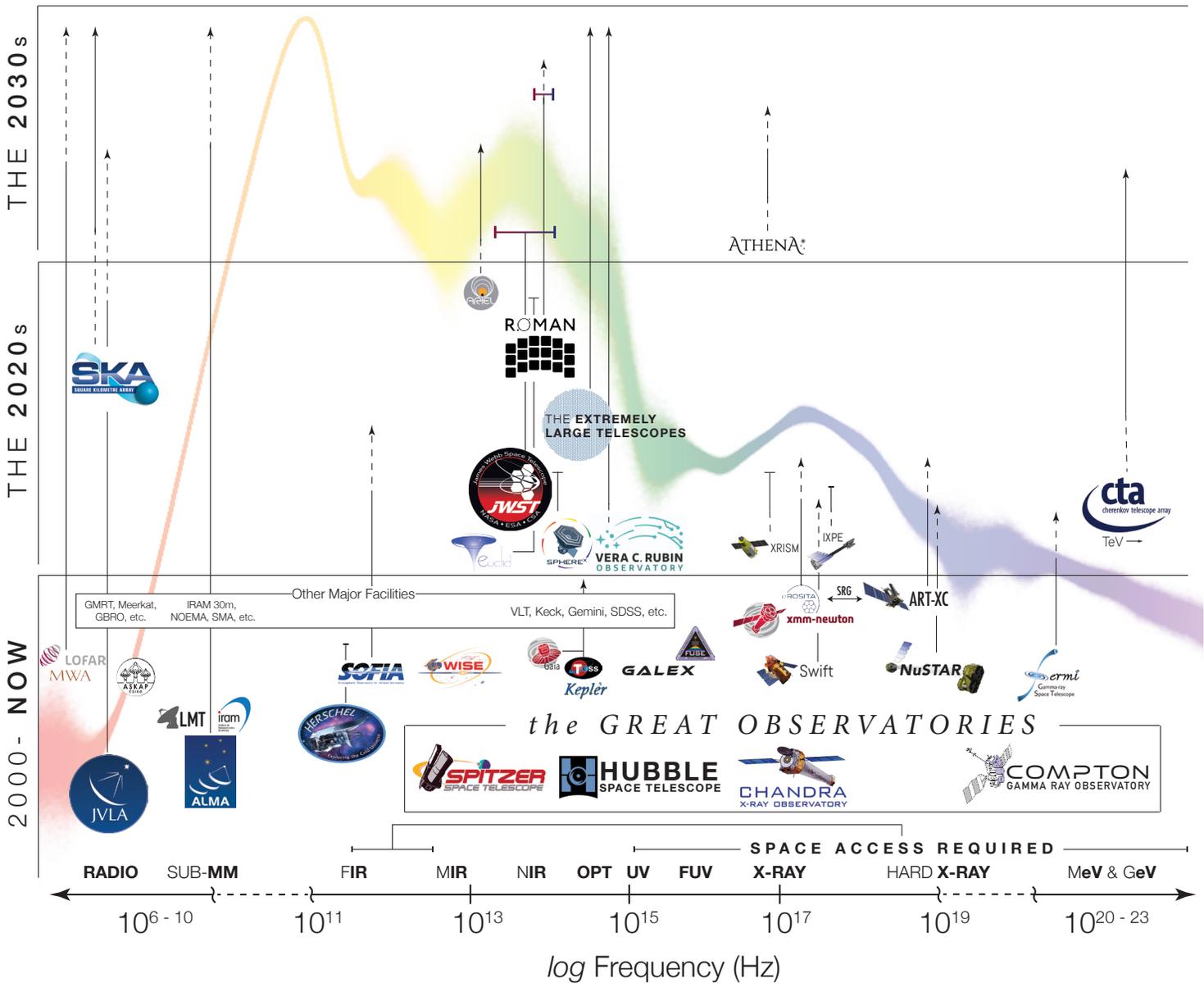

**Fig. 3.1. The Current & Future Mission Landscape.** *The current and expected coverage of major NASA, international and ground-based observatories through the 2030s. Colored horizontal bracketed lines are used to indicate spectral coverage for NASA's next flagship missions, JWST and Roman (previously known as WFIRST). To reduce visual clutter, all other black horizontal endpoints to lines are used to indicate approximate end-of-operations. In some cases, dashed vertical lines are used to indicate possible extended missions. The plot shows significant gaps in wavelength coverage in the far-IR and in the UV through the X-ray and gamma-ray regimes for the coming decade and beyond. The total integrated spectrum of the Universe (Hill et al. 2018) is shown to emphasize the broad wavelength coverage needed to study most astrophysical phenomena.*

## 3.2. Costing *the* Loss *of the* Great Observatories

In Chapter 2 we have shown the benefits of concurrent wavelength coverage with commensurate capabilities in spectral resolution, spatial resolution and overall sensitivity across the electromagnetic spectrum, brought about, in large part, by the Great Observatories. We have also given examples of science in the next decade(s) that will rely on similar capabilities, and where progress will be inhibited without these capabilities. It is clear that gaps in panchromatic coverage will inhibit progress in astronomy across a broad range of topics from star formation, to exoplanets, to galaxy evolution to fundamental physics of the cosmos. This synergy is particularly important during the discovery phase of new phenomenon, as they often raise new questions to be addressed, and new avenues to explore. The interest, excitement and insight engendered by rapidly following up gravitational wave and other transient sources at all wavelengths in the past few years directly highlights the importance of having a suite of agile facilities in space to move the field forward.

### 3.2.1. Scientific Loss

The most obvious cost of the impending gaps in coverage is the cost to science. Major questions will have to wait decades for resolution, for want of multi-wavelength follow-up. For example, while the extreme far-infrared energy outputs of Ultraluminous Infrared Galaxies (ULIRGs) were first realized as a result of the IRAS survey the nature of these enigmatic galaxies was not followed up in the mid and far-infrared until decades later, first with ISO and then with *Spitzer* and *Herschel*. Vigorous science demands healthy debate, and multiple views of the same phenomenon push the field forward. At any one time a theory may be supported by one set of observations but cannot be further tested by independent techniques using other wavelengths. Viewing the Universe through a single window needlessly limits the questions we can ask and the discoveries we can make. Examples of important future science questions that require multi-wavelength observations from space are discussed in detail in sections 2.1-2.4.

Table 3.1 summarizes the wavelengths and capabilities needed to address the future astrophysical questions laid out in sections 2.1-2.4. The vast majority of these questions require not only a wide range in wavelengths, but also a diverse set of capabilities from wide field imaging to high-resolution spectroscopy. The importance of multi-wavelength observations was discussed for similar science themes in the Astrophysics Roadmap. Table 3.1 extends and expands upon this analysis, illustrating the advantage of concurrent coverage in advancing these goals and answering these questions. Some of these capabilities will be met with the planned flagship missions, specifically high-resolution imaging and spectroscopy in the near and mid-infrared with *JWST*, together with high resolution and wide-field imaging in the optical and near-IR with *Roman*. However, even with these highly capable observatories, there exist large regions of wavelength and measurement phase space that will be inaccessible and unexplored in the coming decades. The far-infrared, UV, and X-ray regimes are extremely important for nearly all of the science discussed in this report, but they are notably absent from planned, commensurate coverage. So too are the twin capabilities of wide field coverage and/or rapid follow-up in, for example, the mid-infrared, which feature prominently in the

| Science Cases | Wavelength Regimes | | | | | | | | | |
|---|---|---|---|---|---|---|---|---|---|---|
| | mm/smm | Far-IR | Mid-IR | Near-IR | Visual | Near-UV | Far-UV/EUV | Soft X-ray | Hard X-ray | γ-ray |
| **2.1 Galactic Processes and Stellar Evolution:** | | | | | | | | | | |
| Life Cycle of Dust in Different Galactic Environments | A,B,P | A,B,D | A,B,C,D | A,B,C,D | A,B,C,D | A,B,C,D | A,B,C,D | A,E+ | | |
| Microphysics of Feedback | A,B,D | A,B,D | A,B,D | A,B,C,E | A,B,C,E | A,B,C,E | A,B,C | A,B,C,E | A,D+,E | A,B,G |
| Star formation in different environments | A,B,C,E,P | A,B,C,E,P | A,B,C,D | A,B,C | A,B,C | A,B,C | A,B,C | A,B,C,D | | A,B |
| Stellar evolution in real time | A,C | A,C | A,D,G | A,C,D,G | A,D,G | A,D,G | A,D,G | A,D,G | D+,G | A,B,G |
| **2.2 Astrophysics of Galaxy Evolution:** | | | | | | | | | | |
| The First Generation of Stars | A,B,D | A,B,D | A,B,D | A,B,D | A,B,D | A,B,D | | | | |
| Formation of the first Supermassive Black Holes | | A,B,F | A,B,F | A,B,F | | | A,B,F | A,B,F | | |
| SNe- and AGN-driven Regulation of Galaxy Growth | A,B,D | A,B,D,F | A,B,D,F | A,B,D | A,B,D,G | A,B,D,G | A,B,D,G | A,B,D,E,F,G | A,B,D,E,F,G | |
| The Baryon Cycle of Galactic Ecosystems | A,B,D,P | A,B,D | A,B,D | A,B,D | A,B,D | A,B,D | A,B,D+,E+ | C, D, E | C, D, E | |
| **2.3 Origin of Life and Planets:** | | | | | | | | | | |
| Circumstellar disks and planet formation | A,C,E,P | A,C,E,P | A,C,E | A,C,D,E,P | A,C,D,E,P | D,E | D,E | A,D,G | | |
| Discovery and Characterization of Planets | | | A,D,G | A,C,D,G | A,D,G,P | | | | | |
| Star/planet connection | | A,C,D,G | A,C,D,G | A,C,D,G | A,C,D,G | A,C,D,G | A,C,D,G | A,C,D,G | | |
| Evolution of Complex Organics Needed for Life | A,C,D,E | A,C,D,E | A,C,D,E | A,C,D,E | A,E | | A,E | A,E | | |
| **2.4 Fundamental Physics:** | | | | | | | | | | |
| Multi-Messenger Astrophysics | | | | A,D,G | A,C,D,G | A,G | A,G | B,D | B,D | B,D |
| Exploring the Transient Universe | | A,B,D,G | A,B,D,G | A,B,D,G | A,C,D,G | A,B,C,D,G | A,B,G | A,B,G | A,B | |
| Dark Matter Studies | A,B,D,F | A,B,D,F | | | | D+,F | D+,F | B,E,F | | A,B,D,F |
| Dark Energy | A,B,D,F | A,B,D,F | A,B,C | A,B,C | A,B,C | | | | | |
| Black Hole Seeds and Baryon Evolution | A,B,C,E | | A,B,C,D,G | A,B,C,D,G | A,D,G | E,F | E,F | A,B,C,D,E,F,G | A,B,G | |

A: Broadband imaging (arcmin field)  
B: Wide field imaging or Large Area Mapping Speed  
C: High Spatial Resolution  
D: Spectroscopy Low Resolution (+spatially resolved)  
E: Spectroscopy High Resolution (+spatially resolved)  
F: High sensitivity  
G: Temporal resolution/rapid response  
P: Polarimetry  

mm/smm ~ 3000-300 μm  
Far-IR ~ 300-30 μm  
Mid-IR ~ 30-3 μm  
Near-IR ~ 3- 1 μm  
Visual ~ 1 - 0.35 μm (10,000 - 3,500 Å)  
Near-UV ~ 0.35 - 0.19 μm (3,500 - 1,900 Å)  
Far-UV ~ 1,900 - 912 Å  

EUV ~ 912 - 5 Å (13.6 - 250 eV)  
Soft X-ray ~ 50 - 1 Å (0.25 - 10 keV)  
Hard X-ray ~ 10 - 300 keV  
Soft-γray ~ 0.3 - 30 MeV  
Hard-γray ~ 30 MeV - 300 GeV  

**Table 3.1. Scientific Requirements across the EM Spectrum.** . *Capabilities needed to address the future science priorities outlined in Section 2. Wavelengths are listed along the top, photometric, imaging or spectroscopic requirements are indicated by letters. The broad range of capabilities and wavelengths needed to make progress in each of the key science areas highlights the need for a new generation of Great Observatories in space.*

science of stellar and galactic evolution, the formation and evolution of planets, and the discovery and characterization of transient high energy events. No single observatory, even of flagship class can encompass all these needs, and a traditional cadence for flagship missions will surely result in serial access to different parts of the electromagnetic spectrum spread over decades. However, a set of observatories designed to be responsive as a group, and possibly consisting of a range of sizes and mission classes, could enable rapid progress in these fields. There is a great opportunity to learn from the Great Observatories, and use emerging technologies to expand access to the electromagnetic spectrum from space to tackle some of the most pressing astrophysical questions of the next decade. We outline some of these options below.

### 3.2.2. Cost *to the* Supporting Community

Science moves forward only if there is a vibrant community to make it happen. This community needs to encompass theorists, observers, software engineers and instrument developers, all at varying stages in their careers. A gap in panchromatic coverage on decade timescales will have a major dampening effect on all of these areas.

In decades wherein access to broad swaths of the electromagnetic spectrum is not available, students will have little incentive to enter the field most directly associated with those particular spectral regimes. For example, without the ability to obtain high spectral resolution X-ray spectra from space, there will be little incentive to continue developing theories describing supermassive black hole accretion or feedback in galaxy clusters. As a result, band-specific data analysis techniques and deep knowledge of the field are not passed on to junior researchers. Moreover, the prospects for future instrumentation development then become greatly enfeebled, producing a loss in hard won core-competency that will take decades to reverse. While future development and leadership in any one particular area is obviously an important goal in selecting missions to move forward, this is best done within the context of a broader programmatic goal to provide the entire community with opportunities to access space and foster the next generation of young scientists.

### 3.2.3. Types *of* Panchromatic Capabilities

Based upon the analysis this SAG has performed, we find that a program aiming to emphasize broad wavelength coverage will provide maximum science return in the coming decades by re-establishing the long-term strategic goal of a panchromatic, community-driven, suite of space observatories. We recognize two basic classes of conceptual capability defining such a suite:

**Concurrency**: Overlap of the operational lifetimes of multiple facilities, to the greatest extent possible. The ability to quickly follow emerging threads of discovery at multiple wavelengths was an essential element of the scientific successes of the Great Observatories. Even when operational overlap is impossible, holding down temporal gaps between facilities to shorter than a decade (the typical time between major mission selections) is required to propel the field forward. The advantages of concurrency will only grow in the future, and become key enabling factors to meet the major scientific challenges envisioned in the coming decades. For example, some of the most interesting emerging scientific questions this study has identified will require simultaneous or near-simultaneous observations. These questions will be best addressed by emphasizing large fields of regard and rapid follow-up capabilities in the design and selection of future missions.

**Commensurability**: Offering complementary and comparable capabilities that can jointly address the most pressing scientific questions. Panchromatic coverage alone is a necessary but insufficient condition to address the major scientific needs of the coming decades. Facilities within a panchromatic suite must also offer commensurate scientific capabilities. These capabilities include sensitivity, mapping speed and sky coverage, as well as spatial, spectral, and temporal resolution. Although the degree of commensurability required varies substantially between scientific areas, and will continue to evolve in the next decade, the historical successes of the Great Observatories provide a powerful lesson — all delivered roughly commensurate capabilities (in one or more forms) that enabled the science outlined in the previous sections. For example, measuring the properties of some of the earliest galaxies through gravitational lensing (see **Section 2.2.1**) has provided a valuable example of commensurate capabilities among the Great Observatories that allowed researchers to infer star formation histories during the first few hundred Myr after the Big Bang. This and other examples demonstrate that it is not enough to cover wavelength space. The sensitivities, and/or spatial and spectral resolutions and mapping speeds must also be matched to the properties (spectral energy distribution, time variability, areal distribution, etc.) of individual sources that comprise the samples under study.

Establishing mission overlap across a panchromatic space observatory suite would be extremely challenging under even the best of circumstances. But the major scientific returns of the Great Observatories clearly indicate that a group of observatories providing some key elements of concurrency and commensurability can be highly responsive to new discoveries and the ever-changing landscape of astronomical research.

In order to meet these goals, a suite of community driven, "Giant Leap Observatories" (GLOs) that can make order of magnitude leaps in performance for a broad range of astrophysics questions could, like the original Great Observatories, consist of a mix of flagship and probe missions, augmented by explorers. In order to maximize scientific return, these observatories should be driven, to the greatest extent possible, by peer-reviewed, General Observer (GO) programs solicited from the astronomical community, and be supported by long-lived, interoperable archives and by observation planning and data analysis tools that evolve with the needs of the community.

## 3.3. Development Timescales & Costs: *the* Lessons *of the* Great Observatories

It is instructive to look at development timescales and costs of the original Great Observatories as they provide boundary conditions for the future development of a next generation of GLO's. The timescales and costs are summarized in Table 3.2 where we show, the name of the observatory, the year of the Decadal Survey recommendation, the year of launch, the cost accrued up to time of launch, and the cost inflated to 2019 dollars.

The original Great Observatories were a mix of one very large, one large, and two probe-sized mis-sions. The total cost of all the missions, in 2019 dollars, is about $14.4B. The time from Decadal Approval to launch does not account for the concept development time (pre-phase A), for which major efforts (~ 5 years) were mounted throughout the 1970's and 1980's. Based on these examples, we take two decades to be a canonical timescale for development of a strategic mission. Averaging the total cost over 20 years yields about $0.7B per year. This is slightly less than the peak spending rate during *JWST* phases C/D, which is almost exactly 1/2 the current annual astrophysics budget.

| Observatory | Endorsement | Launch | Development (years) | Cost at time of Launch | 2019 Cost |
|---|---|---|---|---|---|
| *Hubble* (*LST*) | 1972 | 1990 | 18 | $4.7B | $9.2B |
| *Compton* (*GRO*) | 1977 | 1991 | 14 | $0.6B | $1.2B |
| *Chandra* (*AXAF*) | 1982 | 1999 | 17 | $1.9B | $3.0B |
| *Spitzer* (*SIRTF*) | 1991/92 | 2003 | 11 | $0.7B | $1.0B |

Table 3.2. NASA's Great Observatories: Timescales & Costs. *The inflation-adjusted 2019 cost of each mission has only been roughly estimated using the average Consumer Price Index for each calendar year since the mission's launch. Costs for Chandra (AXAF) and Spitzer (SIRTF), respectively, are courtesy R. Holcombe and M. Werner (pvt. comm.). Note that although SIRTF was formally top-ranked in the 1991 Decadal Survey, it was canceled later that same year, and restarted in 1992 (Rieke 2006). The Hubble Space Telescope 2019 cost does not account for five subsequent servicing missions that have extended its mission beyond 30 years.*

Both *Chandra* and *Spitzer* saw major configuration changes during their development, before and after endorsement by the National Academy of Science, precipitated by evolving budget pressures and decisions within NASA. For *Chandra*, the number of grazing incidence mirror shells was reduced from 6 to

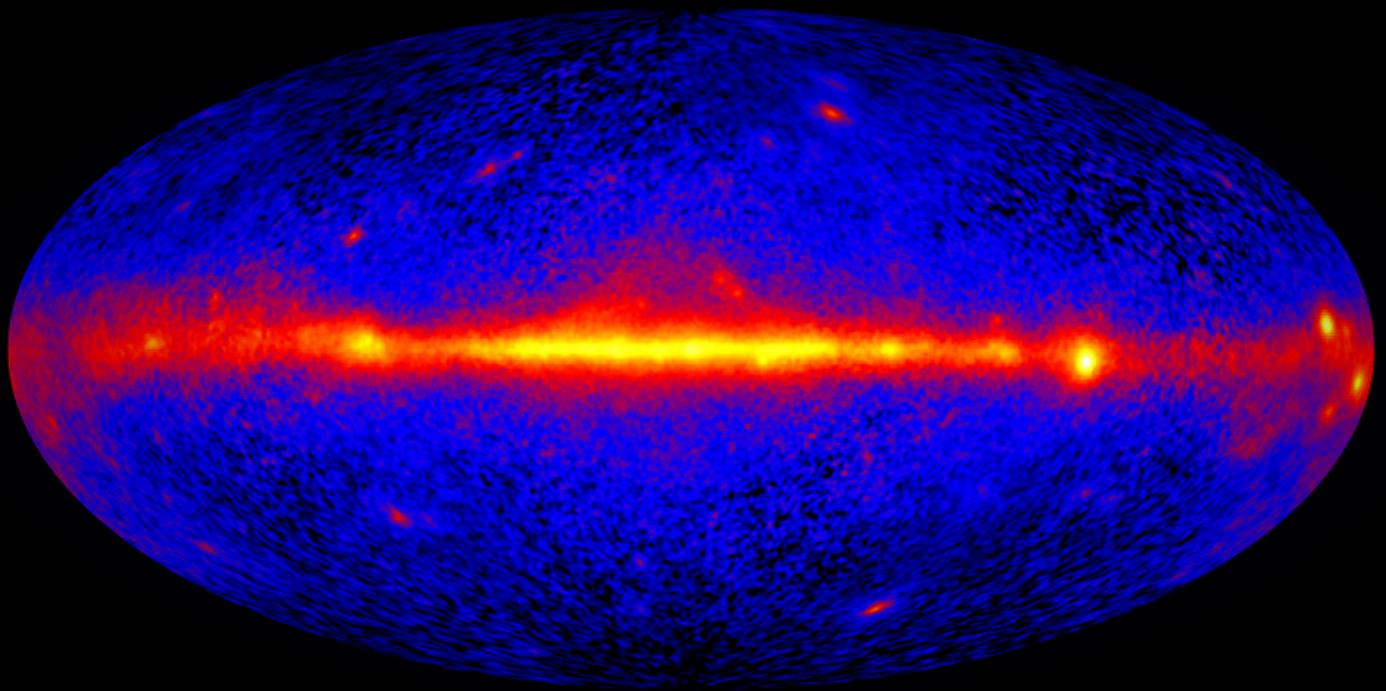

*Compton Gamma Ray Observatory EGRET 100 MeV All-Sky Map*
NASA & the EGRET TEAM

4, the design lifetime was reduced from 15 years to 5, and plans for servicing were dropped. For *Spitzer*, deployment was moved from a space shuttle to an expendable launch vehicle, the size of the primary was reduced to 0.85 m, plans for replenishment of the helium cryogen were dropped, and a warm launch and Earth trailing orbit were chosen to extend cryogenic lifetime (Rieke 2006). Nevertheless, both missions have exceeded their planned lifetime by healthy margins and delivered ground-breaking science, a tribute to the ingenuity of the development and operations teams.

Compelling strategic science goals in the next decades will invariably require an equally ambitious combination of large collecting area and sophisticated, state of the art instrumentation, which is likely to be expensive. The synergistic success of the Great Observatories shows the power of a unified program made up of diverse elements. Mission concepts that yield ambitious, order-of-magnitude advances at relatively modest cost are possible, as the Probe studies have demonstrated. In each field, or each wavelength regime, it is critical to identify the areas ripe for large advances, and target them for development and investment. A successful set of future GLO's would take advantage of these investments, potentially implemented within a number of missions of varied class.

The program-wide approach taken with the Great Observatories was a new way of organizing in the 1980s. With the final 2003 launch in the series NASA continued to plan in the spirit of the Great Observatories. The "Beyond Einstein" program, which took a less broad approach than the original Great Observatories and was focused on high-energy phenomena, was initiated following a 2003 study . Beyond Einstein consisted of two flagship missions, *Constellation-X* and *LISA* (the *Laser Interferometer Space Antenna*), and three "Einstein Probes": *Black Hole Finder Probe*, *Inflation Probe*, and the *Joint Dark Energy Mission* (JDEM). A funding wedge anticipated for 2009 did not materialize and consequently the individual missions were submitted to the Astro2010 Decadal, where many parts were supported but the concept of the original, coordinated program was lost. NASA Astrophysics has continued to support individual strategic missions like *JWST* and *Roman*, but an overarching science case, or set of cases, that directly supports a unified panchromatic program for the development of a next generation of GLOs has yet to be articulated.

A coherent, long-term science plan is an essential element to enable panchromatic exploration, the development of which could leverage the long history of successful international cooperation, such as the ESA partnership with NASA in *HST* as well as *JWST*. As our scientific ambitions grow, so do flagship mission costs and in resource-constrained environments, it makes sense to consider the role international partnerships can play in making our finite resources go farther. Indeed, NASA has been very successful in providing significant access to a wide range of international space science capabilities for modest investments (e.g. *Herschel*, *XMM-Newton*, *Planck*, *Euclid*, *LISA*). By bringing new, complementary capabilities, these observatories often enabled investigators to leverage the scientific output of the Great Observatories to make new discoveries (see [Chapter 2](#)). Some of our most reliable partners, such as ESA and JAXA, engage in planning processes that stretch over decades. The science outlined in the previous sections will be difficult to achieve, and US leadership is many areas of astrophysics may be lost, unless NASA participates in setting long-term, international priorities with its partners.

## 3.4.  Mitigating the Loss of Science & Community Viability in the Coming Decades

The astronomy community is composed of various disciplines that are organized naturally according the detection technologies required for the different wavelength intervals. These technical divisions lead to competition for resources, which can have a fracturing effect on community priorities. Yet, it is clear that there is an abundance of compelling science to be gained from concurrent, commensurate, panchromatic capabilities, as outlined in the previous sections. Indeed, the experience of the Great Observatories was that the intense competition between X-ray and infrared astronomers was greatly reduced by having a program designed to deliver groundbreaking observatories in both wavelength regimes (Harwit 2013).

The dilemma is how to strike a judicious balance that maintains panchromatic capabilities, as well as core scientific and technical competency across disciplines, while simultaneously expanding our capabilities to deliver the most compelling science. Broad community articulation of compelling, overarching science goals that acknowledge the importance of a panchromatic approach is a first step. Establishing a scientific consensus can initiate a flow-down of the scientific and technical requirements needed for achieving sustained panchromatic capabilities.

At current funding levels, NASA clearly cannot develop three ~$9B strategic missions simultaneously in the next decade, or even two. It might be possible, however, to achieve a transformational suite of observatories with a mix of costs like the Great Observatories. While there is great value in having facility class observatories with wide wavelength coverage and multiple observing modes, each component of the program need not "do it all". A program of missions would also naturally eliminate the "too big to fail" problem, as the failure of any one individual component, while certainly reducing the capability of the suite of observatories, might not offer an unacceptable risk.

The SAG-10 group has explored a number of options that could help NASA achieve the goal of sustained panchromatic capability and ultimately establish a set of new panchromatic observatories. The due diligence required to examine each option is well beyond the scope of this study. Instead, we describe them briefly here, along with some basic rationale. The options can be divided into several categories: (1) Mission Classes (2) Organizational Decisions (3) Leveraging New Technologies for Deploying Missions.

### 3.4.1 Mission Classes & Longevity

As with the Great Observatories, not all major advances require extremely expensive missions. Sometimes innovative technologies allow for large improvements in sensitivity (or other capabilities), within a much smaller cost envelope. Hence it makes both scientific and budgetary sense to plan for a range of mission costs. Currently approximately half of the annual astrophysics budget is devoted to the development of strategic missions, with the other devoted to existing mission support and the development of PI class missions (explorers, smallsats, cubesats, balloons, sounding rockets, etc.).

Developing guideline budget shares for large, medium and small missions may help enable future panchromatic coverage. To some degree ESA follows this path with L and M class missions, and the recently initiated S-class missions. NASA already has the Explorer budget line for missions up to $250M. Larger missions, exceeding ~$1B in cost, are considered strategic, to be defined by the Decadal survey. However, this leaves a sizeable gap. NASA has begun exploring ways to fill this gap with the initiation of Probe-class studies.

The Great Observatories were deployed in a staggered cadence spanning ~13 years. They successfully provided an unprecedented panchromatic presence and, due in large part to their long lifetimes, enabled concurrent operations. Although their costs varied by almost an order of magnitude, and their spatial resolutions, fields-of-view, and mapping speeds were not altogether commensurate, they did share many im-portant commensurate capabilities, particularly in terms of imaging and spectroscopic sensitivity, dynamic range, and the ability to measure the SEDs of astronomical objects over a huge range of cosmic time. While simultaneous operations were possible with the Great Observatories, they required a significant effort. The next generation of GLO's will have to deal robustly with increased demand for simultaneity, driven by the growth of time-domain astronomy from, for example, Rubin Observatory and GW alerts. These new science thrusts will likely require a nimble deployment of resources to maximize discovery space.

The number of operating missions at any given time is the product of the rate of launch and the mission lifetime. Given the slow launch rate of strategic missions, longevity has proven to be essential for maintaining commensurate and concurrent panchromatic capabilities. Given that development typically takes 10-20 years for strategic missions, and the deployment of flagships or probes may be one to a few per decade at most, mission lifetimes of a decade or more are essential for maintaining concurrency. This is particularly important for flagship missions, where lifetimes of 20 or more years are likely needed to achieve concurrency. For these missions long lifetimes can be achieved with careful planning and relatively modest increases in budgets. Longevity also paves the way for servicing and instrument replacement/upgrades in an otherwise aging observatory – the advantages of which have been repeatedly made clear with *Hubble* (see below). Rapidly deploying suites of small missions might increase concurrency, but would not alone be able to simultaneously offer broad wavelength coverage and paradigm-shifting capabilities demanded by the compelling scientific questions outlined in **Chapter 2**.

Small PI class missions, entrepreneurial and directed technical developments supported by the NASA Astrophysics Research *&* Analysis (APRA) and Strategic Astrophysics Technology (SAT) programs, and support for young investigator fellowships, all play a vital role in the sustenance of panchromatic capability. These programs provide avenues for airborne, suborbital, and space-based validation of new science, enabled by new technologies. Equally importantly, they train scientifically and technically literate workers experienced in negotiating schedule and cost trade-space between science and engineering. They also provide a means to maintain core-competency across disciplines, to establish mission prioritization metrics based on scientific and technical readiness, and to support more rapid buy down of component level risk, allowing for a more mature and credible estimation of the total cost of large missions.

With a range of small, medium and large mission options it might be possible to maintain panchromatic concurrency, while striving to develop a long-term plan to provide the commensurate capabilities embodied in the four flagship and 11 probe mission concept studies recently commissioned by NASA. Including the strategic goal of panchromatic coverage as an element of the Decadal prioritization metric is likely to increase science return while maintaining core-competency across disciplines for the next generation.

### 3.4.2   Organizational Decisions

Two organizational decisions could have a major influence on whether pan-chromatic coverage can be continued for the next 10 – 20 years:

**Mission Choices**:  A large flagship mission can do more science, both quantitatively and qualitatively, than any single, smaller mission. However, if they are designed to operate together, a mixed set of missions, whether they be more modest flagship missions, or a mix of flagships and probe missions supplemented by explorers, can deliver compelling and commensurate, multi-wavelength science at a competitive price. A decision to go forward with any program involves an opportunity cost. A useful methodology for comparing program choices is called "tensioning", where any program selection is made as an explicit choice between equal cost alternatives. Including the goal of panchromatic science as an evaluation criterion for missions in tension would expand the process beyond a comparison between individual missions and greatly increase the likelihood of achieving the level of science demonstrated by the Great Observatories. In this approach, the science gain from one large mission could be weighed against the sum of the science gains from a set of smaller missions at the same total cost, include opportunity costs from losses in panchromatic coverage.

**Cost Control**: Cost control has been a significant problem for flagship missions. Cost growth on this scale renders decadal strategic planning difficult and has in the past, prevented new missions from being started. It is essential that cost growth be contained on all missions, but this is especially true for flagship-class missions, as their overruns can have a large impact. For example, a 10% overrun on a $5B class mission could mean a small Probe-class mission lost, or the next flagship mission delayed, with a direct negative impact on concurrent panchromatic science. Cost realism is heavily reliant on the experience of a cognizant workforce of scientists and engineers, as well as disciplined management and implementation teams. Providing training programs to scientists on project management, systems engineering, and cost estimation methodologies, along with the availability of validated, standardized and non-proprietary cost estimation tools, could help produce greater cost realism.

### 3.4.3 Technological Advances

We are living in an era of dramatic change in our space capabilities, much of it driven by an outburst of commercial activity in space, usually called "NewSpace". NASA has already begun to take advantage of NewSpace e.g. by means of its CLPS (Commercial Lunar Payload Services) program. These technologies, as well as advances in detectors, software and communications, offer numerous opportunities to increase the launch cadence, capabilities and lifetimes of missions. Although exploring these options is beyond the scope of this report, we list several examples below. Leveraging these and other technological developments could help NASA maintain panchromatic coverage in astrophysics. These will require further analysis as they develop.

#### *CHEAPER HIGH CAPACITY LAUNCHERS*
New commercial launchers (e.g., SpaceX's Falcon-9 Heavy, and Blue Origin's New Glenn) are all either now available or will very likely be so by the mid-2020s. These new launchers bring several advantages, including lower cost/kg to orbit and larger diameter fairings. A lower cost to orbit could result in direct savings of order $100M or more per mission. Much lower cost to Low Earth Orbit (LEO) also allows a relaxation of the stringent mass constraints, leading to heavier, more powerful missions. Just as importantly, larger fairings (e.g. SLS Block 2B at 10m) expand the size of telescopes that can accommodated without resorting to complicated, and risky, deployable mechanisms.

#### *ON-ORBIT SERVICING AND ASSEMBLY IN LEO*
It is undeniable that servicing greatly extended the lifetime and dramatically boosted the scientific productivity of *HST*. Without servicing, *HST* would have been stuck with the spherical aberration, aging CCD detectors filled with traps and wracked by cosmic rays, obsolete one-dimensional UV detectors and no UV or IR imaging capability.

The five *Hubble* servicing missions showed that it is possible to not only correct errors in manufac-

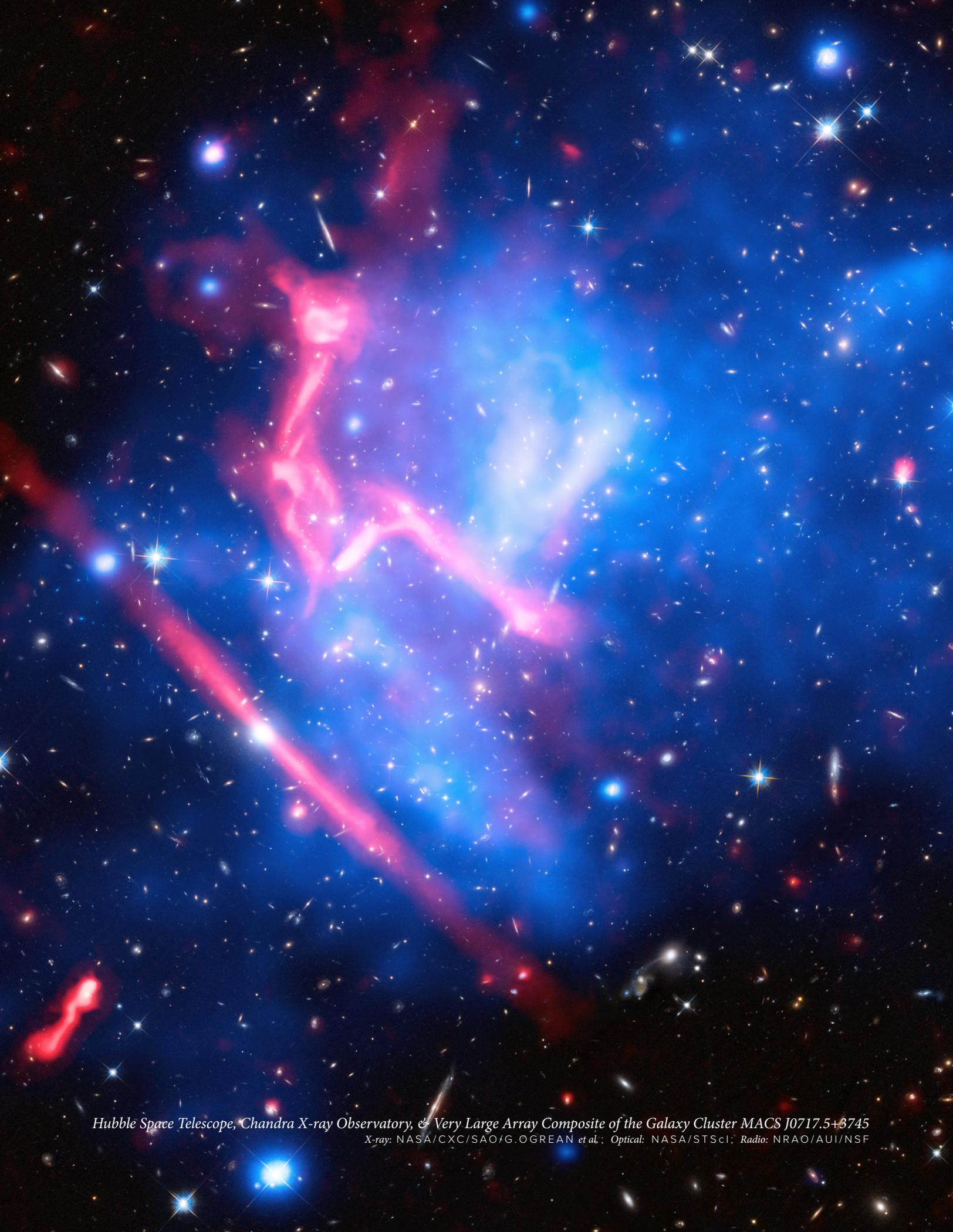

*Hubble Space Telescope, Chandra X-ray Observatory, & Very Large Array Composite of the Galaxy Cluster MACS J0717.5+3745*
X-ray: NASA/CXC/SAO/G.OGREAN et al.; Optical: NASA/STScI; Radio: NRAO/AUI/NSF

turing, but also to increase the power of a telescope and extend its life by decades. However, at over $1B per servicing, including the Shuttle launch, it was unaffordable to extend this model to other missions. Now, however, Congress has mandated that all future NASA missions be serviceable (National Aeronautics and Space Administration Transition Authorization Act of 2017). Building in long term servicing plans for strategic missions that can provide for an extended lifetime once new commercial and other governmental servicing assets are developed, appears prudent. The five servicing missions for *HST* were enabled by a partnership between human spaceflight and the astrophysics division. A similar model would likely have to be employed for future strategic missions.

The NASA Commercial Crew program has created a new, reusable, human spaceflight (HSF) capability that could be used to service LEO missions at roughly super-Explorer-class cost for an enhanced and longer-lived flagship. The advantages to the resulting science would seem clear given the *HST* experience and the declining performance of the remaining GOs. Although LEO has some disadvantages for astronomy, there are missions that could operate there with only minor cost to performance, in return for major gains in longevity and upgraded performance.

Servicing will be far easier if missions are designed for it in advance. For this to happen, modularity of construction is necessary. Robotic servicing, in Earth orbit or beyond (e.g. at L2) could also play a role as the field advances and becomes more adaptable to unforeseen difficulties like those encountered by the astronauts when servicing *Hubble*.

Many of the next generation observatories might be larger than the limits set by the launcher. In-space assembly would then be required. NASA has recognized this coming need by conducting an in-Space Assembled Telescope (iSAT) Study . The iSAT study explores the concept of multiple launches of cargo delivery vehicles and supervised, autonomous robotic arms to assemble 5-20 meter sized UV-NIR telescopes. It is also possible that commercial space stations will provide capable HSF-safe platforms where crews can assemble telescopes. Importantly for project management, on-orbit assembly may also lower the risks of failure on deployment. These stations will be in LEO at first but could well form part or all of the NASA Gateway or of an Earth-Moon L1 station.

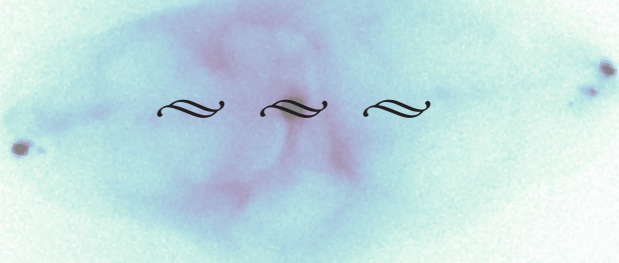

## 3.5. Summary

The NASA Great Observatories program has been an astonishing success, and has played a central role in the present Golden Age of Astronomy. With the aging and decommissioning of the Great Observatories, impending gaps in panchromatic coverage will inhibit progress in astrophysics in the next 1-2 decades. Forthcoming space-based facilities will only partially fill these gaps, and these facilities will be unable to forestall significant loss of scientific capability and progress. The gaps will also erode expertise to develop the technologies needed for future missions of all sizes. A program with a strategic goal of maintaining broad wavelength coverage will provide maximum science return in the coming decades by re-establishing a panchromatic, community-driven, suite of space observatories.

Science across NASA's astrophysics portfolio requires commensurate capabilities across wavelengths, including sensitivity, mapping speed and coverage, and spatial and spectral resolution (Chapter 2). The success of the Great Observatories (combined with other observatories such as *Fermi*, *Herschel* or *XMM-Newton*) was due, in large part, to their remarkable degree of commensurability, with different observatories sharing different combinations of capabilities. In particular, commensurate sensitivities, relative to the spectral energy distributions of astrophysical phenomena, are essential for multi-wavelength science.

Mission concurrency, i.e. overlap in operational lifetimes, is also essential for progress in most of astrophysics (Chapter 2). Concurrency allows discoveries made in one wavelength regime to be applied in multiple wavelength regimes, enabling rapid development and testing of models and leading to deeper astrophysical understanding. Time domain and multi-messenger astronomy, by definition, require concurrency to study rapidly evolving phenomena across the electromagnetic spectrum. Achieving mission concurrency to the largest degree possible will ensure future progress.

Well-supported **General Observer (GO) programs** were essential to the Great Observatories' success, and remain crucial to the success of future missions. These programs enable the community to respond to a changing scientific landscape, and in the case of the Great Observatories, quickly advance into new, rapidly growing areas. GO programs have also proven successful in expanding the range of science achieved on smaller missions.

**Multi-wavelength archives** are also increasingly important, enabling new science, serving as the foundation for future studies with new observatories, and setting baselines needed to characterize time variable phenomena. Continued support for the archives is an important component for maintaining panchromatic science, but not a substitute for the capability of making new panchromatic observations.

Operating multiple concurrent observatories requires a higher rate of deployment. The higher launch rate can come from a **mix of mission sizes** and costs. The Great Observatories spanned nearly an order of magnitude in cost, yet functioned together as a system to redefine astrophysics. Within the current budget envelope, a range of possibilities can be employed to maintain panchromatic coverage and deliver transformational gains in science, including mixing flagship and probe-scale missions, each with GO programs, as well as Explorer missions.

Maintaining panchromatic capability also requires **longevity**. The Great Observatories have demonstrated that missions can be operated effectively over multi-decade timespans, and that servicing could be a valuable way to maintain and upgrade capabilities. The use of servicing, particularly given emerging capabilities for human and robotic servicing, as well as in-orbit construction, may be viable routes for establishing long-term panchromatic capabilities with cutting-edge facilities.

Commensurate and concurrent panchromatic capabilities requires **strategic planning** to set mission sizes and capabilities, rates of mission deployments and mission lifetimes, ensure participation in international missions, and to consider opportunity costs incurred by losing capabilities in parts of the electromagnetic spectrum. Such planning may take advantage of possible routes to lower costs, including new advances in detectors and telescope technologies, higher capacity commercial launch vehicles, and modular spacecraft bus architectures.

The power of NASA's Great Observatories program was that it transcended individual missions and wavelength regimes. Success did not rely on a single flagship mission, but rather a suite of extremely capable observatories acting together to push outward the frontiers of astrophysics. This legacy points the way to a future where panchromatic capabilities are not just maintained, but enhanced, and the remarkable advances in our understanding of the Universe made possible by the Great Observatories are carried forward into the coming decades. A program of "Giant Leap Observatories" that builds on the model set by the Great Observatories can advance our understanding of the Universe far into the future.

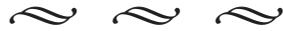

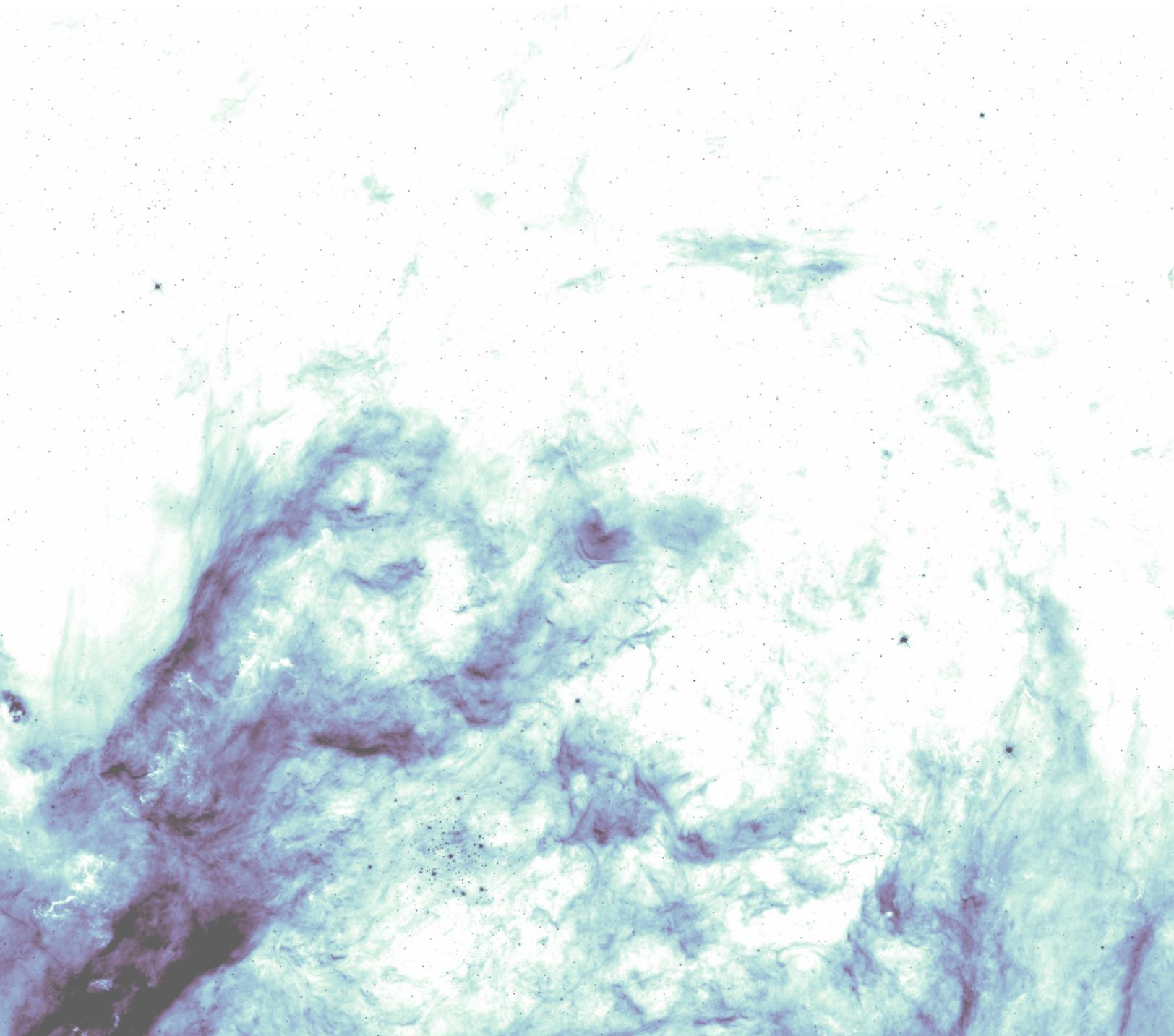

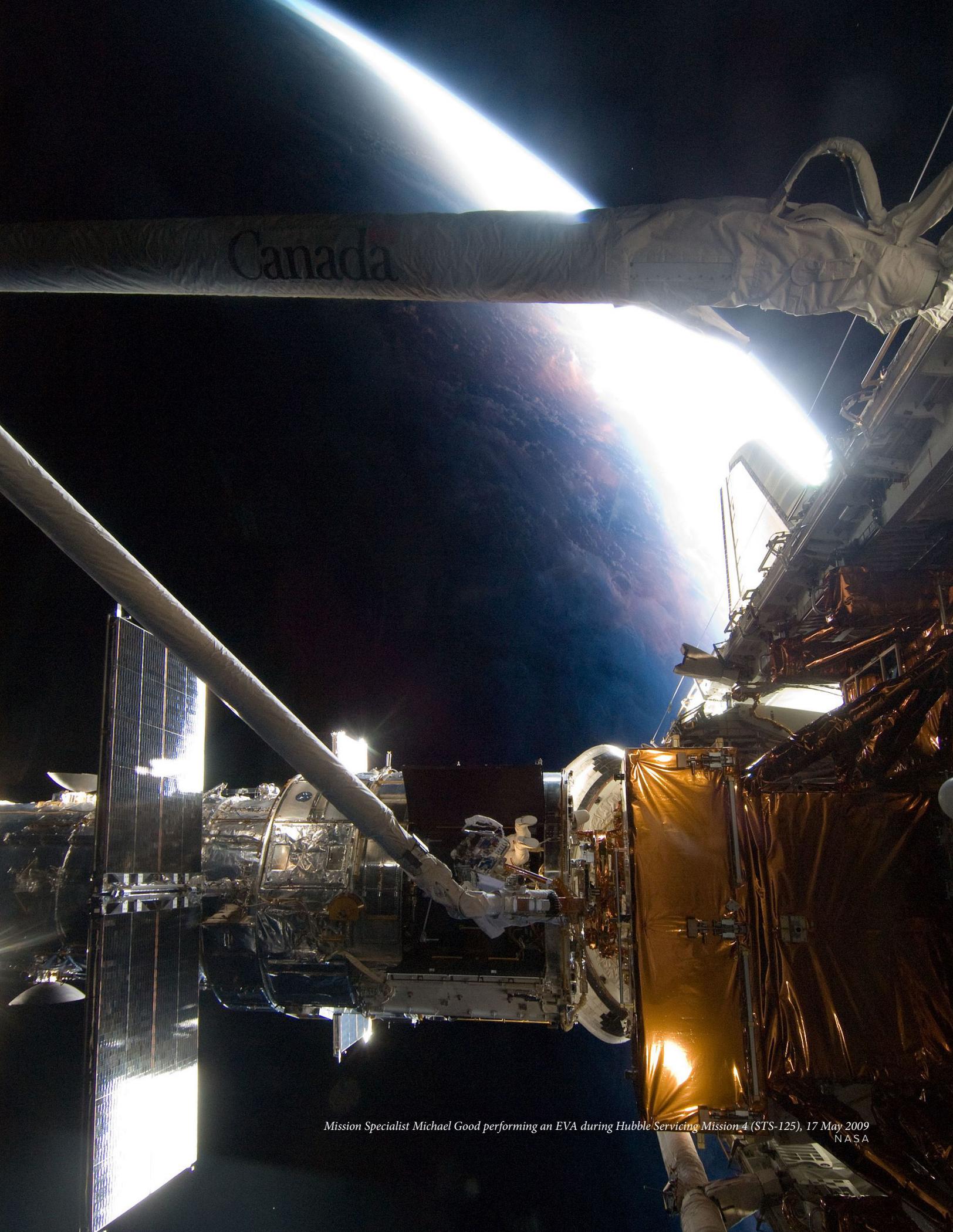

*Mission Specialist Michael Good performing an EVA during Hubble Servicing Mission 4 (STS-125), 17 May 2009*
NASA

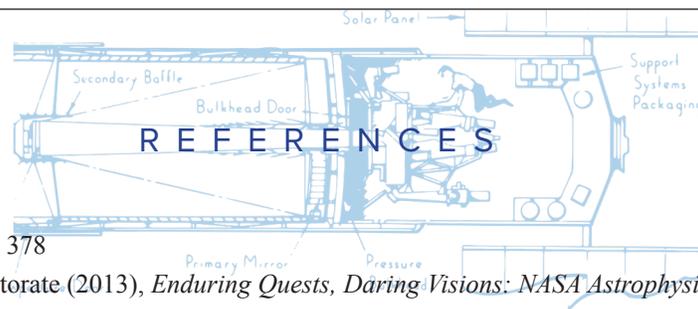

# REFERENCES

**Chapter 1:**
Harwit, M. 1975, QJRAS, 16, 378
NASA Science Mission Directorate (2013), *Enduring Quests, Daring Visions: NASA Astrophysics in the Next Three Decades*

**Section 2.1:**
Ackermann, *et al.* 2013, Science, 339, 807
Aliu, E., *et al.* 2014, ApJ, 770, 93
Andersson, B.-G., Lazarian, A., *&* Villancourt, J. E. 2015, ARA&A, 53, 501
Arzoumanian, D., *et al.* 2019, A&A, 621, 42
Barnes, A. T., *et al.* 2017, MNRAS, 469, 2263
Binder, B., *et al.* 2015, MNRAS, 51, 4471
Binns, W. R., *et al.* 2007, Space Science Reviews, 130, 439
Boyce, H., *et al.* 2019, ApJ, 871, 161
Boyer, M. L., *et al.* 2012, ApJ, 748, 40
Boyer, M. L., *et al.* 2015a, ApJS, 216, 10
Boyer, M. L., *et al.* 2015b, ApJ, 800, 51
Brown, A. G. A., de Geus, E. J., *&* de Zeeuw, P. T. 1994, A&A, 289, 101
Butler, M. J., *&* Tan, J. C. 2009, ApJ, 696, 484
Churazov, E., Khabibullin, I., Sunyaev, R., *&* Ponti, G. 2017, MNRAS, 471, 3293
Chuss, D. T., *et al.* 2019, ApJ, 872, 187
Dong, H., *et al.* 2011, MNRAS, 417, 114
Draine, B. T., Lee, H. M. 1984, ApJ, 285, 89
Draine, B. T. 2009, in ASPC Series, 414, 453
Dunham, M.M., *et al.* 2010, ApJ, 710, 470)
Dwek, E., Arendt, R. G., Bouchet, P., *et al.* 2008, ApJ, 676, 1029
Dwek, E., Galliano, F., *&* Jones, A. 2009, ASPC Series, 414, 183
Dwek, E., *et al.* 2010, ApJ, 722, 425
Elmegreen, B. G., *&* Scalo, J. 2004, ARA&A, 42, 211
Evans, N. J. II, *et al.*, 2009, ApJS, 181, 321
Egan, M. P., Carey, S. J., Price, S. D. 1999, ESA-SP, 427, 671
Federrath, C., *&* Klessen, R. S. 2013, ApJ, 763, 51
Figer, D. F., *et al.* 1999, ApJ, 525, 750
Fischer, W. J., *et al.* 2012, 756, 99
Fischer, W.J., *et al.* 2017, ApJ, 840, 69
Fischer, W. J., Safron, E., Megeath, S. T. 2019, ApJ, 872, 183
Fissel, L. M., Ade, P. A. R., Angilé, F. E. 2019, ApJ, 878, 110
Frank, K. A., *et al.* 2016, ApJ, 829, 40
Fullerton, A. W., Massa, D. L., *&* Prinja, R. K. 2006, ApJ, 637, 1025
Furlan, E., Fischer, W. J., Ali, B., *et al.* 2016, ApJS, 224, 5
Gall, C., Hjorth, J., *&* Andersen, A. C. 2011, A&ARv, 19, 43
Ginsburg, A., Bally, J., Barnes, A., *et al.* 2018, ApJ, 853, 171
Gull, T.R., *et al.* 2016, MNRAS, 462, 3196
Gutermuth, R. A., *et al.* 2009, ApJS, 184, 18
Gutermuth, R. A., *et al.* 2011, ApJ, 739, 84
Gordon, K., *et al.* 2019, BAAS, 51, 397
Hartigan, P. Holcomb, R., *&* Frank, A. 2019, ApJ, 876, 147
Hartmann, L., Herczeg, G., *&* Calvet, N. 2016, ARA&A, 54, 135
Heiderman, A., *et al.* 2010, ApJ, 723, 1019
Henning, Th., *et al.* 2010, A&A, 518, 95

Henshaw, J. D., *et al.* 2017, MNRAS, 464, 31
Hensley, B. S., *et al.* 2019, BAAS, 51, 224
Hirschi, R. 2008, in "Clumping in Hot Star Winds", Hamann, Oskinova, Feldmeier (eds), p. 9
Hoang, T., *&* Lazarian, A. 2016, ApJ, 831, 159
Hosek., M. W. Jr., *et al.* 2019, ApJ, 870, 44
Hsu, W.-H., *et al.* 2012, ApJ, 752, 59
Hsu, W.-H., *et al.* 2013, ApJ, 764, 114
Indebetouw, R., *et al.* 2014, ApJ, 782, L2
Ingleby, L., *et al.* 2014, ApJ, 790, 47
Jenkins, E. B. 2009, ApJ, 700, 1299
Johnson, S. P., *et al.*, MNRAS, 431, 662
Jones, A. P., *et al.* 2013, A&A, 558, 62
Kastner, J., H., *et al.* 2004, Nature, 430, 6998
Koepferl, C. M., Tobitaille, T. P., *&* Dale, J. E. 2017, ApJ, 849, 3
Karnath, N., *et al.* 2019, ApJ, 871, 46
Koyama, K., *et al.* 1996, PASJ, 48, 249
Kruijssen, J. M. D., *&* Longmore, S. N. 2013, MNRAS, 435, 2598
Krumholz, M. R., *et al.*, 2018, MNRAS, 477, 2716
Kuhn, M. A., *et al.* 2014, ApJ, 787, 107
Kuhn, M. A., *et al.* 2019, ApJ, 870, 32
Lada, C. J., Lombardi, M., Roman-Zuniga, C., *et al.* ApJ, 778, 133
Langer, W. D., *et al.* 2015, A&A, 576, 1
Larson, R. L., Evans, N. J. II, Green, J. D., *&* Yang, Y.-L. 2015, ApJ, 806, 70
Laycock, S. G. T., Maccarone, T. J., *&* Christodoulou, D. M. 2015, MNRAS, 452, 31
Lau, R.M., *et al.* 2019, ApJ, 878, 71
Lee, J. C., *&* Ravel, B. 2005, ApJ, 622, 970
Lombardi, M., Bouy, H., Alves, J., *&* Lada, C. J. 2014, A&A, 566, 45
Lopez, L., *et al.* 2019, BAAB, 51, 454
McCray, R., *&* Fransson, C. 2016, ARAA, 54, 19
Mac Low, M.-M., *&* Klessen, 2004, Reviews of Modern Physics, 76, 125
Manoj, P., Green, J. D., Megeath, S. T. 2016, ApJ, 831, 69
Maret, S., *et al.* 2009, ApJ, 698, 1244
Matsuura, M., *et al.* 2011, Science, 333, 1258
Matsuura, M., *et al.* 2015, ApJ, 800, 50
Megeath, S. T., *et al.* 2012, AJ, 144, 192
Megeath, S. T., *et al.* 2016, AJ, 151, 5
Meixner, M., *et al.* 2006, AJ, 132, 2268
Meixner, M., *et al.* 2010, A&A, 518, 71
Molinari, S., *et al.* 2010, A&A, 518, 100
Molinari, S., *et al.* 2011, ApJ, 735, 33
Moore, C.J., *et al.* 2019, MNRAS, 488, L94
Neufeld, D. A., *et al.* 2009, ApJ, 706, 170
Nowak, M. A., *et al.* 2012, ApJ, 759, 95
Nichols, J., *et al.* 2015, ApJ, 809, 133
Pabst, C., *et al.* 2019, Nature, 565, 618
Patnaude, D. J., *&* Fesen, R. A. 2009, ApJ, 697, 535
Pillai, T., Wyrowski, F., Carey, S. J., *et al.* 2006, A&A, 447, 929
Pokhrel, R., Gutermuth, R., Ali, B., *et al.* 2016, MNRAS, 461, 22
Ponti, G., De Marco, B., Morris, M. R. *et al.* 2015, MNRAS, 454, 1525
Povich, M. S., Townsley, L. K., Robitaille, T. P. 2016, 825, 125
Puls, J., Vink, J. R., *&* Najarro, F. 2008, A&ARv, 16, 209
Rui, N. Z., *et al.* 2019, ApJ., 877, 37
Russell, C. M. P., Wang, Q. D., *&* Cuadra, J. 2017, MNRAS, 464, 4958
Sana, H., *et al.* 2012, Science, 337, 444
Schneider, N., *et al.* 2013, ApJ, 766, 17

Stolovy, S., *et al.* 2006, Journal of Physics, Conf. Series, 54, 176
Stutz, A. M., Tobin, J. J., Stanke, T., *et al.* 2013, ApJ, 767, 36
Stutz, A. M., & Kainulainen, J. 2015, A&A, 577, 6
Stutz, A. M., & Gould, A. 2016, A&A, 590, 2
Wang, Q. D., Gotthelf, E. V., & Lang, C. C. 2002, Nature, 415, 148
Wang, Q. D., *et al.* 2010, MNRAS, 402, 895
Wang, Q. D., *et al.* 2013, Science, 341, 981
Weinstein, A. 2014, Nuclear Physics B, 256, 136
Yuan, Q., *et al.* 2018, MNRAS, 473, 306
Zeegers, S. T., Costantini, E., de Vries, C. P. 2017, A&A, 599, 117
Zubko, V., Dwek, E., Arendt, R. G. 2004, ApJS, 152, 211

**Section 2.2:**
Adelberger, K. L., *et al.* 2003, ApJ, 584, 45
Aird, J., *et al.* 2010, MNRAS, 401, 2531
Armus, L., *et al.* 1990, ApJ, 364, 471
Baldry, I. K., *et al.* 2008, MNRAS.388, 945
Begelman, M., Rossi, E. M., and Armitage, P,J., 2008, MNRAS, 387, 1649
Behroozi, P. S., *et al.* 2013, ApJ, 762L, 31
Bolatto, A. D., *et al.* 2008, ApJ, 686, 948
Bouwens, R. J., *et al.* 2014, ApJ, 793, 115
Bouwens, R. J., *et al.* 2015, ApJ, 803, 34
Bouwens, R. J., *et al.* 2016, ApJ, 831, 176
Bouwens, R. J., *et al.* 2017, ApJ, 843, 129
Casey, C. M., *et al.* 2018, ApJ, 862, 77
Casey, C. M., *et al.* 2018, ApJ, 862, 78
Cisternas, M., *et al.* 2011, ApJ, 726, 57
Civano, F., *et al.*, 2019, arXiv:1903.11091
Coe, D., *et al.* 2013, ApJ, 762, 32
Coe, D., *et al.* 2015, ApJ, 800, 84
Croton, D. J., *et al.* 2006, MNRAS, 365, 11
Croxall, K. V., *et al.* 2013, ApJ, 777, 96
Delvecchio, I., *et al.* 2014, MNRAS, 439, 2736
di Matteo, T., *et al.* 2005, Nature, 433, 604
Diaz-Santos, T., *et al.* 2017, ApJ, 846, 32
Duncan, K. & Conselice, C. J. 2015, MNRAS, 451, 2030
Elbaz, D., *et al.* 2010, A&A, 518L, 29
Elbaz, D., *et al.* 2011, A&A, 533A.119
Fabian, A. C. 2012, ARA&A, 50, 455
Fernandez-Ontiveros, J.A., *et al.* 2016, ApJS, 226, 19
Ferkinhoff, C., *et al.* 2015, ApJ, 806, 260
Ferrarese, L. & Merritt, D. 2000, ApJ, 539, L9
Finkelstein, S. L., *et al.* 2012, ApJ, 756, 164
Finkelstein, S. L., *et al.* 2013, Nature, 502, 524
Finkelstein, S. L., *et al.* 2015, ApJ, 810, 71
Fischer, T. C., *et al.* 2010, AJ, 140, 577
Fornasini, F. M., *et al.* 2018, ApJ, 865, 43
Gebhardt, K., *et al.* 2000, ApJ, 539, L13
Granato, G. L, *et al.* 2004, ApJ, 600, 580
Hashimoto, T., *et al.* 2018, Nature, 557, 392
Hayward, C. C. & Hopkins, P. F. 2017, MNRAS, 465, 1682
Heckman, T. L., *et al.* 1990, ApJS, 74, 833
Heckman, T. L., *et al.* 2000, ApJS, 129, 493
Henden *et al.*, 2019, MNRAS, 479, 5385.

Henriques, B. M. B., *et al.* 2015, MNRAS, 451, 2663
Henriques, B. M. B., *et al.* 2019, MNRAS, 485, 3446
Hickox, R. C. *&* Alexander, D. M. 2018, ARA&A, 56, 625
Hopkins, P. F., *et al.* 2006, ApJS, 163, 1
Hopkins, P. F., *et al.* 2014, MNRAS, 445, 581
Infante, L., *et al.* 2015, ApJ, 815, 18
Izotov, Y. I., *et al.* 2016, MNRAS, 461, 3683
Izotov, Y.I., *et al.* 2018, MNRAS, 474, 4514
Johnson, T. L., Rigby, J. R., Sharon, K., *et al.* 2017, ApJL, 843, 21
Kauffmann, G. *&* Haenalt, M. 2000, MNRAS, 311, 576
Kewley, L.J., *&* Ellingson, S.L. 2008, ApJ, 681, 1183
Kirkpatrick, A., *et al.* 2017, ApJ, 849, 111
Kormendy, J. *&* Richstone, D. 1995, ARA&A, 33, 581
Kornei, K. A., *et al.* 2012, ApJ, 758, 135
Laha, S., *et al.* 2016, MNRAS, 457, 3896
Livermore, R. C., *et al.* 2017, ApJ, 835, 113
McConnell, N. J. *&* Ma, C.-P. 2013, ApJ, 764, 184
Madau, P. *&* Dickinson, M. 2014, ARA&A, 52, 415
Magorrian, J., *et al.* 1998, AJ, 115, 2285
Martin, C. L. 1999, ApJ, 513, 156
Martin, C. L., *et al.* 2012, ApJ, 760, 127
Mezcua, M. *et al.* 2016, ApJ, 817, 20
Moster, B. P., *et al.* 2010, ApJ, 710, 903
Nestor, D. B., *et al.* 2011, ApJ, 736, 18
Noeske, K. G., *et al.* 2007, ApJ, 660, 43
Oesch, P. A., *et al.* 2014, ApJ, 786, 108
Oesch, P. A., *et al.* 2016, ApJ, 819, 129
Paggi, A., *et al.* 2016, ApJ, 823, 112
Peeples, M. S., *et al.* 2014, ApJ, 786, 54
Ponti, G., *et al.* 2019, Natur, 567, 347
Rudie, G. C., *et al.* 2012, ApJ, 750, 67
Salmon, B., *et al.* 2015, ApJ, 799, 183
Sanders, R. L., *et al.* 2015, ApJ, 799, 138
Sandstrom, K. M., *et al.* 2010, ApJ, 715, 701
Shankar, F., *et al.* 2009, ApJ, 690, 20
Shapley, A. E., *et al.* 2003, ApJ, 588, 65
Siana, B., *et al.* 2010, ApJ, 723, 241
Smit, R., *et al.* 2012, ApJ, 756, 14
Smit, R., *et al.* 2014, ApJ, 784, 58
Smit, R., *et al.* 2015, ApJ, 801, 122
Smith, J.D.T., *et al.* 2019, BAAS, 51, 400
Smith, Sijacki *&* Sijing, 2019 MNRAS, 485, 3317
Somerville, R. S. *&* Dave, R. 2015, ARA&A, 53, 51
Somerville, R. S., *et al.* 2015, MNRAS, 453, 4337
Song, M., *et al.* 2016, ApJ, 826, 113
Spilker, J. S., *et al.* 2018, Science, 361, 1016
Springel, V., *et al.* 2005, MNRAS, 361, 776
Stark, D. P., *et al.* 2013, ApJ, 763, 129
Steidel, C., *et al.* 1994, ApJ, 437L, 75
Steidel, C. C., *et al.* 2018, ApJ, 869, 123
Strickland, D. K., *et al.* 2000, AJ, 120, 2965
Strickland, D. K., Heckman, T. M., Weaver, K. A., Dahlem, M. 2000, AJ, 120, 2965
Tombesi, F., Sambruna, R. M., Reeves, J. N., *et al.* 2010, ApJ, 719, 700.
Sturm, E., *et al.* 2011, ApJ, 733L, 16
Tombesi, F., *et al.* 2010, ApJ, 719, 700

Trakhtenbrot, B., Netzer, H., Lira, P., and Shemmer, O., 2011, ApJ, 730, 7
Tremonti, C. A., *et al.* 2004, ApJ, 613, 898
Tumlinson, J., *et al.* 2017, ARA&A, 55, 389
Turner, M. L., *et al.* 2014, MNRAS.445, 794
Vanzella, E., *et al.* 2012, ApJ, 751, 70
Vanzella, E., *et al.* 2016, ApJ, 825, 41
Veilleux, S., *et al.* 2005, ARA&A, 43, 769
Veilleux, S., *et al.* 2013, ApJ, 776, 27
Volonteri, M., 2010, ARA&A, 18, 279
Werk, J. K., *et al.* 2014, ApJ, 792, 8
Whitaker, K. E., *et al.* 2017, ApJ, 850, 208
Xue, Y. Q., *et al.* 2011, ApJS, 195, 10
Yung, L. Y., *et al.* 2019, MNRAS, 483, 2983
Yung, L. Y., *et al.* 2019, arXiv, 1901.05964
Zavala, J. A., 2018, ApJ, 869, 71
Zheng, W., *et al.* 2014, ApJ, 795, 93
Zitrin, A., *et al.* 2015, ApJ, 810, L12

**Section 2.3:**

Andrews, S., *et al.* 2011, ApJ, 732, 42
Andrews, S., *et al.* 2019, ApJ, 869, L41
Avenhaus, H., *et al.* 2018, ApJ, 863, 44
Bae, J., *et al.* 2018, ApJ, 864, L26
Bergin, E., *et al.* 2013, ASPC, 476, 185
Brittain, S., *et al.* 2014, ApJ, 791, 136
Charbonneau, D., *et al.* 2002, ApJ, 568, 377
Charbonneau, D., *et al.* 2005, ApJ, 626, 523
Deming, D., *et al.* 2005, Nature, 434, 740
Espaillat, C., *et al.*, 2014, Protostars *&* Planets VI, 497 (U. Arizona: Tucson)
Fischer, D., *et al.* 2014, Protostars *&* Planets VI, 715 (U. Arizona: Tucson)
France, K., *et al.* 2017, ApJ, 844, 169
Fulton, B.J., *et al.* 2017, AJ, 154, 109
Gail, H.-P., 2002, A&A, 390, 253
Grant, S., *et al.* 2018, ApJ, 863, 13
Janson, M., *et al.* 2012, ApJ, 747, 116
Kalas, P., *et al.* 2008, Science, 322, 1345
Knutson, H., *et al.* 2007, Nature, 447, 183
Kreidberg, L., *et al.* 2014, 505, 69
Lagrange, A.-M., *et al.* 2009, A&A, 493, L21
Lichtenberg, T., *et al.* 2019, Nature Ast, 3, 307
Lowrance, P., *et al.* 1999, ApJ, 512, L1
Marois, C., *et al.* 2008, Science, 322, 1348
McClure, M., *et al.* 2013, ApJ, 775, 114
Meyer, M., *et al.* 2007, Protostars *&* Planets V, 573
Miotello, A., *et al.* 2017, A&A, 599, 113
Neufeld, D., *et al.* 1994, ApJ, 428, 170
Neugebauer, G., *et al.* 1984, Science, 224, 14
Pontoppidan, K. *et al.* 2010, ApJ, 720, 887
Schneider, G. *et al.* 2014, AJ, 148, 4
Sing, D. *et al.* 2016, Nature, 529, 59
Teague, R. *et al.* 2018, ApJ, 860, L12
Tinetti G., *et al.* 2007, Nature, 448, 169
Wyatt, M. 2008, ARAA, 46, 339
Zhang, Z. *et al.* 2018, AJ, 156, 178

Zhu, Z. *et al.* 2012, ApJ, 755, 6

**Section 2.4:**
Abazajian, K., Fuller, G. M., & Patel, M. 2001, PhRv, 64, 023501
Abbott, B. P., *et al.*, 2017a, ApJ, 848, 12
Abbott, B. P., *et al.*, 2017b, Nature, 551, 85
Abdo, A. A., *et al.*, 2009, Nature, 462, 331
Abeysekara, A. U., *et al.* 2017, ApJ, 842, 85
Ackermann, M., *et al.*, 2015, PhRvL, 115, 1301
Ackermann, M., *et al.*, 2017, ApJ, 840, 43
Allen, S. W., Evrard, A. E., & Mantz, A. B., 2011, ARA&A, 49, 409
Antonioli, P., *et al.*, 2004, NJP, 6, 114
Barkana, R. & Loeb, A., PhR, 349, 125 (2001), ArXiv: astro-ph/0010468
Bertone, G., Hooper, D., & Silk, J. 2005, PhR, 405, 279
Bloom, J. S., Kulkarni, S. R., & Djorgovski, S. G., 2002, AJ, 123, 1111
Bulbul, E., *et al.* 2014, ApJ, 789, 23
Boyarsky, A., Ruchayskiy, O., Iakubovskyi, D., & Franse, J., 2014, PhRvL, 113, 1301
Burrows, D. B., *et al.*, 2011, Nature, 476, 421
Camera, S., *et al.*, 2017, MNRAS, 464, 4747
Campana, S., *et al.*, 2006, Nature, 442, 1008
Chornock, R., *et al.* 2013, ApJ, 774, 26
Chornock, R., Berger, E., Fox, D. B., Fong, W., Laskar, T., & Roth, K. C., 2014, arXiv: 1405.7400
Civano, F. *et al.* 2019, arXiv 1903.11091.
Clowe, D., *et al.* 2006, ApJL, 648, L109
Conlon, J.P., Day, F., Jennings, N., Krippendorf, S., & Muia, F. 2018, MNRAS, 473, 4
Crocker, R. M., & Aharonian, F., 2011, PhRvL, 106, 1102
Cucchiara, A., *et al.* 2015, ApJ, 804, 51
Dobler, G., Finkbeiner, D. P., Cholis, I., Slatyer, T., & Weiner, N., 2010, ApJ, 717, 825
Dodelson, S., & Widrow, L. M. 1994, PhRvL, 72, 17
Finkbeiner, D. P., 2004, ApJ, 614, 186
Fong, W. F., Berger, E., & Fox, D. B., 2010, ApJ, 708, 9
Fox, D. B., *et al.*, 2005, Nature, 437, 845
Galama, T., *et al.*, 1998, Nature, 395, 6703
Gezari, S., *et al.*, 2012, Nature, 485, 217
Guo, F., Mathews, W. G., Dobler, G., & Oh, S. P., 2012, ApJ, 756, 182
Harvey, D., Massey, R., Kitching, T., Taylor, A., & Tittley, E., 2015, Science, 347, 1462
Hatt, D., *et al.*, 2018, ApJ, 866, 145
Hooper, D., & Goodenough, L., 2010, PhLB, 697, 412
Huang, C.-D., *et al.*, 2018, ApJ, 857, 67
IceCube Collaboration, 2018, Science, 361, 147
IceCube Collaboration, *et al.*, 2018, Science, 361, 1378
IceCube Gen-2 Collaboration, 2014, arXiv: 1412:5106
Jang, I. S., & Lee, M. G., 2017, ApJ, 836, 74
Jeltema, T. E., & Profumo, S., 2008, ApJ, 686, 1045
Jiang, N., Dou, L., Wang, T., Yang, C., Lyu, J., & Zhou, H., 2016, ApJL, 828, L14
KM3NeT Collaboration, 2018, arXiv: 1810.08499
Kataoka, J., *et al.* 2018, Galaxies 2018, 6, 27
Keivani, A., *et al.*, 2018, ApJ, 864, 84
Kelly, P. L., *et al.*, 2015, Science, 347, 1123
Kouveliotou, C., *et al.*, 1998, Nature, 393, 235
Krolik, J., Piran, T., Svirski, G., & Cheng, R. M., 2016, ApJ, 827, 127
LIGO Lab, 2019, https://www.ligo.caltech.edu/news/ligo20190214
Lacki, B. C., 2014, MNRAS, 444, L39
Lee, S. K., *et al.* 2016, PRL, 116, 051102.

Magorrian, J. *et al.* 1998, AJ, 115, 2285
Mao, S. *&* Schneider, P., 1998, MNRAS, 295, 587
Marsh, D. J., 2016, PhR, 643, 1
Massey, R., *et al.*, 2007, Nature, 445, 286
Metcalf, R. B., *&* Madau, P., 2001, ApJ, 563, 9
Morselli, A. *et al.*, 2011, Il Nuovo Cimento (2011) 34:311
Oh, S.-P. *&* Z. Haiman, 2002, ApJ, 569, 558
Oukbir, J. *&* Blanchard, A. 1992, A&A, 262, L21
Perlmutter, S., *et al.*, 1999, ApJ, 517, 565
Petrovic, J *et al.*, 2014, JCAP, 10, 52.
Planck Collaboration, 2015, A&A, 594, A25
Planck Collaboration, 2018, arXiv: 1807.06209
Ponti, G., *et al.*, 2019, Nature, 567, 347
Radice, D., Perego, A., Zappa, F., *&* Bernuzzi, S., 2018, ApJL, 852, L29
Razzaque, S., 2013, PhRvD, 88, 081302
Rees, M. J., 1984, ARA&A 22, 471
Riess, A. G., *et al.*, 1998, AJ, 116, 1009
Riess, A. G., *et al.*, 2004, ApJ, 607, 665
Riess, A. G., Casertano, S., Yuan, W., Macri, L. M., *&* Scolnic, D. 2019, ApJ in press, ArXiv: 1903.07603
Sathyaprakash, B., *et al.*, 2012, CQG, 29, 124013
Shang, C., Bryan, G. L., *&* Haiman, Z., 2010, MNRAS 402, 1249
Soderberg, A. M., *et al.*, 2008, Nature, 453, 469
Su, M., Slatyer, T. R., *&* Finkbeiner, D. P., 2010, ApJ, 724, 1044
Totani, T., *et al.* 2006, PASJ, 58, 485
Umetsu, K., *et al.*, 2014, ApJ, 795, 163
Vikhlinin, A. 2010, PNAS, 107, 7179

**Chapter 3:**

Harwit, M. (2013). *In Search of the True Universe: The Tools, Shaping and Cost of Cosmological Thought*. New York: Cambridge University Press
NASA Science Mission Directorate (2013), Enduring Quests, Daring Visions: NASA Astrophysics in the Next Three Decades
Rieke, G.H. (2006). *The Last of the Great Observatories: Spitzer and the Era of Faster, Better, Cheaper at NASA*. Tucson: The University of Arizona Press

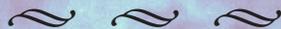

# ADDITIONAL CREDITS

COVER IMAGE
*The Great Observatories image of Messier 1, the Crab Nebula. The Crab is the remnant of a supernova explosion in the constellation of Taurus, recorded by Chinese astronomers in 1054 AD. In this composite image, high energy X-rays seen with the Chandra X-Ray Observatory are shown in blue, optical emission lines from ionized oxygen and hydrogen seen with the Hubble Space Telescope are shown in red and yellow, and infrared emission from ionized oxygen and warm dust seen with the Spitzer Space Telescope is shown in purple.*

SPECIAL THANKS *to*
*Martin Harwit, George Helou, Margaret Meixner, Harvey Tananbaum, & Mike Werner*

REPORT GRAPHIC DESIGN *&* LAYOUT *by* GRANT TREMBLAY

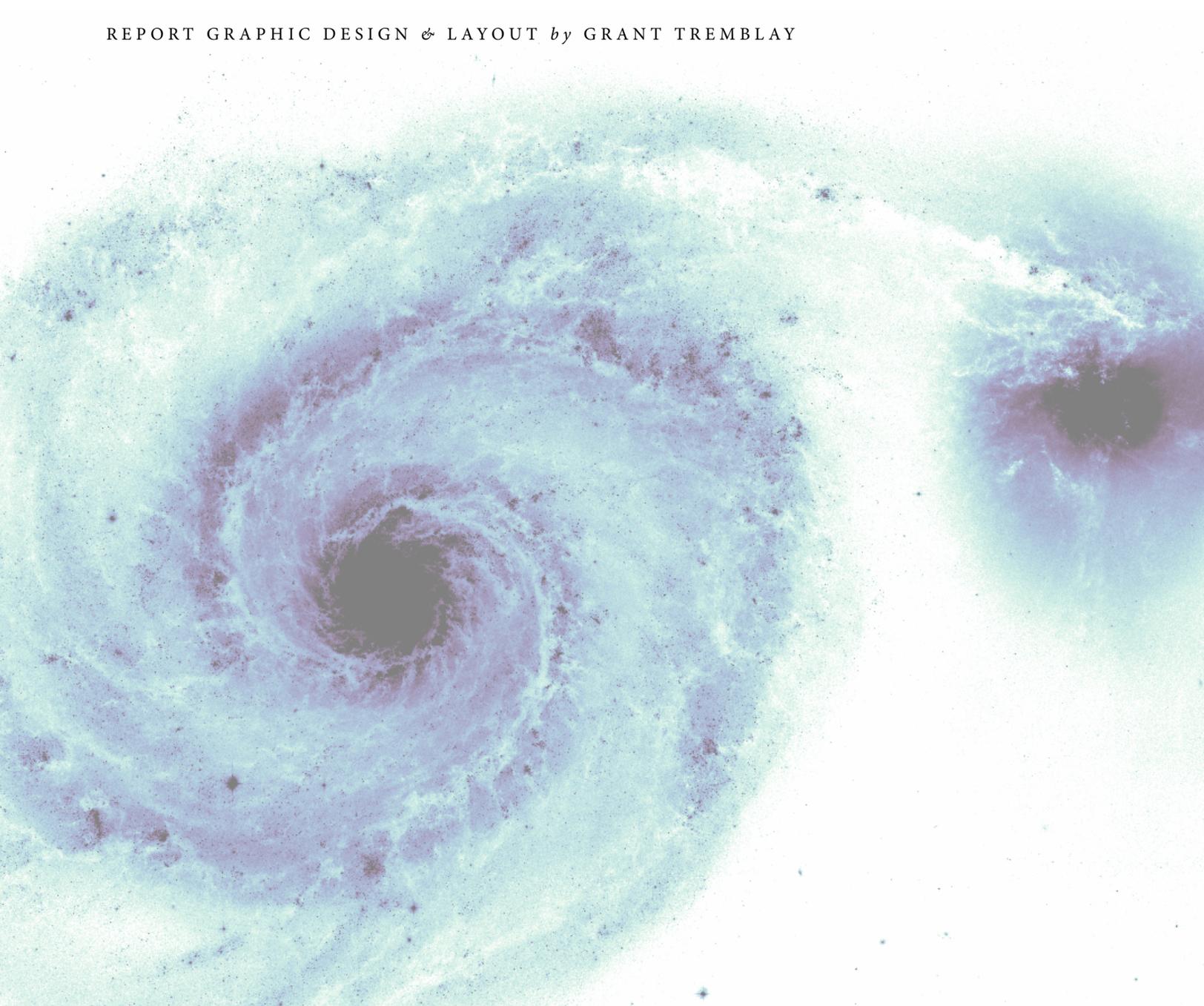

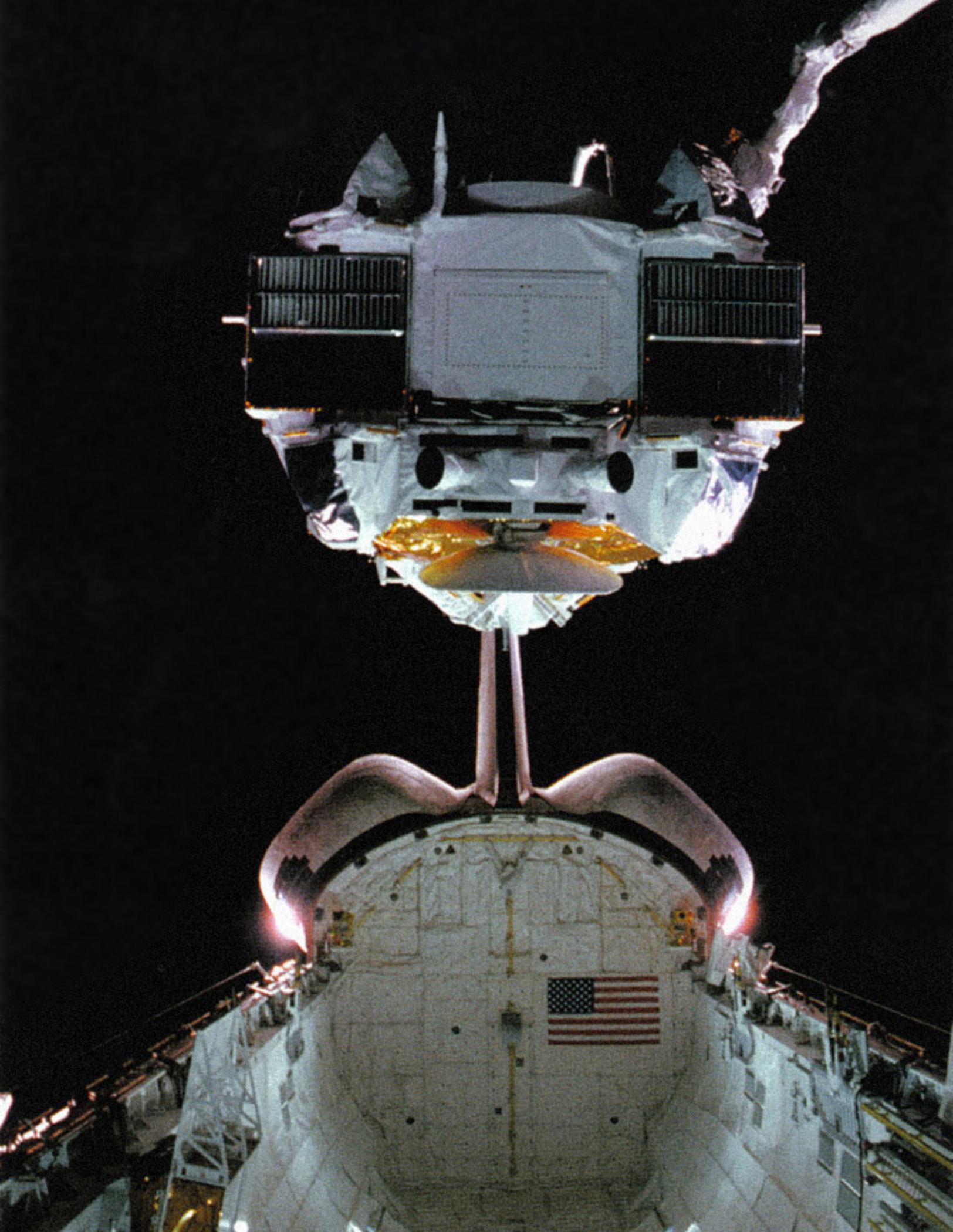